%% file: phdfalessiarxiv.tex
\documentclass[a4paper,11pt, openright]{book}
\usepackage[utf8]{inputenc}
\usepackage[T1]{fontenc}
\usepackage{etex}
\usepackage{mhchem}
\usepackage{bm}
\usepackage{stmaryrd}
\input{./TeX_style}
\usepackage{lipsum}
\newcommand{\gav}[1]{\left \langle #1\right \rangle}
\newcommand{\sav}[1]{\left \langle #1\right \rangle_{\psi}}
\newcommand{\vav}[1]{\left \langle #1\right \rangle_{v}}
\newcommand{\wb}[1]{\bm #1}
\newcommand{\bs}[1]{\hat{I}_{#1}}
\begin{document}
\begin{titlepage}
\begin{center}
\begin{figure}[htbp]
\begin{center}
\includegraphics[scale=0.2]{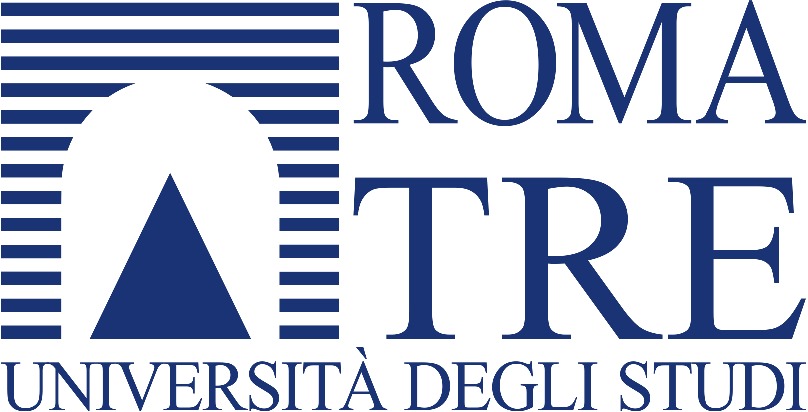}
\end{center}
\end{figure}
\rule[0.1cm]{15.8cm}{0.1mm}
\rule[0.5cm]{15.8cm}{0.6mm}
{\small{\bf UNIVERSIT\`A DI ROMA TRE\\
FACOLT\`A DI SCIENZE MATEMATICHE, FISICHE E NATURALI\\
Dipartimento di Matematica e Fisica}}
\end{center}
\vspace{15mm}
\begin{center}
{\Huge{ \textsc{Gyrokinetic theory for particle transport in fusion plasmas}}}\\
\vspace{10mm} {\large{\bf Ph.D. Thesis}}
\end{center}
\vspace{80mm}
\par
\noindent
\begin{minipage}[t]{0.45\textwidth}
{\large{\bf Relatore:\\
Dott. Fulvio Zonca}}
\end{minipage}
\hfill
\begin{minipage}[t]{0.45\textwidth}\raggedleft
{\large{\bf Studente:\\
Matteo Valerio Falessi}}
\end{minipage}
\vspace{30mm}
\begin{center}
{\large{
XXIX Ciclo di Dottorato}}
\end{center}
\end{titlepage}
\pagenumbering{roman}

\begin{flushright}
  \emph{to my parents \dots } \\
\vspace{10mm}
  \emph{\dots to Paolo.}

\end{flushright}
\newpage
\begin{flushright}
 \textcolor{white}{1line} \\
 \textcolor{white}{1line} \\
\emph{Geometry is nothing more than a branch of physics; geometrical truth are not essentially different from physical ones in any aspect and are established in the same way}
 \medskip\\
\textsc{David Hilbert (1862-1943)}
\end{flushright}



\setcounter{page}{1}
\tableofcontents

\clearpage
\pagenumbering{arabic}
\chapter{Introduction}
\label{sec-1}
\section{Aim of the thesis}
\label{sec-1-2}
The aim of this thesis is the self-consistent study of transport processes in a thermonuclear plasma on the energy confinement time scale.
A precise and quantitative definition of what this means essentially implies discussing and understanding the spatiotemporal scales
of nonlinear evolution of plasma profiles; that is, of what common wisdom defines as plasma equilibrium. 
A practical and effective approach to this problem intuitively yields to defining suitable time and spatial averages.
In fact, plasma equilibrium should evolve slowly in some sense, while cross field transport (across equilibrium
magnetic flux surfaces) should cause the distortion of plasma profiles on a sufficiently long length scale.
In this thesis work, we will show that important aspects of underlying physics processes arise in providing
answer to questions such as ``how slow (or fast) does the plasma equilibrium actually evolve"; and
``what are the characteristic length scales of equilibrium distortions''. 

Transport processes are due to Coulomb collisions between particles and to plasma turbulence which is spontaneously excited in magnetic confinement experiments typically from drift wave instabilities. Although the results discussed in this work are derived assuming drift wave turbulence in magnetized fusion plasmas, they are applicable and readily generalized to a broad class of electromagnetic fluctuations, including magnetohydrodynamic instabilities and Alfv\'en waves  \cite{chen2016physics}. Transport processes are characterized by length and time scales which can be very different. Transport characteristic scales of variation, meanwhile, can be ordered using the drift parameter, i.e. $\delta = \rho /L$, where $\rho$ is the Larmor radius of the particles constituting the plasma and $L$ is the length-scale of equilibrium profiles. The turbulent spectrum is dominated by frequencies $\sim \delta \Omega$, where we have introduced the cyclotron frequency $\Omega = e B / m c$, and by the    characteristic non linear evolution time $\sim \delta^{-2} \Omega^{-1}$ which is usually shorter with respect to the transport time scale $\sim \delta^{-3} \Omega^{-1}$:\cite{frieman1982nonlinear,hinton1976theory}. For this reason, the theory of plasma transport deals mainly with turbulence and induced fluxes which modify the equilibrium profiles on short time-scales, while the study of collisional fluxes, which give significant modifications of the profiles on longer time-scales, usually is approached by means of a phenomenological model. A limited amount of works \cite{abel2013multiscale,plunk2009theory,sugama1996transport,balescu1990anomalous,shaing1988neoclassical} have studied in a self-consistent way collisional transport \cite{hinton1976theory,balescu1988transport,sonnino2008nonlinear} and fluctuation induced transport \cite{frieman1982nonlinear,brizard2007foundations,sugama1996transport}, anyway these studies are respectively characterized by a simplified approach in the description of turbulence or collisions. The motivation of this thesis work stems from the fundamental importance of the self-consistency of the adopted description in order to understand transport processes on the energy confinement (transport) time scale because of the mutual interaction between collisions and turbulence. Therefore it is crucial in order to predict fluxes of particle and energy in a modern magnetic confinement experiment. Again, as anticipated above, this thesis work will illuminate the importance of spatiotemporal scales of nonlinear
distortions of plasma equilibrium connected with transport processes. In particular, the emphasis will be on those structures
that are long lived, i.e. are not rapidly damped by collisionless processes such as Landau damping, and are 
dissipated only by collisions. These structures, dubbed zonal structures, can be considered as corrugations of 
plasma profiles \cite{chen2007theory} and, generally, have a phase space counterpart \cite{zonca2015nonlinear}  that amplifies the system deviation from a local
thermodynamic equilibrium. In this thesis work, we will show that the interplay between collisional and fluctuation induced transport is particularly evident in the study of zonal structures \cite{zonca2015nonlinear}.
\section{Contents \& outline of the thesis}
\label{sec-1-3}
The first part of this thesis is devoted to an introduction to the physics of magnetic confinement fusion with a detailed derivation of the curvilinear set of coordinates that will be used throughout the work. Using these coordinates and the drift ordering \cite{hinton1976theory}, we derive a set of evolution equations for the number of particles and the energy density on the transport time scale. These equations generalize the expressions derived in \cite{hinton1976theory} taking into account the role of fluctuation induced transport and show the interplay between collisions and fluctuations. In particular we show that fluctuations may enhance collisional transport while the collisions can damp long lived structures formed by saturated instabilities, i.e. zonal structures \cite{zonca2015nonlinear}. As novel result derived in this work, the fluctuation induced fluxes are described using gyrokinetic field theory \cite{brizard2007foundations}, \cite{zonca2014energetic} and are expressed in terms of the gyrocenter distribution function. The main advantage of studying the evolutive equations for the moments instead of the kinetic equation is that we derive the equations for the fluctuations induced fluxes up to the transport time scale using standard first order gyrokinetic theory. Our results, in the appropriate limit, recover those obtained in Refs. \cite{plunk2009theory,abel2013multiscale,barnes2009trinity} by systematic
spatiotemporal scale separation between the fast turbulence response and the slow equilibrium evolution. 
In addition to the novelty of the approach, extending the classical work reviewed in Ref. \cite{hinton1976theory} to fluctuation induced transport, our analysis
based on gyrokinetic field theory makes possible a comparison with the theory of phase space zonal structures \cite{zonca2015nonlinear,chen2016physics}, revealing that the fluctuations induced part of the transport equations can be obtained by taking the proper moment of the long length scale limit of the equation governing the evolution of phase space zonal structures. In other words, as another novel result of this thesis, we demonstrate that a self-consistent gyrokinetic theory of plasma transport on the ${\cal O}(\delta^{-3}
\Omega^{-1})$ time scale can be formulated treating collisional and fluctuation induced transport on the same footing; and it consists of the
collisional evolution equation for phase space zonal structures, which are the phase space counterpart of the corrugation of plasma profiles due to 
equilibrium modifications \cite{chen2007theory,zonca2015nonlinear}. We show that plasma nonlinear evolution can yield to structures formation that are characterized by mesoscales, intermediate between the typical ones of plasma turbulence and those of the reference plasma equilibrium. Thus, nonlinear dynamics can eventually invalidate the scale separation between turbulent length scale and the equilibrium ones which is usually taken as granted. Further to this, plasma nonlinear evolution modifies the particle distribution function, that is the phase space zonal structures. Unlike in neoclassical transport theory, phase space zonal structures bear information about turbulence and, thus, are determined by processes which, in principle, are non local in space and time. This eventually changes the standard neoclassical closure scheme \cite{hinton1976theory}, spontaneously leading to non-linear closure relations. 
In the second part of the thesis, we introduce the Thermodynamic Field Theory, proposed by G. Sonnino \cite{sonnino2009nonlinear}, as an effective tool for the derivation of nonlinear closure relations leading to a macroscopic evolution of the plasma which respect the Thermodynamics principles. In this part, we focus on collisional transport only, as a self-consistent nonlinear closure theory including fluctuations and collisions on the same footing
is not available to date and is beyond the scope of this thesis work. Following Ref. \cite{sonnino2009nonlinear}, non-linear closure relations are derived with the constraint that they reduce to the Onsager linear relations when we neglect the effect of fluctuations.
As an original application of the developed transport theory, we calculate classical and neoclassical collisional fluxes from their analytic expression for one of the proposed reference scenarios Divertor Tokamak Test Facility \cite{pizzuto2016dtt} (DTT), which is the flagship Italian proposal for experimental studies of the power exhaust issues in next step burning plasma experiments, including the International Thermonuclear Experimental Reactor \cite{iter91,aymar97} and the DEMOstration power plant \cite{chapman11}.
\vspace*{1\baselineskip}

An outline of the thesis is displayed below:
\begin{itemize}
\item In \autoref{cap:intro} we provide the definition of a weakly coupled plasma and we introduce the Klimontovich kinetic equation as an exact equation governing its dynamics. The Landau kinetic equation, which will be used in this work, can be obtained taking the proper scale limit of this equation. Furthermore we introduce the phenomenology of nuclear processes in order to illustrate and discuss the range of parameters of a thermonuclear plasma. Finally we briefly describe the physics of toroidal confinement systems and pay particular attention to the derivation of the set of curvilinear coordinates;
\item In \autoref{cap:gyrotransp} we use the Landau kinetic equation in order to derive the evolutive equations for the moments of the distribution function, we introduce the drift ordering and the flux surface average. With these tools, we derive a set of transport equations for particles and for energy density, as original result of this thesis. In particular we use gyrokinetic field theory to express fluctuation induced fluxes through the push-forward representation of the moments of the distribution function;
\item In \autoref{cap:zonal}, we introduce phase space zonal structure theory and we show that, by taking the moments of the long wave length scale limit of the evolutive equation for phase space zonal structures we obtain the fluctuation induced part of the transport equation derived in the previous chapter. This is another original result of this thesis, and illuminates the characteristic spatiotemporal scales arising in the phase space as 
counterpart of corrugation of plasma profiles in the nonlinear plasma evolution;
\item In \autoref{cap:coll}, as original application of the developed transport theory, we explicitly calculate the collisional fluxes for the DTT reference scenario in various collisionality regimes;
\item In \autoref{cap:longer} we address the problem of extending the transport theory derived in this thesis on a time-scale $\sim \delta^{-1} \tau_{transp}$. This is intended to illustrate the ongoing work and possible future developments of the theoretical framework, developed in this thesis work, beyond the mere application of transport equations derived in Chapter 3 and 4 to cases of practical interest. We also introduce the Thermodynamic field Theory as an effective tool to deal with non linear closure relations. 
\end{itemize}
Concluding remarks and comments on future developments are finally given in \autoref{cap:concl}.

In order to make the conceptual flow and narrative smoother, many of the technical details and derivations are omitted 
from the main text. However, plenty of details are provided in three Appendixes: one on 
{\em Noncanonical Hamiltonian methods for particle motion in a strong magnetic field}, another one
on the {\em Push-forward representation of the moments of the distribution function}, and, finally, one
on the {\em Derivation of the fluctuation induced particle flux}.
\chapter{Introduction to toroidally confined systems}
\label{cap:intro}
In this chapter, we briefly introduce the physics of toroidally confined systems and of nuclear reactions. The aim of this part of the thesis is to connect the rest of the work with the background of  a generic reader, i.e. a physicist not specialized in magnetic confinement fusion. 
\section{Definition of a plasma}
For the sake of clarity, we begin with the definition of a plasma. More precisely, a weakly coupled plasma  is an ensemble of charged particles where the potential energy of a typical particle due to its nearest neighbor is much smaller than its kinetic energy \cite{nicholson1983introduction}, i.e. $n_{0}^{1/3}e^{2}\ll T$, where $n_{0}$ is the density of each species, $e$ is the charge of an electron (without loss of generality we consider a plasma with $Z=1$) and the temperature is expressed in energy units. This condition implies that:
\begin{equation}
\label{eq:K1}
\Lambda \propto n_{0} \lambda_{D}^{3} \gg 1
\; ,
\end{equation}
where $\lambda_{D}= (T/4 \pi n_{0}e^{2})^{1/2}$ is the Debye length. Thus, the definition of weakly coupled plasma requires that the number of particles inside a sphere with radius $\lambda_{D}$ is much greater than one. For a plasma at given temperature, this condition is satisfied for sufficiently low density while, for a plasma at given density, it holds for sufficiently high temperatures. Using this definition, it is possible to derive a kinetic equation describing the dynamics of the plasma \cite{bogoliubov1962studies,balescu1963statistical} which is the Landau kinetic equation. The interested reader can find rigorous results on the derivation of the Landau kinetic equation, which has been proposed for the first time in \cite{landau1965collected}, in the review \cite{alexandre2004landau}. In the next section, we will show how the Vlasov equation can be obtained from kinetic theory by taking the proper scale limit.
\section{Klimontovich equation}
\label{sec-2-2}
Following \cite{nicholson1983introduction,klimontovich2012statistical} we can use the kinetic theory in order to formally describe the motion of individual particles. We introduce the density of particles in the phase space with coordinates $({\bm x}, {\bm v})$:
\begin{equation}
\label{eq:K187}
N({\bm x},{\bm v},t)=\delta[{\bm x}-{\bm X}_{1}(t)]\delta[{\bm v}-{\bm V}_{1}(t)]
\; ,
\end{equation}
where $\delta[{\bm x}-{\bm X}_{1}]\equiv \delta(x-X_{1}) \delta(y-Y_{1})\delta(z-Z_{1})$ and we have indicated with ${\bm X}_{1}, {\bm V}_{1}$ the Lagrangian coordinates of the particle. In the same way we can describe an ensemble of $N_{0}$ particles:
\begin{equation}
\label{eq:K188}
N({\bm x},{\bm v},t)= \sum_{i=1}^{N_{0}}\delta[{\bm x}-{\bm X}_{i}(t)]\delta[{\bm v}-{\bm V}_{i}(t)]
\end{equation}
and the whole plasma by summing over the different particle species. The equations of motion for the single particles are the following:
\begin{eqnarray}
\label{eq:K190}
  {\bm \dot{X}}_{i}(t)& = & {\bm V}_{i}(t)
  ; ,\\
\label{eq:K999}m_{s} {\bm \dot{V}}_{i}(t) & = & q_{s} {\bm E}^{m}[{\bm X}_{i}(t),t]+ \frac{q_{s}}{c} {\bm V}_{i}(t) \times {\bm B}^{m}[{\bm X}_{i}(t),t]
\; ,
\end{eqnarray}
where we have indicated the (microscopic) electric and magnetic fields, sum of the fields produced by the particles and the fields imposed externally, with an $m$ superscript. These fields must satisfy the Maxwell equations in the void with the source terms given by:
\begin{equation}
\label{eq:K191}
\rho^{m}({\bm x},t)= \sum_{e,i}q_{s} \int d {\bm v} \,N_{s}({\bm x},{\bm v},t) \quad J^{m}({\bm x},t) = \sum_{e,i}q_{s} \int d {\bm v} \,{\bm v}N_{s}({\bm x},{\bm v},t).
\end{equation}
The equations governing the motion of the particles and the Maxwell equations, i.e. :
\begin{align*}
\label{eq:maxwellmicro}
&{\bm \nabla}\cdot {\bm E}^{m}({\bm x},t)= 4 \pi \rho^{m}({\bm x},t)\\
&{\bm \nabla}\cdot {\bm B}^{m}({\bm x},t)= 0\\
&{\bm \nabla} \times {\bm E}^{m}({\bm x},t)= -\frac{1}{c}\frac{\partial {\bm B}^{m}({\bm x},t)}{\partial t}\\
&{\bm \nabla} \times {\bm B}^{m}({\bm x},t)=\frac{4 \pi}{c} {\bm J}^{m}({\bm x},t)+ \frac{1}{c}\frac{\partial {\bm E}^{m}({\bm x},t)}{\partial t}
\end{align*}
form a closed set of equations:
\begin{itemize}
\item from Eq. (\ref{eq:K190}), the positions of the particles at the next time step are known once the fields ${\bm E}^{m}, {\bm B}^{m}$ are given;
\item solving the Maxwell equations with the sources given by Eq. (\ref{eq:K191}) gives the value of the fields at the next time step.
\end{itemize}
Taking the time derivative of Eq. (\ref{eq:K188}), we obtain the following expression:
\begin{equation}
\label{eq:K192}
\partial_{t}N_{s}({\bm x},{\bm v},t) = - \sum_{i=1}^{N_{0}}{\bm \dot{X_{i}}} \cdot \partial_{{\bm x}}\delta[{\bm x}-{\bm X}_{i}(t)]\delta[{\bm v}-{\bm V}_{i}(t)] - \sum_{i=1}^{N_{0}}{\bm \dot{V_{i}}} \cdot \partial_{{\bm v}} \delta[{\bm x}-{\bm X}_{i}(t)]\delta[{\bm v}-{\bm V}_{i}(t)].
\end{equation}
Substituting Eq. (\ref{eq:K190}) and Eq. (\ref{eq:K999}) into Eq. (\ref{eq:K192}), we obtain:
\begin{eqnarray*}
\label{eq:K193}
\partial_{t}N_{s}({\bm x},{\bm v},t) = & - & \sum_{i=1}^{N_{0}}{\bm {\bm V}_{i}} \cdot \partial_{{\bm x}}\delta[{\bm x}-{\bm X}_{i}(t)]\delta[{\bm v}-{\bm V}_{i}(t)] + \\ &-& \sum_{i=1}^{N_{0}} \left\{{\bm q_{s} {\bm E}^{m}[{\bm X}_{i}(t),t]+ \frac{q_{s}}{c} {\bm V}_{i}(t) \times {\bm B}^{m}[{\bm X}_{i}(t),t] }\right \} \cdot \partial_{{\bm v}} \delta[{\bm x}-{\bm X}_{i}(t)]\delta[{\bm v}-{\bm V}_{i}(t)]
\; ,
\end{eqnarray*}
which can be cast in a slightly different form by using the properties of the delta function:
\begin{eqnarray}
\label{eq:K194}
\partial_{t}N_{s}({\bm x},{\bm v},t) = & - & {\bm v} \cdot \partial_{{\bm x}}\sum_{i=1}^{N_{0}}\delta[{\bm x}-{\bm X}_{i}(t)]\delta[{\bm v}-{\bm V}_{i}(t)] + \\\nonumber &-& \left\{q_{s} {\bm E}^{m}[{\bm x},t]+ \frac{q_{s}}{c} {\bm v} \times {\bm B}^{m}[{\bm x},t]\right \}  \cdot \partial_{{\bm v}} \sum_{i=1}^{N_{0}} \delta[{\bm x}-{\bm X}_{i}(t)]\delta[{\bm v}-{\bm V}_{i}(t)]
\; ,
\end{eqnarray}
which is the Klimontovich equation. This equation, together with the Maxwell's equations for the microscopic fields, constitutes an exact description of the microscopic dynamics of the plasma. The solution to this equation contains information about the trajectories of all the particles of the plasma. However, we are interested in a less detailed information. In particular we want to describe the macroscopic behavior of the plasma. For this reason we define the smooth function $f_{s}$:
\begin{equation}
\label{eq:K194a} f_{s}({\bm x},{\bm v},t) \equiv \langle N_{s} \rangle
\; ,
\end{equation}
where the $\langle \rangle$ operation is the average over all the different copies of the original system created by respecting a prescription, e.g. that the plasma is at thermal equilibrium. We now introduce another definition of this average operation, which is heuristic but useful in order to grasp its meaning: by the definition of plasma, we know that there is a huge number of particles inside the Debye sphere and, therefore, we can imagine to take a box, with the characteristic length much larger than the inter-particle distance and much smaller than the Debye length. We can now count the particles inside this box which will have small statistical fluctuations around $f({\bm x},v,t)$ and, in the limit in which the number of particles becomes very high, the fluctuations must disappear by virtue of the central limit theorem.

In order to write down the evolutive equations for $f$, we need to separate the average and the fluctuating part of the fields:
\begin{eqnarray}
\label{eq:K195}
  N_{s} & = & f_{s} + \delta N_{s}
  \; ,\\
  {\bm E}^{m} & = & {\bm E} + \delta {\bm E}
  \; ,\\
  {\bm B}^{m} & = & {\bm B} + \delta {\bm B}
                    \; ,
\end{eqnarray}
where $\langle N_{s}\rangle = f_{s},\, \langle {\bm E^{m}}\rangle={\bm E}, \dots$ We now substitute these expressions into Eq. (\ref{eq:K194}) and we take the ensemble average obtaining:
\begin{eqnarray}
\label{eq:K196}
\partial_{t}f_{s}({\bm x},{\bm v},t) + {\bm v} \cdot \partial_{{\bm x}} f_{s} + \left\{q_{s} {\bm E}({\bm x},t)+ \frac{q_{s}}{c} {\bm v} \times {\bm B}({\bm x},t) \right \}  \cdot \partial_{{\bm v}} f_{s} = \\ \nonumber - \frac{q_s}{m_{s}} \langle (\delta {\bm E} + {\bm v}/c \times \delta {\bm B})\cdot \partial_{v}\delta N_{s} \rangle.
\end{eqnarray}
In this equation, we see that all the effects due to the discrete particle nature of the plasma are on the right hand side of the equation. In order to clarify this point, we may imagine to break each electron into an infinite number of pieces so that $n_{0} \rightarrow \infty$, $m_{e} \rightarrow 0$ and $e \rightarrow 0$ obtaining the so called ``mush limit''. On the right hand side we have $\delta N_{s} \sim N_{0}^{1/2} \sim \Lambda_{e}^{1/2}$,  where we have indicted with $\Lambda_{e}$ the number of particles inside the Debye sphere. We have that $ \delta E\sim e \delta N_{s} \sim N_{0}^{-1}N_{0}^{1/2} \sim N_{0}^{-1/2} \sim \Lambda_{e}^{-1/2}$ so that the ratio between the magnitude of the right hand side and the left side scales as $N_{0}^{-1}\sim \Lambda_{e}^{-1}$. In the mush limit, i.e. $\Lambda_{e}\rightarrow  \infty$, we obtain:
\begin{equation}
\label{eq:K197}
\partial_{t}f_{s}({\bm x},{\bm v},t) + {\bm v} \cdot \partial_{{\bm x}} f_{s} + \left\{q_{s} {\bm E}({\bm x},t)+ \frac{q_{s}}{c} {\bm v} \times {\bm B}({\bm x},t) \right \}  \cdot \partial_{{\bm v}} f_{s} = 0
\; ,
\end{equation}
which is the Vlasov equation \cite{spohn2012large} describing the dynamics of a collisionless plasma. For this reason it is quite natural to associate the right hand side of (\ref{eq:K196}) with statistical fluctuations due to collisional effects. 

The dynamics of a plasma is completely described by its kinetic equation. Anyway this is not the only way to study its evolution. In the next chapter we will show how it is possible, starting from the kinetic equation, to obtain a set of evolution equations for the moments of the distribution function such as density, temperature... In general solving this system of equations is equivalent to study the original kinetic problem but, in the particular case of magnetized plasmas which are studied in this thesis, the moment approach can be very convenient. For this reason in the next chapter we will use this methodology systematically.
\section{Nuclear fusion}
\subsection{Historical remarks}
\label{sec-3-1-2}
The first researches in the field of nuclear reactions started in the twenties and were focused, in particular, on the investigation of the mechanism responsible for the generation of energy inside the stars. The British astronomer Atkinson and the Austrian physicist Houtermans in 1929 made the hypothesis that this energy could be originated by the fusion between light atoms nuclei and was described by the mass defect introduced by Albert Einstein:
\begin{equation}
\label{eq:141}
\Delta E = \Delta m c^{2}.
\end{equation}
In order to undergo a fusion reaction, two nuclei need to be at distances of the order of $10^{-13} cm$ because attractive nuclear forces are effective on this length-scale. In order to reach this distance, they must have a sufficiently high kinetic energy to overcome Coulomb repulsion.
\begin{figure}[htb]
\centering
\includegraphics[width=.5\linewidth]{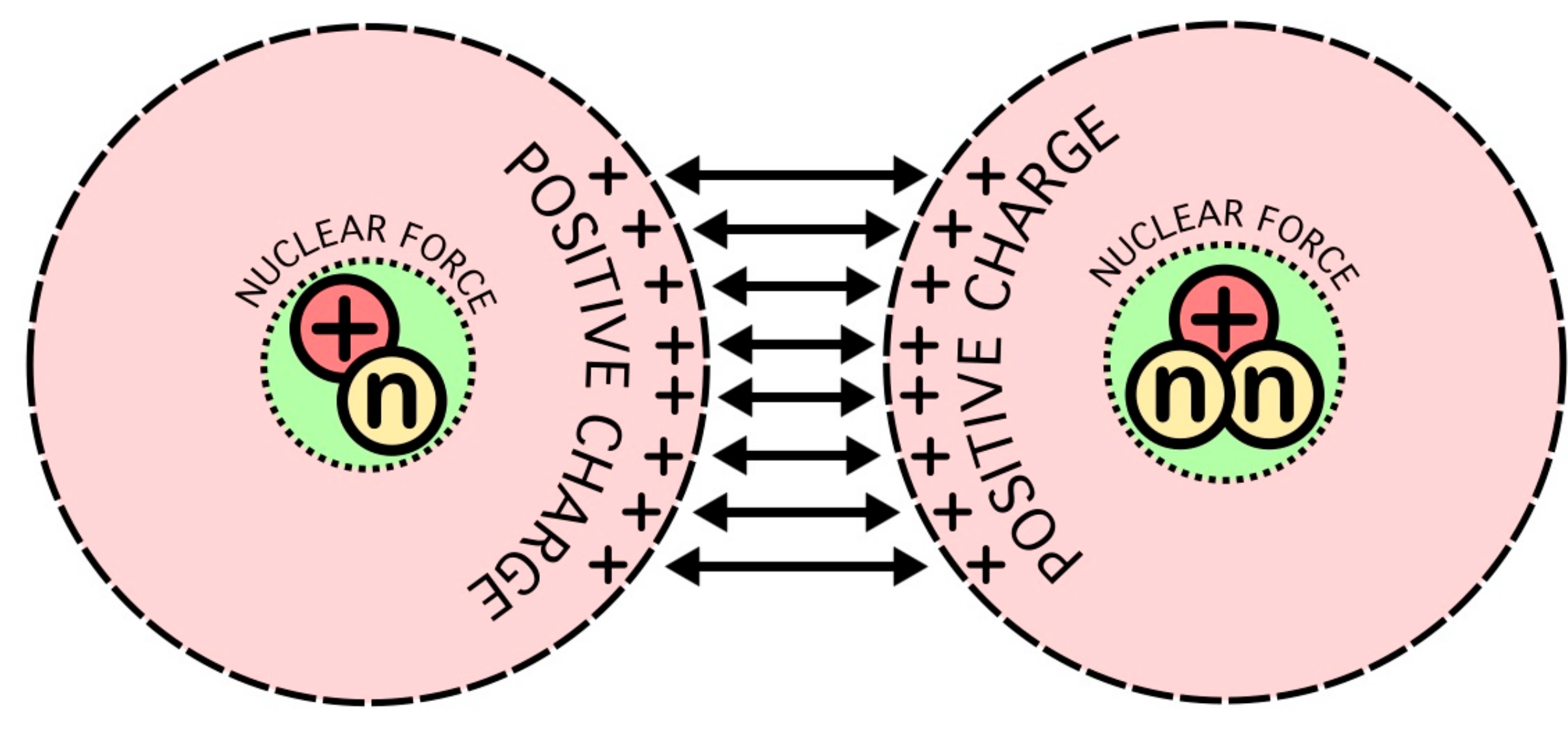}
\caption{Cartoon describing the competition between nuclear and electromagnetic forces.}
\end{figure}
Atkinson and Houtermans showed that the Coulomb barrier could be overcome without reaching the expected and extremely high values of kinetic energy \cite{atkinson1929frage}. The tunneling effect allowed a small percentage of the particles without the required kinetic energy to interact via fusion reactions. The proton-proton chain, identified by Weizsacher and Bethe in 1938 \cite{von1938uber}, and the carbon cycle were the first mechanisms to be proposed in order to describe stellar dynamics. After the studies on nuclear fission and the construction of the first reactor  (Fermi, Chicago 1942), the research in the field of nuclear fusion started again and, under the supervision of Teller, in 1952, the first nuclear explosion took place. In the fifties, the deuterium-tritium and deuterium-deuterium reactions were identified as the best candidates in order to obtain controlled thermonuclear fusion. From that moment the main research goal has been the confinement of the plasma in order to prevent it from touching the material walls of the experiment. Initially, open ended confining configurations have been proposed but, for the reasons that we will explain in the next sections, these machines were soon abandoned in favor of toroidal configurations: Stellarators and Tokamaks.
\subsection{Nuclear processes}
\label{sec-3-1-3}
Nuclear fusion occurs when two light nuclei merge forming an heavier nucleus with a mass that is lower with respect to the sum of the initial masses. We can say that the particle obtained is more stable because energy is required to realize from it the initial particles. The energy is released in the kinetic form and it is called nuclear binding energy because it is the same amount of energy required to disassemble the nucleus of an atom into its component parts. The amount of energy is given by Eq. (\ref{eq:141}). As an example, let's consider the Helium-4, which is composed by two protons and two neutrons. The mass defect $\Delta m$ can be calculated as $\Delta m = 2 (m_{p}+m_{n})- m_{a}$ where $m_{p}$ and $m_{n}$ are respectively the proton and the neutron mass and $m_{a}$ is the mass of the Helium-4. Plugging in the correct values for the masses, we obtain: $\Delta m = 0.5042 \times 10^{-28} Kg$ and the corresponding energy by means of Eq. (\ref{eq:141}). Dividing the binding energy by the mass number $A = N+ Z$, where $N$ is the number of neutrons in the nucleus and $Z$ is the atomic number, we obtain the binding energy per nucleon, which can be plotted as a function of $A$.
\begin{figure}
\centering
\includegraphics[width=.7\linewidth]{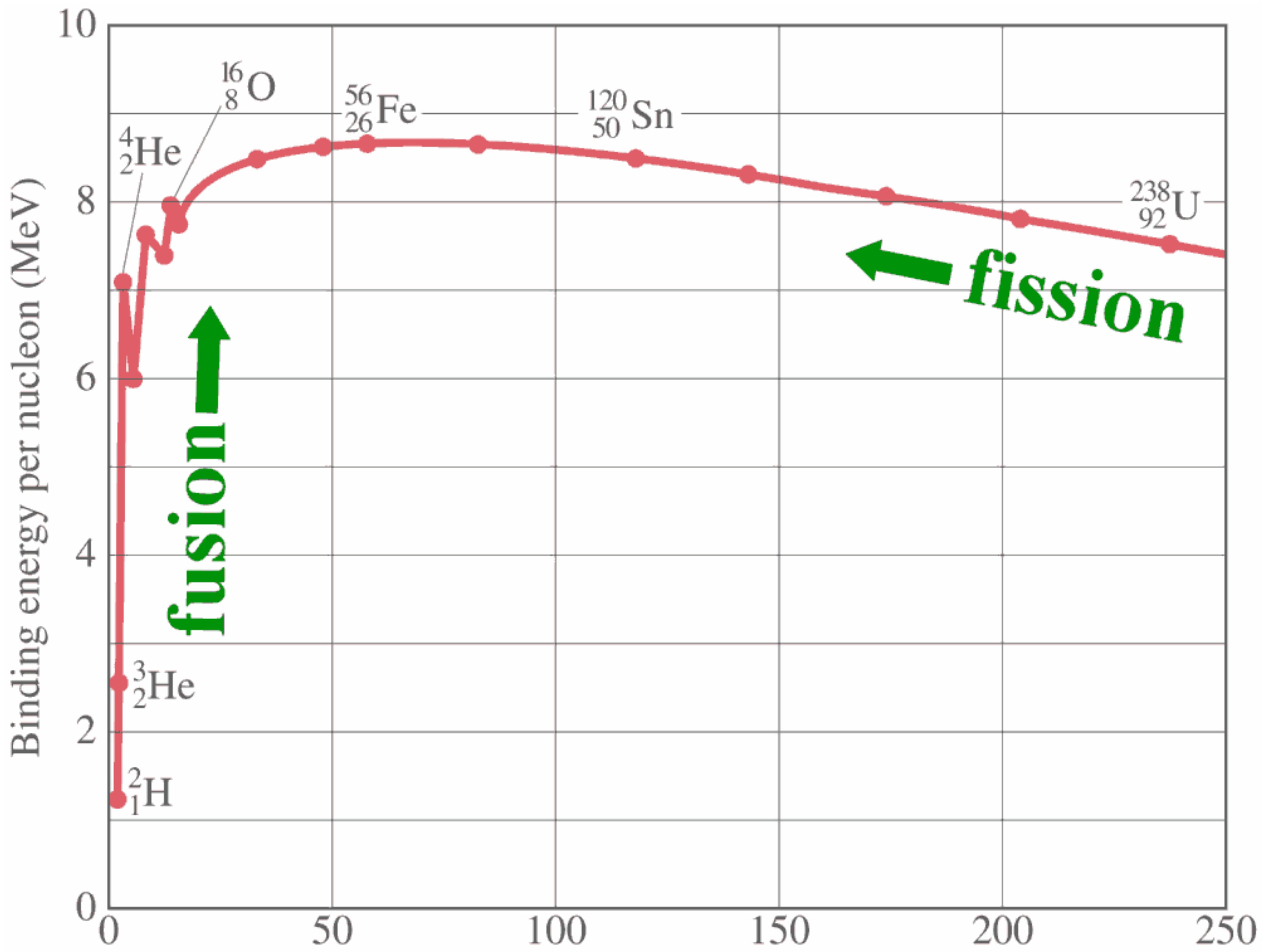}
\caption{Plot of the binding energy per nucleon (Courtesy of Pearson Prentice Hall).}
\end{figure}
Nuclear fission happens when heavy nuclei, such as Uranium, split into smaller parts with higher (in absolute value) binding energy per nucleon. This energy is extracted in the kinetic form. Using the previous plot, we can verify that the production of energy per nucleon in a fusion reaction is bigger than its analogue for a fission reaction, i.e. 7.07 MeV/nucleon for Helium fusion versus 0.851 MeV/nucleon for Uranium fission. The formation of Helium-4 through the fusion of 4 nucleons is very unlikely to happen because it involves the simultaneous interaction of four particles. However, Helium-4 can be produced by fusion of $\ce{^{2}H}$ and $\ce{^{3}H}$, which are respectively known has Deuterium (D) and Tritium (T). For example, the following reaction is possible:
\begin{equation}
\label{eq:117}
\ce{D + T \rightarrow ^{4}He}+n+17.586 \, MeV.
\end{equation}
\begin{figure}[htb]
\centering
\includegraphics[width=.3\linewidth]{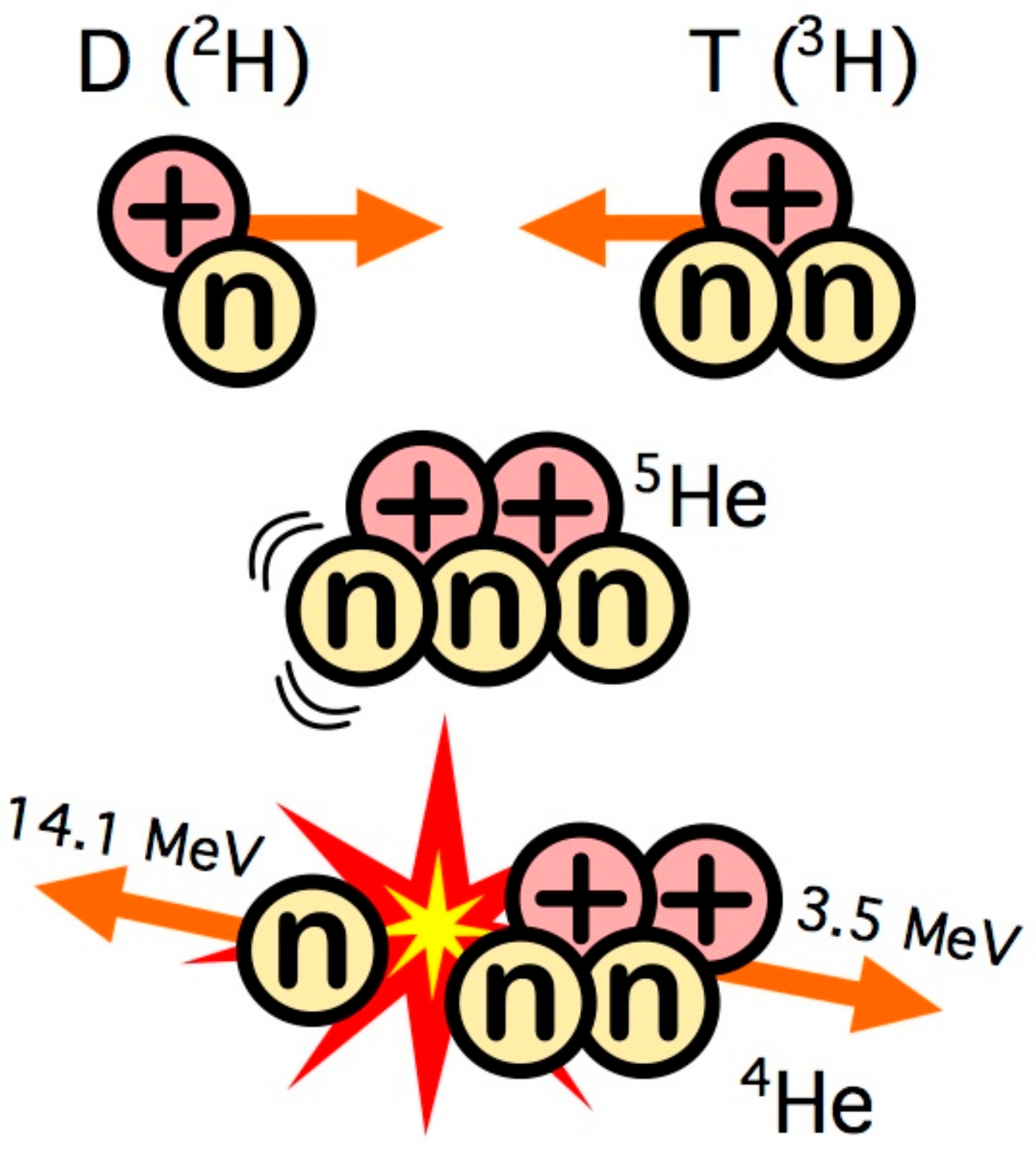}
\caption{Cartoon describing the $D-T$ fusion reaction.}
\end{figure}

It is well known that nuclear energy released per unit of mass is huge compared with chemical reactions. For example, the ratio between the energy per unit of mass obtained by means of a $D-T$ reaction and by means of the combustion of carbon is: $10^{7}$.
\subsection{Interaction processes and fusion cross-section}
\label{sec-3-1-4}
Without loss of generality, in this chapter we consider fusion reactions between two different species $N$ and $M$ composing the plasma. The number of fusion reactions taking place in the plasma per unit of time and volume is:
\begin{equation}
\label{eq:118}
R=\sigma_{f} n_{N}n_{M}v
\; ,
\end{equation}
where $v$ is the relative velocity between the colliding particles and $\sigma_{f}$ is the cross-section of the fusion reaction. The cross-section $\sigma_{f}$ can be written as:
\begin{equation}
\label{eq:119}
\sigma_{f}= \frac{A}{E}e^{-B E^{-1/2}}
\; ,
\end{equation}
where $A$ and $B$ are constants depending on the reaction. We show here a plot of the total cross section $\sigma$ for different fusion reactions:

\begin{figure}[htb]
\centering
\includegraphics[width=.45\linewidth]{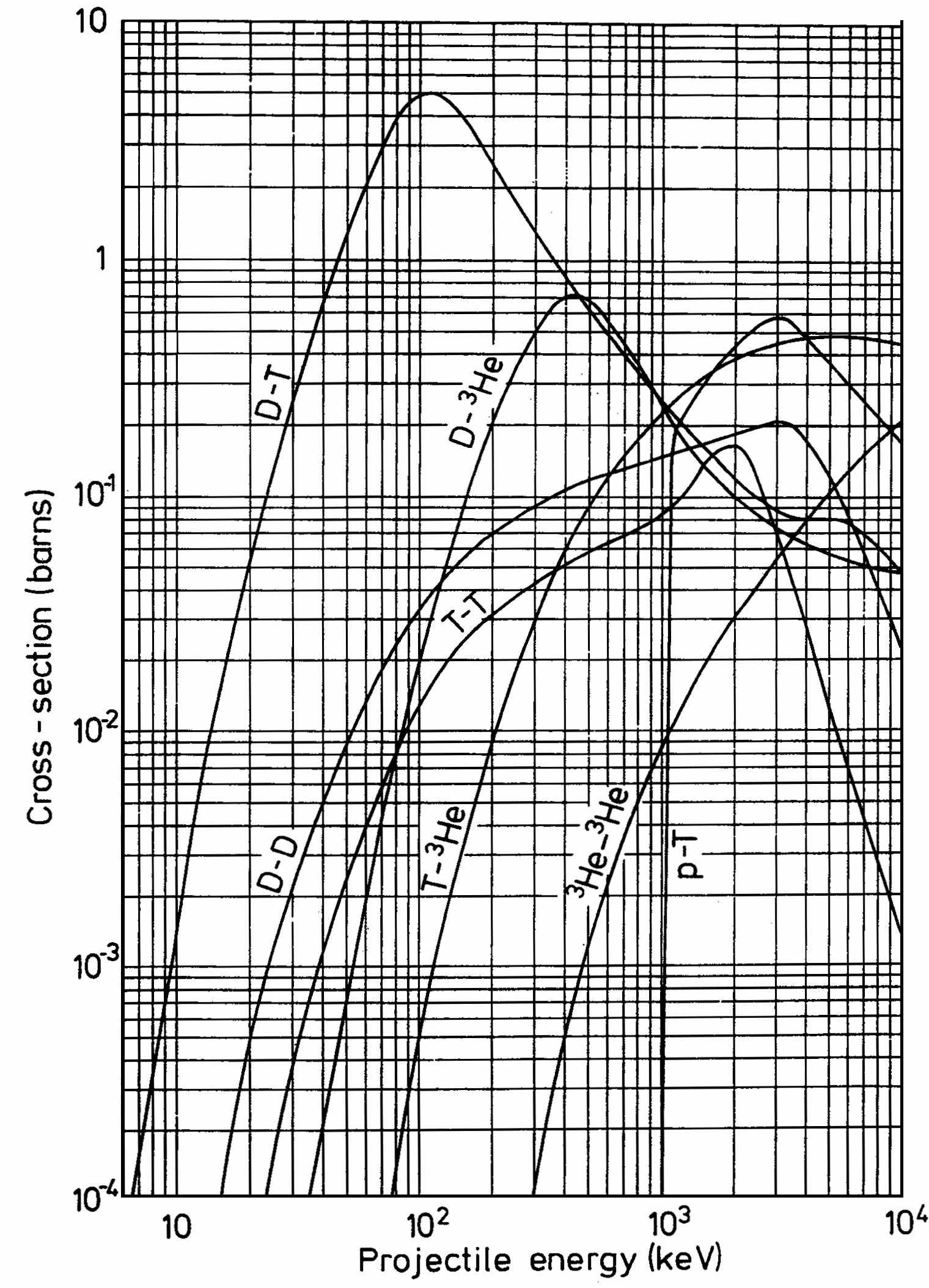}
\caption{Plot of the cross section $\sigma$ for several fusion reactions as a function of the projectile energy. We recall that $1 Barn = 10^{-24}cm^{2}$. Courtesy of Wesson, Campbell: \cite{wesson2011tokamaks}.}
\end{figure}
\clearpage
Under energies of $500 KeV$ the elastic electromagnetic Coulomb interaction is dominating with a cross-section:
\begin{equation}
\label{eq:120}
\sigma_{c}\approx C(Z_{1}Z_{2})^{2}\frac{1}{E^{2}}.
\end{equation}
For magnetically confined systems, we can assume that the bulk plasma is well described by local Maxwellian distributions in the velocity space and, therefore, we can calculate the relevant physical quantity which is the average number of fusion reactions per unit of time and volume:
\begin{equation}
\label{eq:121}
\bar{R}= n_{N}n_{M}\langle \sigma_f v \rangle
\; ,
\end{equation}
where $\langle \sigma_{f} v \rangle$ is the reactivity and depends only on the temperature once $\sigma_{f}(v)$ has been calculated. Multiplying Eq. (\ref{eq:121}) by the energy produced by the fusion reaction, we obtain the density of power:
\begin{equation}
\label{eq:122}
q^{'''}= \bar{R}E_{fus}= n_{N}n_{M}\langle \sigma_{f} v\rangle E_{fus}.
\end{equation}
If the system of two particles is confined with fixed pressure $P$ we can apply the ideal gas law obtaining:
\begin{equation}
\label{eq:123}
q^{'''} \propto \frac{\langle \sigma_{f} v\rangle}{T^{2}}E_{fus}.
\end{equation}
We show here a plot of the density of power $q^{'''}$ for different fusion reactions:

\begin{figure}[htb]
\centering
\includegraphics[width=.6\linewidth]{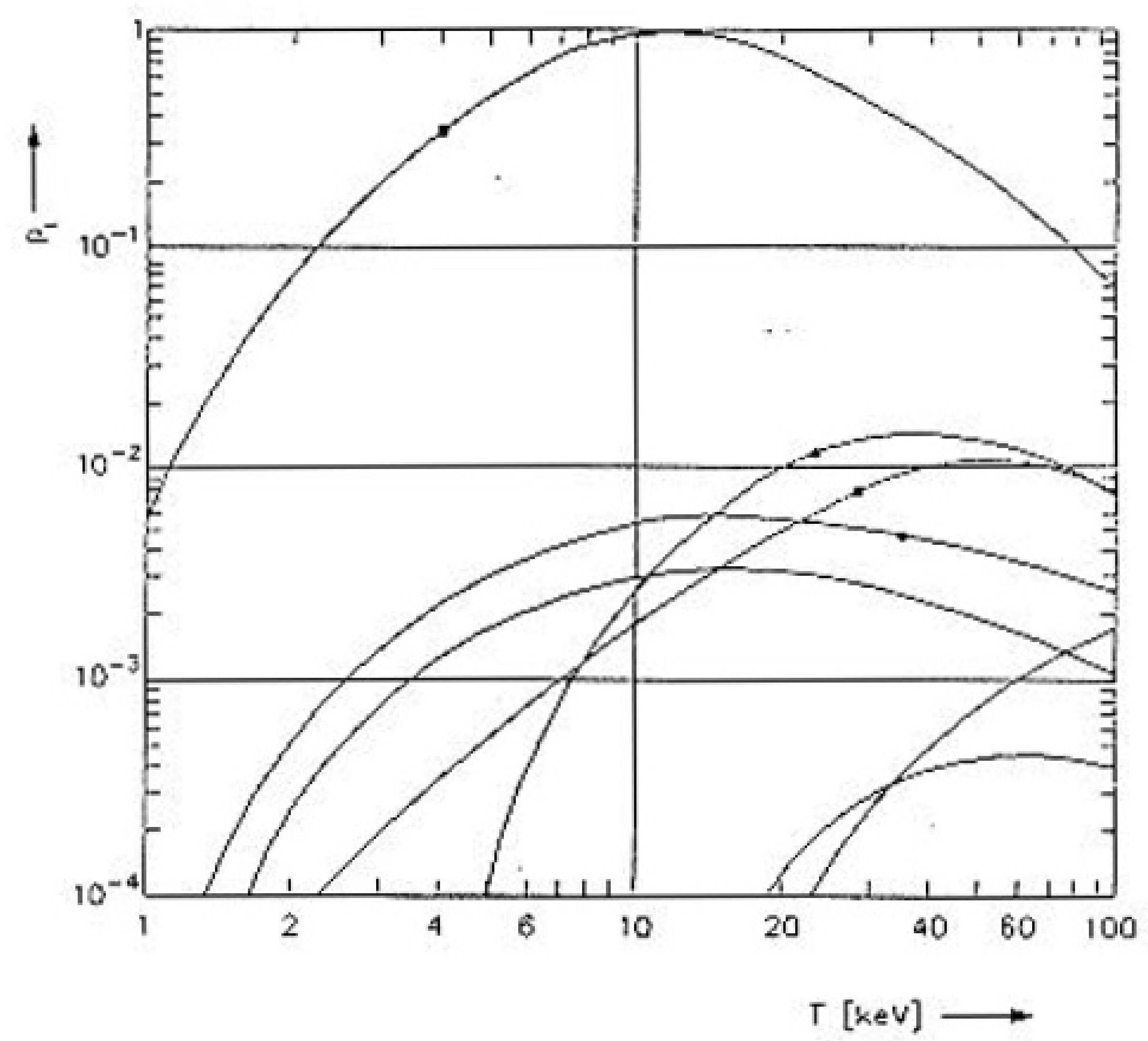}
\caption{Plot of the density of power for various fusion reactions. The peak for the  $D-T$ reaction is around $13.5 KeV$. Courtesy of Wesson, Campbell: \cite{wesson2011tokamaks}.}
\end{figure}
\clearpage
\subsection{Thermonuclear fusion reactions}
\label{sec-3-1-5}
Nuclear reactions that take place on the stars have almost no interest for the controlled thermonuclear fusion. The reactions of interests need a faster reaction kinematics and either the reactivity $\langle \sigma_{f} v \rangle$ or the power density $q^{'''}$ must have values bigger by different orders of magnitude than their stellar analogues. The reactions that actually are the most promising are based on the deuterium, which can be extracted from water and, therefore is abundant in nature. These are:
\begin{align}
\label{eq:125}
&\ce{D + T \rightarrow  ^{4}He + n +} 17.586 \, MeV, \\
\nonumber &\ce{D + D \rightarrow   ^{3}He +} n + 3.267 \, MeV, \\
\nonumber &\ce{D + D \rightarrow   T }+ p + 4.032 \, MeV, \\
\nonumber &\ce{D + ^{3}He \rightarrow   ^{4}He +} p + 18.351 \, MeV \\
\nonumber &\ce{T + T \rightarrow ^{4}He + }2 p + 11.327 \, MeV.
\end{align}
We show here a plot $\langle \sigma_{f} v\rangle$ for some of these reactions as a function of the temperature:
\begin{figure}[htb]
\centering
\includegraphics[width=.5\linewidth]{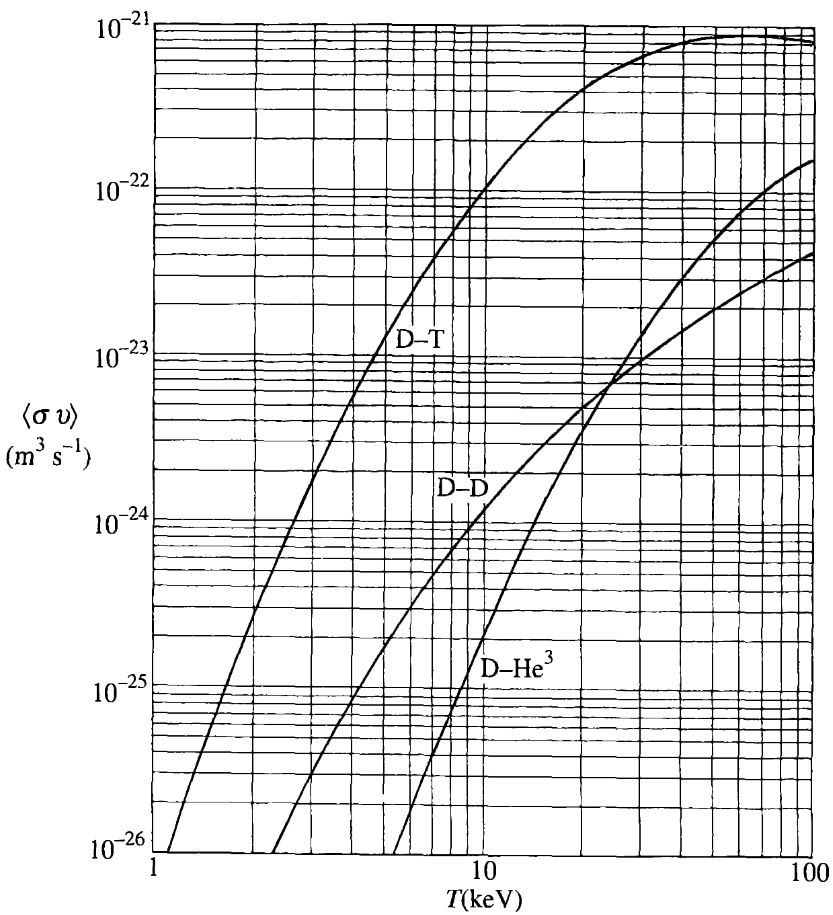}
\caption{Plot of the reaction parameter $\langle \sigma v \rangle$ as a function of the temperature. Courtesy of Wesson, Campbell: \cite{wesson2011tokamaks}.}
\end{figure}

The second, the third and the fourth equations of (\ref{eq:125}) do not produce neutrons but only charged particles. Therefore, they do not require a neutron moderator in order to extract fusion energy and can be confined with magnetic fields. The first equation of (\ref{eq:125}) is the most convenient from an energetic point of view because it has the highest cross-section under $100 KeV$. Tritium production and neutron moderation can be simultaneously solved with a Lithium moderator. The Lithium is present in nature with two isotopes: $\ce{^{6}Li}$ and $\ce{^{7}Li}$. The $\ce{^{7}Li}$ can be used in order to capture a fast neutron ($14.07 MeV$) produced by the $D-T$ reaction:
\begin{equation}
\label{eq:126}
\ce{^{7}Li + n \rightarrow ^{4}He + T +} n'- 2.47 \, MeV
\; ,
\end{equation}
and, again,  the slow neutron $n'$ is captured with the following reaction:
\begin{equation}
\label{eq:127}
\ce{^{6}Li} + n' \ce{\rightarrow ^{4}He + T +} n'+ 4.78 \, MeV.
\end{equation}
Using these processes the Tritium required for the $D-T$ reaction, which is difficult to find in nature, can be drastically reduced. Only an initial small quantity is required to operate. 
\subsection{Ignition}
\label{sec-3-1-6}
In 1957, John D. Lawson derived a criterion for the production of energy from a $D-T$ thermonuclear plasma as a function of the plasma parameters and of the efficiency of the thermodynamic cycle $\eta$ \cite{lawson1957some}. Today the main goal is to reach a state where all the energy losses are compensated by the $\alpha$ particles generated by the fusion process which, slowing down, transfer their energy to the thermal plasma. This state is called ignition. All the transport processes (losses) are described by means of a confinement time $\tau_{E}$, such that the total power is:
\begin{equation}
\label{eq:115}
P_{L}= \frac{3 n T}{\tau_{E}}
\end{equation}
where $3 n T$ is the thermal energy of the plasma. The Lawson criterion \cite{lawson1957some} reads:
\begin{equation}
\label{eq:124}
n \tau_{E}= \frac{3 T}{\frac{\eta}{1- \eta}\frac{Q_{T}}{4}\langle \sigma v\rangle - \alpha T^{1/2}}
\end{equation}
where $\alpha T^{1/2}$ is a loss term due to the radiative processes and $Q_{T}$ is the energy produced by the nuclear reaction, i.e. for the $D-T$ reactions $Q_{T}=22.4 MeV$. As stated already, it can be shown that $\langle \sigma v\rangle$ is a function only of the temperature and can be plotted:
\begin{figure}[htb]
\centering
\includegraphics[width=.5\linewidth]{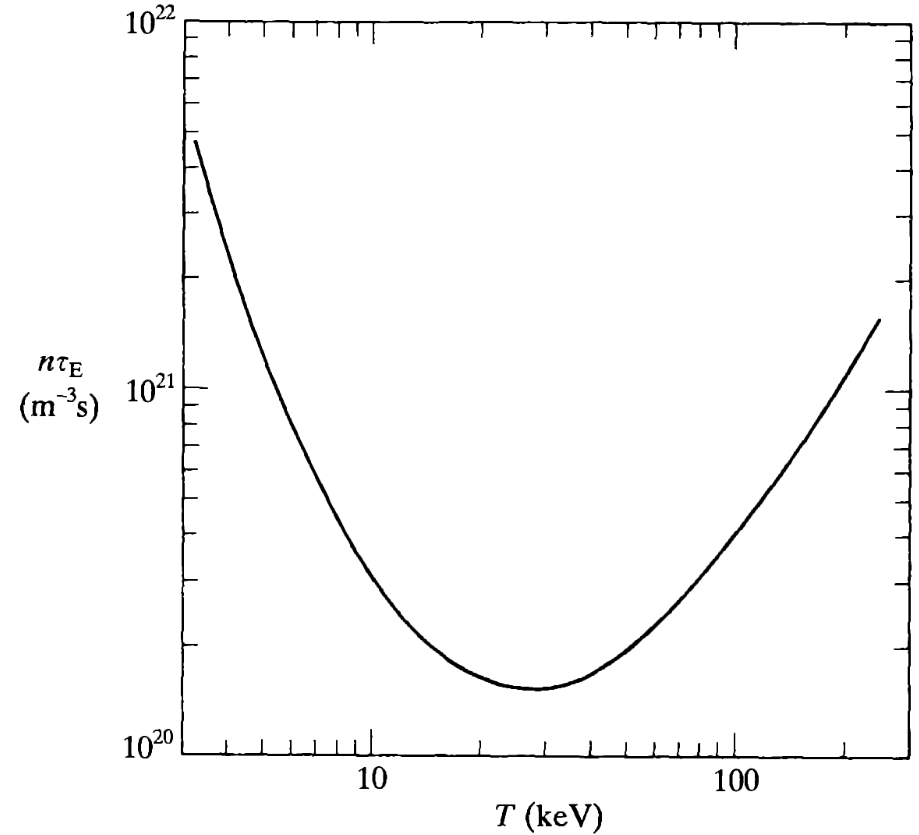}
\caption{Plot of the necessary condition for ignition as stated by the Lawson criterion.}
\end{figure}

A necessary condition for ignition is that the plasma reaches a state (point) above the curve plotted in the previous image. When the density of power $P_{\alpha}$ produced by the $\alpha$ particles:
\begin{equation}
\label{eq:128}
P_{\alpha}= \frac{n^{2}}{4}\langle \sigma v\rangle Q_{\alpha}
\; ,
\end{equation}
where $Q_{\alpha}$ is the energy produced in the fusion reaction, is equal or greater than the loss by radiation and conduction, $P_{\alpha}=P_{L}+P_{irr}$, the process is self sustainable without providing any energy with heating systems. For $\eta = 0.136$ and a temperature of the order of $20 KeV$, the ignition condition requires that $n \tau_{E} > 1.5 \times 10^{20}m^{-3}$ in order to obtain energy from the reactor.
\section{Introduction to Tokamak physics}
\label{sec-3-2}
In the previous section, we have shown that, in order to achieve ignition, the macroscopic parameters of the plasma should lay inside a certain region. The challenge is to drive the plasma into this region and, then, to keep it in this state. Modern magnetic confinement experiments rely on strong magnetic fields in order to reach this goal. From the microscopic point of view, strong magnetic fields confine the plasma by keeping the charged particles moving with narrow spirals along magnetic field lines. This is reflected, from a macroscopic point of view, into the fact that the magnetic pressure, i.e. $B^{2}/8 \pi$, can balance the kinetic pressure of the plasma. Heuristically, equilibrium can be achieved when the ratio $\beta$ between magnetic and kinetic pressures:
\begin{equation}
\label{eq:130}
\beta = \frac{n T}{B^{2}/8\pi}
\end{equation}
is lower than $1$. A commercial reactor will work with $\beta \approx 0.5$.

Different machine configurations have been studied by the magnetic confinement community:
\begin{itemize}
\item open systems, where the magnetic field lines exit the confinement region and can touch the material walls of the system;
\item toroidal systems, where the magnetic fields lines are completely enclosed in the confinement region.
\end{itemize}
Most of the magnetic confinement machines belong to the second class. This is due to the fact that transport processes in a strongly magnetized plasma are faster in the direction parallel to the magnetic field because of the free streaming of the particles along the magnetic field lines. In an open system, this can cause severe losses of particles and energy, which are avoided in a toroidal system thanks to the topology of the magnetic field.

\begin{figure}[htb]
\centering
\includegraphics[width=.55\linewidth]{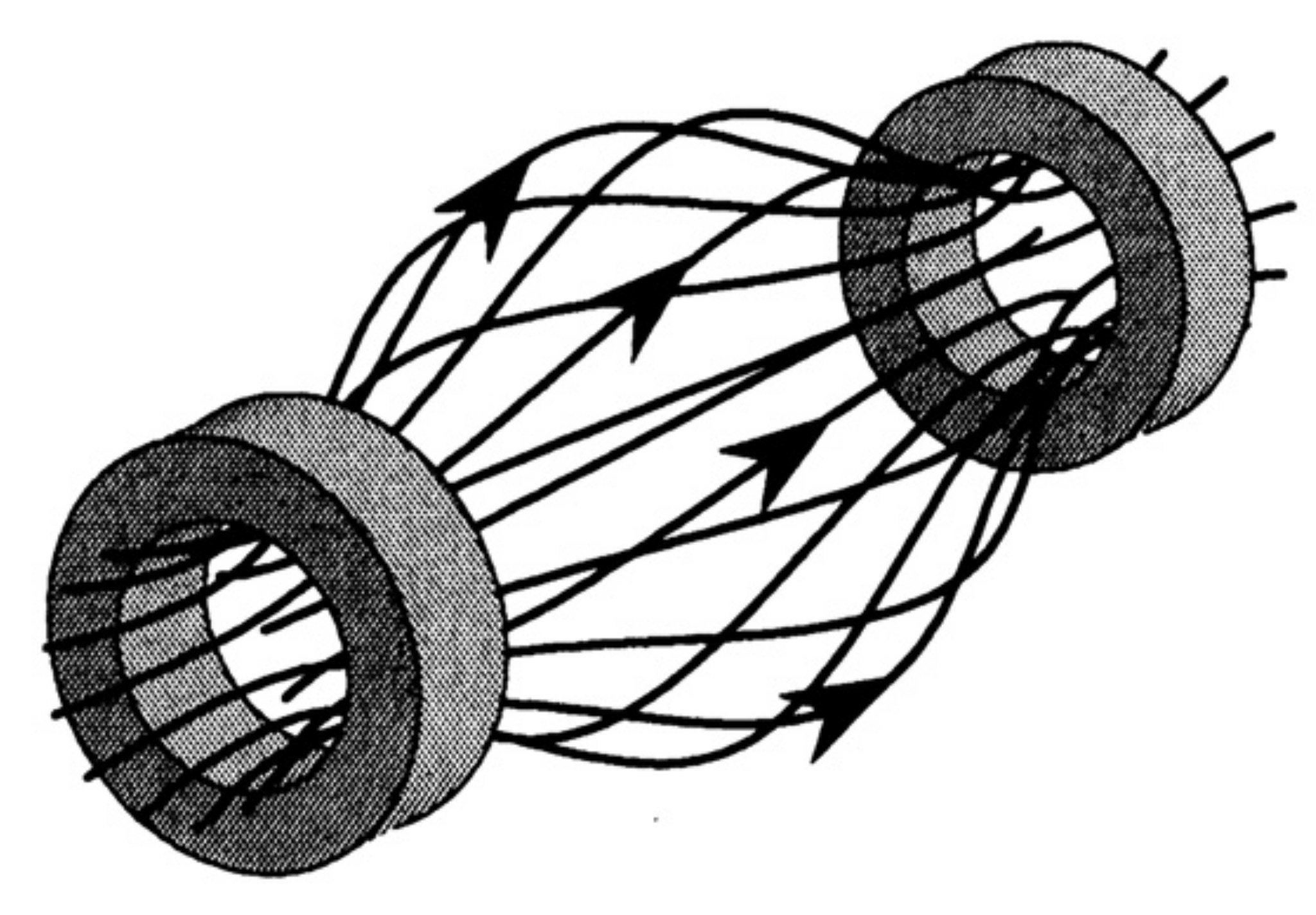}
\caption{Magnetic field lines in an open system.}
\end{figure}
A magnetically confined plasma should satisfy the magneto-hydrodynamic force balance equation \cite{freidberg1982ideal,boozer2005physics}:
\begin{equation}
\label{eq:131}
{\bm j}\times {\bm B}= c {\bm \nabla}P
\end{equation}
where ${\bm j}$ is the plasma current, $P$ is the kinetic pressure, $c$ is the speed of light and ${\bm B}$ is the magnetic field. From this equation, we know that magnetic field lines wind up over surfaces with constant pressure which are called magnetic surfaces.
\begin{figure}[htb]
\centering
\includegraphics[width=.7\linewidth]{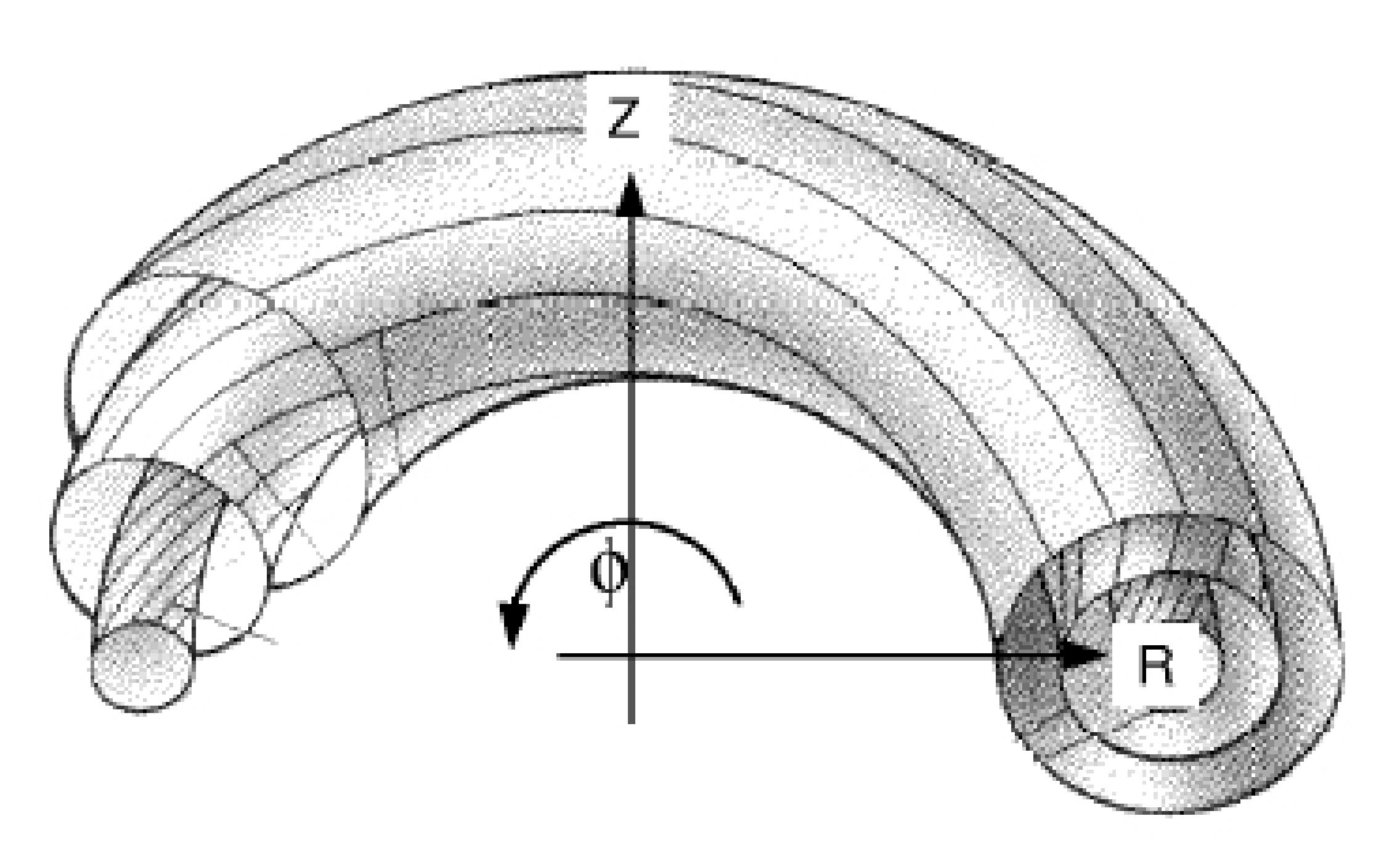}
\caption{Magnetic field lines winding over different magnetic surfaces in a toroidal system.}
\end{figure}
In this thesis, we will treat in detail axisymmetric configurations, i.e. configurations which are symmetric under rotation around the symmetry axis of the torus, where magnetic surfaces are topologically nested tori. The most internal surface, which is degenerate, is called magnetic axis. Due to this geometrical structure, in this work, we will introduce a radial coordinate to label the magnetic surfaces and two angular coordinates to describe the points on each surface. It has been shown \cite{boozer2005physics} that a pure toroidal magnetic field is not enough in order to achieve magnetic confinement and that a poloidal component of ${\bm B}$ is necessary. Therefore magnetic field lines have helical shape over the magnetic surfaces.
In the following  we will focus on the Tokamak configuration, which is the most successful magnetic confinement configuration realized for the first time in $1952$ by L. Artsimovich. The name is an acronym for the Russian wording Toroidalnaya Kamera Magnitnymi Katushkami, or "toroidal chamber with axial magnetic field" in English.  This configuration is characterized by a strong toroidal magnetic field generated by the toroidal coils and a poloidal field due to the plasma current, which is induced by the transformer. This current causes also the ionization of the plasma by Ohmic heating.
\begin{figure}[htb]
\centering
\includegraphics[width=.9\linewidth]{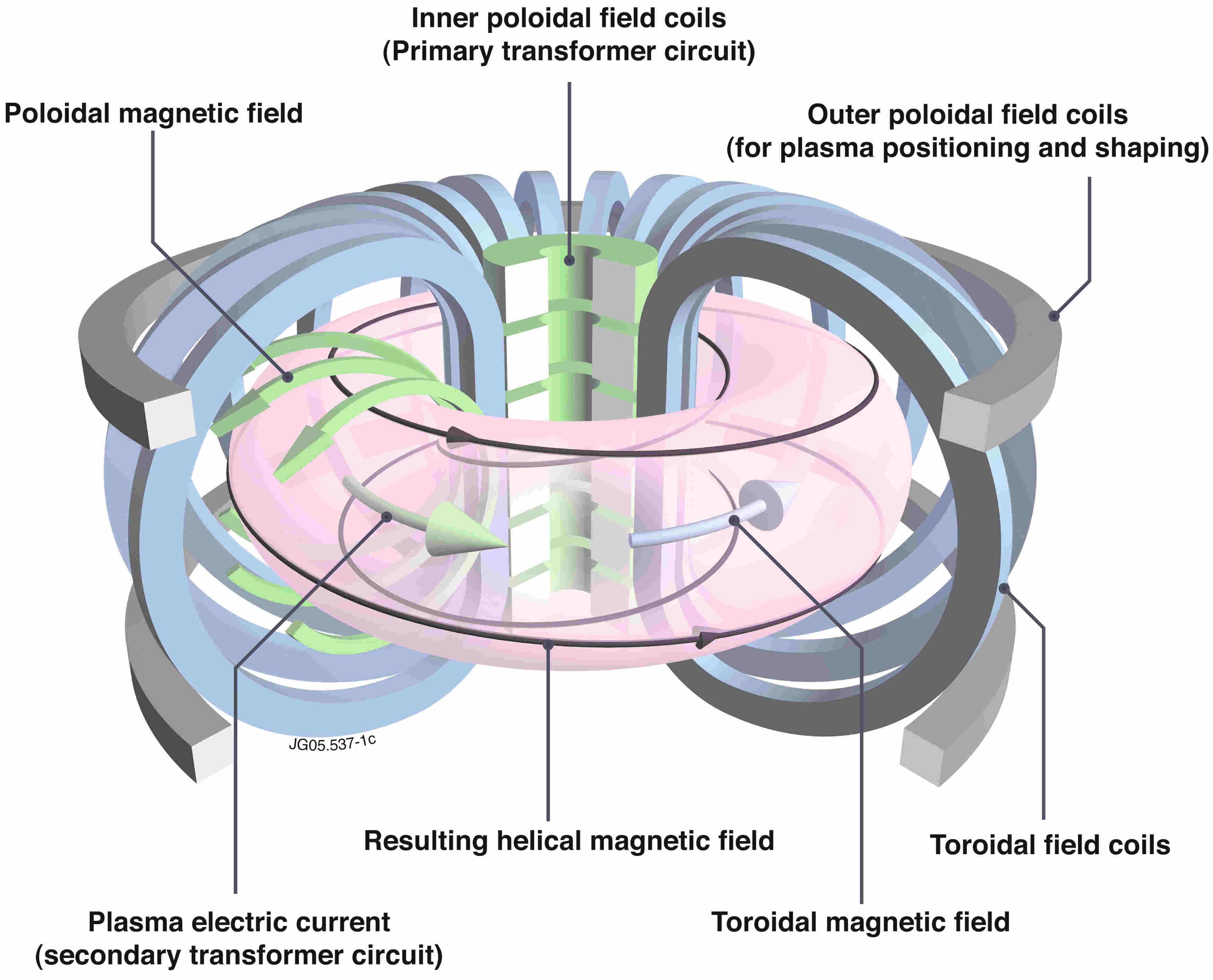}
\caption{Tokamak configuration (Courtesy of EUROfusion).}
\end{figure}
In order to understand some key features of the magnetic field in a Tokamak, we can approximate it with the field generated by the coils in the vacuum chamber:
\begin{equation}
\label{eq:132}
{\bm B} \approx \frac{B_{0}R_0}{R}\hat{\varphi}
\; ,
\end{equation}
where $B_{0}$ is the magnitude of the toroidal field on the magnetic axis, $R$ is the distance from the symmetry axis of the torus, $R_{0}$ is the distance of the magnetic axis from the symmetry axis and $\hat{\varphi}$ is the versor relative to the toroidal coordinate.
In a uniform magnetic field, charged particles are moving on a plane orthogonal to ${\bm B}$ along circular orbits with radius:
\begin{equation}
\label{eq:133}
\rho_{j}= \sqrt{\frac{T_{j}}{m_{j}\Omega_{j}^{2}}}
\; ,
\end{equation}
where $\Omega_{j}= q_{j}B/m_{j}c$ is the cyclotron frequency of the $j$ species in a magnetic field with magnitude $B$, $q_{j}$ is the charge and $m_{j}$ is the mass. Particles undergoing to this motion are confined if $\rho_{j}\ll a$, where $a$ is the minor radius of the Tokamak. However, even using Eq. (\ref{eq:132}), the magnetic field inside a Tokamak is not uniform and magnetic field lines are not straight. These effects are responsible of a force acting on the charged particles, which cause them to slowly drift across different magnetic field lines. It can be shown that a generic force ${\bm F}$ acting on a charged particle inside a magnetic field is responsible of a drift velocity of the "center of gyration" of the particle \cite{nicholson1983introduction}:
\begin{equation}
\label{eq:133a}
{\bm v}_{F}= \frac{c/q}{B^{2}}{\bm F}\times {\bm B}.
\end{equation}
In particular the following drift is always present in a magnetic configuration with curved magnetic field lines because of the centrifugal force:
\begin{equation}
\label{eq:134}
{\bm F}= - \mathcal{K}m v_{\parallel}^{2}
\end{equation}
where $v_{\parallel}$ is the component of the velocity parallel to the magnetic field ${\bm B}$ and $\mathcal{K} \equiv ({\bm b}\cdot {\bm \nabla}){\bm b}$. Therefore:
\begin{equation}
\label{eq:135}
v_{\mathcal{K}}= \left( \frac{{\bm B}\times \mathcal{K}}{B \Omega} \right)v_{\parallel}^{2}.
\end{equation}
It can be shown \cite{nicholson1983introduction} that the non-uniformity of the magnetic field act in the same way and that the total magnetic drift velocity is:
\begin{equation}
\label{eq:136}
{\bm v}_{D}= \frac{1}{\Omega}{\bm b}\times \left( \frac{v_{\perp}^2}{2}\frac{{\bm \nabla}B}{B} + \mathcal{K} v_{\parallel}^{2}\right).
\end{equation}
We stress the fact that the magnetic drift velocity introduced here is a consequence of the motion of the particles in the equilibrium magnetic field and that fluctuations will produce additional drifts such as the ${\bm E}\times{\bm B}$ drift \cite{nicholson1983introduction}.

In the next sections, we will study these topics with much more details but, for the moment, we remark the fact that the unperturbed motion of the particles in an axisymmetric device allows the existence of two integrals of motion which are the toroidal momentum $p_{\phi}$, the kinetic energy $E$ and one adiabatic integral, i.e. a quantity that is conserved by the single particle motion on time scales, which are long compared with the particle gyration frequency: the magnetic moment $\mu= \frac{1}{2}v_{\perp}^{2}$. From the conservation of the last two quantities, it follows that particles approaching a region with increasing magnetic field can "bounce back" being effectively trapped, if we neglect perpendicular dynamics, on a certain region of the field line, where the magnetic field intensity is low. The trapping condition \cite{helander2005collisional} for a particle is the following:
\begin{equation}
\label{eq:137}
\frac{1}{2}m v_{\parallel 0}^{2} < \mu B_{max}-\mu B_{min}
\; , 
\end{equation}
where $v_{\parallel 0}$ is the parallel velocity, when the particle passes through the point where $B=B_{min}$. It can be shown that the trapped particle fraction $f_{t}$ for an isotropic distribution function in a large aspect ratio machine with circular magnetic surfaces is such that, up to the leading order,: $f_{t} \approx \sqrt{\epsilon_{B}}$ with $\epsilon_{B} = r / R_{0} \ll 1$. Trapped particles are subject to magnetic drifts between collisions and, therefore, have a small deviation from the magnetic field line, which gives to their trajectories a characteristic banana shape:
\begin{figure}[htb]
\centering
\includegraphics[width=.9\linewidth]{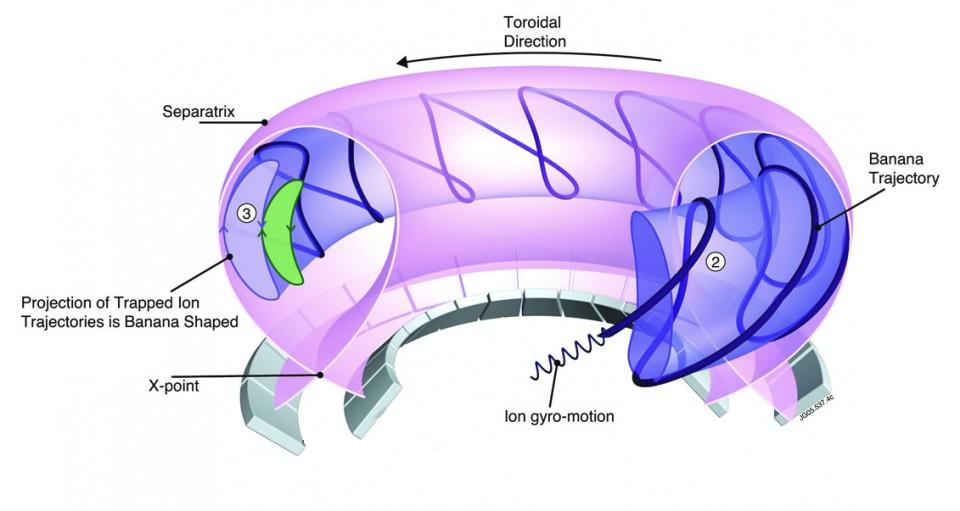}
\caption{Cartoon with the shape of the banana trajectories inside a Tokamak (Courtesy of EUROfusion).}
\end{figure}

In the next sections, we will describe the radial transport processes in detail. For the moment, we give some estimates based on a random walk model. Because of collisions, particles are randomly displaced by a step $\Delta x$.  The time between two collisions is $\Delta t$. The resulting diffusion coefficient describing the dynamics of the profiles such as density and temperature is $D \approx (\Delta x)^{2}/\Delta t$. Classical transport theory \cite{balescu1988transport} predicts that the step-size is the Larmor radius $\rho_{j}$ while the time step is the inverse of the collision frequency between the particles. As an example, the diffusion coefficient related to the collision between ions is:
\begin{equation}
\label{eq:138}
D_{c}\approx\rho_{i}^{2}\nu_{ii}.
\end{equation}
Classical transport theory is valid only if the mean free path is short compared with the characteristic length-scale of variation of the magnetic field and of the other equilibrium quantities. In this case the radial extension of an orbit is the Larmor radius of that particle, which is also the characteristic step size of particle random walk. Otherwise, particles are subject to magnetic drifts and, therefore, the radial extensions of their trajectory becomes larger. For this reason, neoclassical transport theory \cite{hinton1976theory} has been developed taking into account the particular geometry of the magnetic field and the existence of trapped particles. For the particles moving on banana orbits, we can define the bounce frequency $\nu_{b}\approx v_{th}/q R$ and the effective collision frequency $\nu^{eff}_{j}\approx \nu_{ej}/2 \epsilon$. If the former is bigger than the latter, the trapped particles will complete several times their banana trajectories in the time between two collisions and, therefore, the width of the banana orbit $r_{b}$ can be taken as the step-size of the random walk. Meanwhile, the time step is the inverse of the  effective collision frequency. Following \cite{hinton1976theory}, for the circular surfaces and large aspect ratio case, we can write:
\begin{equation}
\label{eq:139}
\Delta x \approx r_{b} \approx\frac{q \rho_{j}}{\sqrt{\epsilon}}.
\; ,
\end{equation}
We can compare the classical diffusion coefficient with the neoclassical one:
\begin{equation}
\label{eq:140}
D_{nc}\approx f_{t}r_{b}^{2}\nu^{eff}\approx \frac{q^{2}\rho_{j}^{2}\nu_{ej}}{\epsilon^{3/2}}\approx D_{cl} \frac{q^{2}}{\epsilon^{3/2}}
\end{equation}
where $f_{t}$ is the effective trapped particle fraction and $j$ is the subscript denoting the particle species. The same estimates can be done with different assumptions on the collisionality: the Pfirsch Schl\"uter regime, where $\nu_{b} \ll \nu^{eff}$ and the Plateau regime, which is is characterized by an intermediate collisionality.

The values of diffusion coefficients measured experimentally are one order of magnitude higher with respect to the those predicted by the collisional transport theory. More than this from the previous estimate we would expect $D_{i}/D_{e}$ of the order of the square root of the ratio between the masses of electrons and ions. This is not confirmed experimentally and, instead, the two coefficients appear to be almost identical. For these reasons, the plasma physicists community introduced the wording "anomalous transport" when dealing with transport processes which are not correctly described by classical and neoclassical transport theory. In order to understand these experimental observations, several phenomenological relations have been extracted from the experimental databases \cite{yushmanov1990scalings} but their understanding in terms of fundamental physics is not satisfactory. Values for the diffusion coefficients in agreement with the neoclassical predictions are observed in some restricted regions of the plasma, e.g. those enclosed by Internal Transport Barriers (ITBs) \cite{koide1994internal} which split the plasma into macro-regions with fast mixing process happening inside them. The mechanism that lead to the formation of ITBs is not clear. The stochasticity of the magnetic field lines has been proposed as responsible for the formation of the ITBs \cite{borgogno2011barriers,falessi2015lagrangian} governing electrons dynamics.
\section{Magnetic coordinates}
\subsubsection{Clebsch coordinates}
\label{sec-3-3-2}
Following \cite{d2012flux}, we can write the vector potential in terms of two scalar functions $\alpha,\beta$:
\begin{equation}
\label{eq:14}
{\bm A} = \alpha {\bm \nabla} \beta
\end{equation}
and, therefore, the well known relation between the magnetic field and the vector potential:
\begin{equation}
\label{eq:13}
{\bm B} = {\bm \nabla}  \times {\bm A}
\; ,
\end{equation}
becomes:
\begin{equation}
\label{eq:15}
{\bm B} = {\bm\nabla}\alpha\times{\bm\nabla}\beta.
\end{equation}
This expression has an interesting geometrical meaning: the magnetic field ${\bm B}$ lies in the intersection between the two surfaces $\alpha = {\mathrm const}$ and $\beta = {\mathrm const}$. 
We will use $(\alpha,\beta, \mathit{l} )$  as a set of coordinates obtaining:
\begin{equation}
\label{eq:16}
\tilde{ {\bm e}}^{(1)} = {\bm \nabla }\alpha, \,\, \tilde{ \bm e}^{(2)} = {\bm\nabla}\beta,\,\, \tilde{ \bm e}^{(3)} = {\bm\nabla}\mathit{l}
\; ,
\end{equation}
where we have indicated with $\tilde{ \bm e}^{(i)}$ the i-th vector of the coordinate basis. We choose the coordinate $\mathit{l}$ such that ${\bm b}= {\bm B}/B$ is the third vector of the reciprocal basis:
\begin{equation}
\label{eq:17}
\tilde{\bm e}_{(3)}= \frac{\bm B}{B}.
\end{equation}
The vectors of the reciprocal basis can be expressed in terms of the cross product between $\tilde{\bm e}^{(i)}$ and the determinant of the change of coordinates from the Cartesian ones, i.e. $J^{-1}=\left({\bm \nabla} \alpha \times {\bm \nabla}\beta\right)\cdot {\bm \nabla}\mathit{l}$:
\begin{eqnarray}
\label{eq:18a}
\tilde{\bm e}_{(1)} & = & J {\bm\nabla\beta}\times {\bm \nabla} \mathit{l} \; , \\
\nonumber \tilde{\bm e}_{(2)} & = & J {\bm\nabla \mathit{l}}\times {\bm \nabla} \alpha \; ,\\
\nonumber \tilde{\bm e}_{(3)} & = & J {\bm\nabla \alpha}\times {\bm \nabla} \beta.
\end{eqnarray}
We can write a generic vector in terms of its covariant or contravariant components:
\begin{equation}
\label{eq:19}
{\bm V} =\tilde{V}^{i}\tilde{\bm e}_{(i)}=\tilde{V}_{i}\tilde{\bm e}^{(i)}.
\end{equation}
For the magnetic field,  in particular, by construction of the set of coordinates $(\alpha,\beta,\mathit{l})$, we have:
\begin{equation}
\label{eq:20a}
{\bm B} =\tilde{B}^{3} \tilde{\bm e}_{(3)} \Rightarrow \tilde{B}^{3} = B
\; .
\end{equation}
Substituting the expression for $J$ into Eq. (\ref{eq:18a}) and using Eq. (\ref{eq:20a}), we obtain the following expression for ${\bm b}$:
\begin{equation}
\label{eq:10}
{\bm b}=\tilde{\bm e}_{(3)}= \frac{{\bm \nabla} \alpha \times {\bm \nabla} \beta}{{\bm \nabla}\mathit{l}\cdot{\bm \nabla} \alpha \times {\bm \nabla} \beta}
\end{equation}
and, therefore, it follows that ${\bm b} \cdot {\bm \nabla}\mathit{l}=1$. Using this result we can calculate $J$:
\begin{equation}
\label{eq:39}
J = \frac{1}{{\bm \nabla}\mathit{l}\cdot{\bm \nabla} \alpha \times {\bm \nabla} \beta}=\frac{1}{{\bm B}\cdot {\bm \nabla} \mathit{l}}=B^{-1}.
\end{equation}
Substituting this expression into Eq. (\ref{eq:18a}), we obtain $\tilde{B}_{i}$:
\begin{eqnarray}
\label{eq:38}
\tilde{B}_{1} & = & {\bm b} \cdot \left( {\bm \nabla}\beta \times {\bm \nabla} l \right) \; ,\\
\nonumber \tilde{B}_{2} & = & {\bm b} \cdot \left( {\bm \nabla}\mathit{l} \times {\bm \nabla} \alpha \right) \; ,\\
\nonumber \tilde{B}_{3} & = & B .
\end{eqnarray}
We now show that the magnetic field lines are horizontal straight lines in the plane $(\beta,l)$:
\begin{equation}
\label{eq:22}
\frac{d \beta}{d \mathit{l}} = \frac{\tilde{B}^{\beta}}{\tilde{B}^{\mathit{l}}}=\frac{\bm B \cdot {\bm \nabla}\beta}{\bm B \cdot {\bm \nabla} \mathit{l}} =\frac{{\bm \nabla} \alpha \times {\bm \nabla} \beta \cdot {\bm \nabla \beta}}{{\bm B} \cdot \nabla \mathit{l}} = 0
\end{equation}
and, therefore, $\beta$ is constant along a magnetic field line.
\subsubsection{Flux coordinates}
\label{sec-3-3-3}
Given a flux function $\rho$, i.e. a scalar function such that:
\begin{equation}
\label{eq:24}
{\bm B} \cdot {\bm \nabla}\rho=0
\; ,
\end{equation}
we will show why it is useful to choose it as a coordinate. Choosing $\rho$ as the radial coordinate from Eq. (\ref{eq:24}), it follows that $\tilde{B}^{1}=0$ and, therefore:
\begin{equation}
\label{eq:25}
{\bm B}= \tilde{B}^{2}J {\bm \nabla}\phi \times {\bm\nabla} \rho + \tilde{B}^{3} J{\bm \nabla}\rho \times {\bm \nabla}\theta
\; ,
\end{equation}
where we are using the notation already introduced and the coordinate set: $(\rho,\theta,\phi)$. The condition on the divergence of ${\bm B}$ can be solved in terms of an unknown function $\nu$:
\begin{equation}
\label{eq:26}
{\bm \nabla \cdot B}=0 \Rightarrow \tilde{B}^{2}= - \frac{1}{J}\frac{\partial \nu}{\partial \tilde{x}^{3}}, \,\, \tilde{B}^{3}=  \frac{1}{J}\frac{\partial \nu}{\partial \tilde{x}^{2}}.
\end{equation}
We can substitute these expressions into Eq. (\ref{eq:25}), obtaining the usual expression for ${\bm{B}}$:

\begin{equation}
\label{eq:27}
{\bm B}= {\bm \nabla} \rho \times {\bm \nabla} \nu.
\end{equation}
We have therefore shown that, given a flux function $\rho$, it is possible, if we are able to obtain an expression for $\nu$ from Eq. (\ref{eq:26}), to express the magnetic field in Clebsch form. In this work, we will not enter into details of the procedure used in order to express $\nu$ in terms of other flux functions, which can be found in Ref.\cite{d2012flux}. Defining the poloidal and toroidal magnetic fluxes respectively as:
\begin{equation}
\label{eq:23}
2 \pi \, \psi \equiv \psi^{r}_{pol}=\iint_{S^{r}_{pol}} {\bm B} \cdot d S,\quad   \psi_{tor}=\iint_{S_{tor}} {\bm B} \cdot d S
\; ,
\end{equation}
with the domain of integration shown in the following figure:
\begin{figure}[htb]
\centering
\includegraphics[width=.9\linewidth]{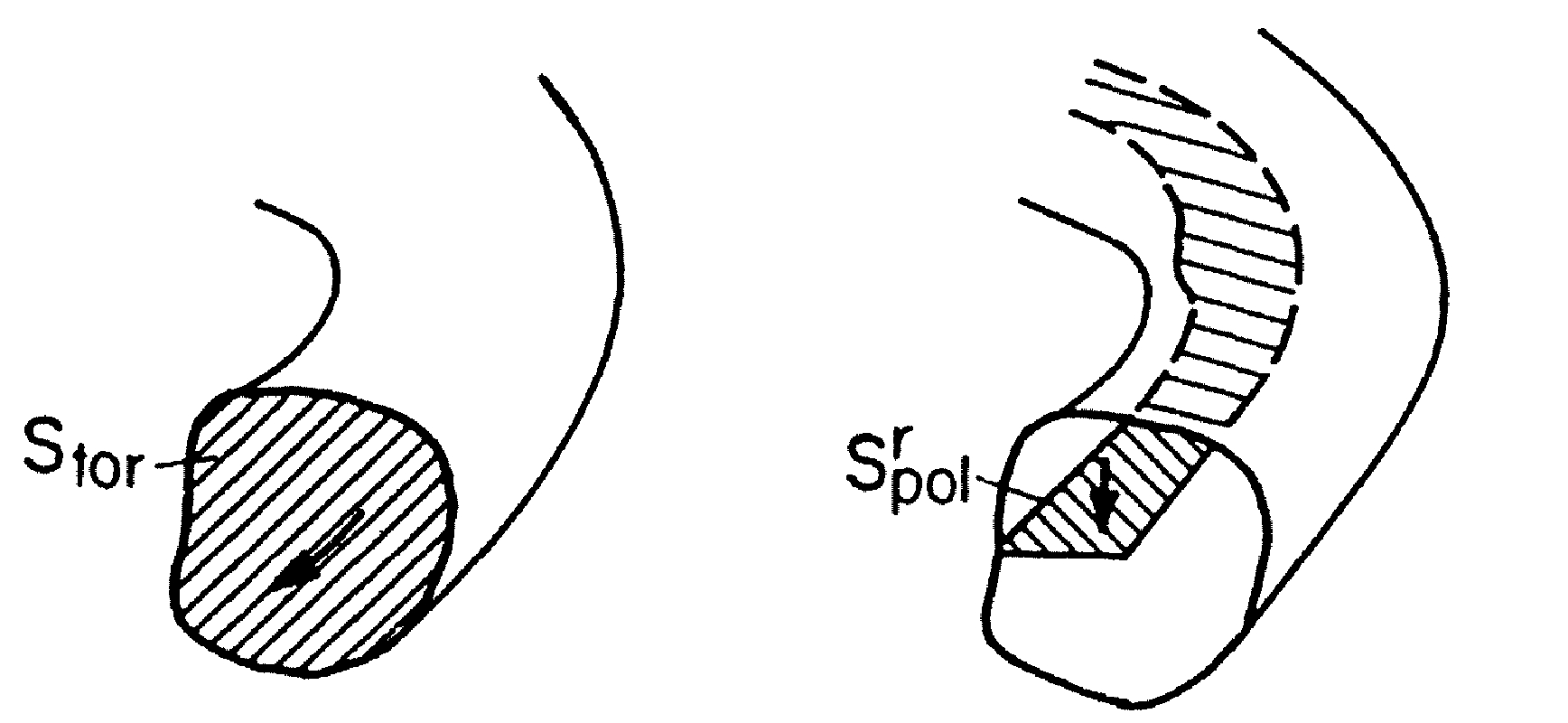}
\caption{Cartoon with $S^{r}_{pol}$ and $S_{tor}$.}
\end{figure}

the function $\nu$ can be expressed with the following formula:
\begin{equation}
\label{eq:28}
\nu = \frac{1}{2\pi}\left( \dot{\psi}_{tor}\theta - \dot{\psi}^{r}_{pol}\phi\right) + \nu^{*}
\; ,
\end{equation}
with the function $\nu^{*}$ being periodic in $\theta$ and $\phi$ and the dot indicating a derivative with respect to the flux function $\rho$. We already showed that $\nu$ should be constant along a magnetic field line. Therefore, if $\nu^{*}$ is a flux function from Eq. (\ref{eq:28}), it immediately follows that, along a magnetic field line:
\begin{equation}
\label{eq:29}
\dot{\psi}_{tor} \theta - \dot{\psi}^{r}_{pol}\phi=cost
\; .
\end{equation}
Therefore, the magnetic field lines are straight lines in the plane $(\theta,\phi)$ because $\dot{\psi}_{tor}^{r}, \dot{\psi}_{pol}^{r}$ are constant on a given flux surface. If $\nu^{*}$ is not a flux function, we need to make a change of coordinates in order to obtain the same property. In this work, we choose to leave $\phi$ unchanged but we stress the fact that this is not the only possibility. The new angular coordinate is:
\begin{equation}
\label{eq:21}
\bar{\theta}=\theta+2\pi \frac{\nu^{*}}{\dot{\psi}_{tor}}.
\end{equation}
In order to have a compact notation, we maintain the notation $\theta$ for this new variable. In these new set of variables $\nu^{*}$ disappears from Eq. (\ref{eq:28}) and, in the plane $(\theta,\phi)$, the magnetic field lines are straight lines:
\begin{equation}
\label{eq:30}
\frac{d \theta}{d \phi}= \frac{{\bm B}\cdot {\bm \nabla}\theta}{{\bm B}\cdot {\bm \nabla}\phi}=\frac{\dot{\psi}^{r}_{pol}}{\dot{\psi}_{tor}}= q^{-1}.
\end{equation}
In this relation a new flux function, i.e. the safety factor $q$, has been introduced. The magnetic field can be written in the usual Clebsch form:
\begin{equation}
\label{eq:32}
{\bm B} = {\bm \nabla} \rho \times {\bm \nabla} \left( \frac{\dot{\psi}_{tor}}{2 \pi}\theta - \frac{\dot{\psi}_{pol}^{r}}{2 \pi}\phi\right)= \frac{1}{2 \pi}\left(\dot{\psi}_{pol}^{r} {\bm \nabla}\phi \times{\bm \nabla}\rho+ \dot{\psi}_{tor}{\bm \nabla}\rho \times{\bm \nabla}\theta \right).
\end{equation}
Recalling Eq. (\ref{eq:25}), we obtain an expression for $J$:
\begin{equation}
\label{eq:40}
J = \frac{\dot{\psi}^{r}_{pol}}{2 \pi}\left({\bm B}\cdot{\bm \nabla}\theta\right)^{-1}.
\end{equation}
From now on, we choose $\rho=\psi$ as it is usually done in the Tokamak community obtaining:

\begin{equation}
\label{eq:33}
{\bm B} = {\bm \nabla \psi}\times {\bm \nabla} \left( q(\psi)\theta - \phi\right), \quad J = \left({\bm B}\cdot{\bm \nabla}\theta\right)^{-1}.
\end{equation}
In this work, we will assume axisymmetry, $\partial_{\phi}\Big|_{\theta,\psi}=0$ and, therefore, the following relations hold:
\begin{equation}
\label{eq:36}
{\bm \nabla \psi} \times {\bm \nabla \theta} = k {\bm \nabla \phi}, \quad k= \frac{1}{J({\bm \nabla \phi})^{2}}.
\end{equation}
Choosing $\phi$ as the toroidal angle, we know that $\left({\bm \nabla} \phi\right)^{2}=R^{-2}$ and we can therefore, finally, write the expression for the magnetic field that will be used in this work:
\begin{equation}
\label{eq:37}
{\bm B}= F {\bm \nabla}\phi + {\bm \nabla}\phi \times {\bm\nabla}\psi, \quad F= \frac{q R^{2}}{J}.
\end{equation}
\subsubsection{Arbitrary coordinates}
\label{sec-3-3-4}
We can make an additional change of coordinates, retaining the same flux function as radial coordinate, without changing Eq. (\ref{eq:29}):
\begin{eqnarray}
\label{eq:43}
  \bar{\theta} & = & \theta - \Pi/q
                     \; , \\
\nonumber \bar{\phi} & = & \phi - \Pi
\; , 
\end{eqnarray}
with $\Pi$ being a periodic function of $\theta$ and $\phi$. Thus the magnetic field lines will be straight in the plane $(\bar{\theta},\bar{\phi})$. Using the axisymmetry of our system, we can show that:

\begin{equation}
\label{eq:44}
\frac{1}{q} \frac{\partial \Pi}{\partial \theta} = 1- \frac{J}{\bar{J}}
\end{equation}
with $\bar{J}= {\bm \nabla}\bar{\theta}\times{\bm \nabla}\psi\cdot {\nabla \bar{\phi}}$. This equation relates $\bar{J}$ to $\Pi$. Once we have chosen $\bar{J}$ as the most convenient for our calculation, we can integrate Eq. (\ref{eq:44}) in order to obtain the corresponding $\Pi$. The Boozer coordinates: \cite{boozer1981plasma,boozer1982establishment} are defined by:
\begin{equation}
\label{eq:45}
\bar{J}= \frac{J_B}{B^{2}}
\; ,
\end{equation}
where $J_{B}$ is a flux function and we recall that $J= qR^{2}/ F$. Hamada coordinates \cite{hamada1962hydromagnetic}, meanwhile, are defined by the following Jacobian:
\begin{equation}
\label{eq:46}
\bar{J}= J_H
\; ,
\end{equation}
where $J_{H}$ is a flux function.
In order to use a compact notation from now on, we will call the new set of coordinates $(\psi,\theta,\phi)$, i.e. removing the bar. In \autoref{cap:coll}, we will use Hamada coordinates to calculate collisional transport in an application of practical interest. It will be useful to find the relations between these coordinates and the set of coordinates, $(\psi,\theta_{0},\phi)$ which is the set of polar coordinates centered around the magnetic axis on a surface with fixed $\phi$. Using the expression for the Jacobians of the two set of coordinates we obtain the following relation:
\begin{equation}
\label{eq:47}
\frac{\partial \theta}{\partial \theta_{0}}= \frac{J_0}{J_{B}}B^{2}
\; , 
\end{equation}
with $J_{0}={\bm \nabla}\theta_{0}\times{\bm \nabla}\psi\cdot {\nabla \phi}$. For the angular coordinate $\theta$, the following relation must be valid:
\begin{equation}
\label{eq:48}
\int_{0}^{2 \pi}d \theta_{0}\frac{\partial \theta}{\partial \theta_{0}}= 2 \pi
\; ,
\end{equation}
and, thus, it follows that:
\begin{equation}
\label{eq:49}
J_{B}= \frac{1}{2 \pi}\int_{0}^{2 \pi}d \theta_{0}J_{0}B^{2}.
\end{equation}
Using these results, we can compute the value of the coordinate $\theta$ at each point of the space:
\begin{equation}
\label{eq:50}
\theta = 2 \pi \frac{\int_{0}^{\theta} d \theta_{0} J_{0}B^{2}}{\int_{0}^{2 \pi} d \theta_{0} J_{0}B^{2}}.
\end{equation}
The Jacobian $J_{0}$ can be expressed in terms of the equilibrium magnetic field:
\begin{equation}
\label{eq:51}
J_{0}^{-1}= \frac{1}{R}\left(\frac{\partial \psi}{\partial R}\frac{\partial \theta_{0}}{\partial z}- \frac{\partial\psi}{\partial z}\frac{\partial \theta_{0}}{\partial R}\right)
\; ,
\end{equation}
and, therefore, Eq. (\ref{eq:50}) can be computed, analytically or numerically.

\chapter{Gyrokinetic transport theory}
\label{cap:gyrotransp}
In this Chapter we lay the foundations of gyrokinetic transport theory as originally formulated in this thesis work. 
We adopt a moment equation approach to transport equations \cite{hinton1976theory}, combined with 
the conceptual framework of nonlinear gyrokinetic theory \cite{frieman1982nonlinear,brizard2007foundations}. In this way, 
we provide a compact and physically transparent derivation of cross-field particle
and energy fluxes, which include collisional and fluctuation induced transport processes on the same footing. 
The resulting transport equations are original
results of this thesis work, which generalize and extend the analysis of Ref. \cite{hinton1976theory} to fluctuation
induced transport. Our analysis also recovers, in the appropriate limit, the results originally proposed in Refs. \cite{plunk2009theory,barnes2009trinity,abel2013multiscale}. As a crucial element of novelty, our results 
demonstrate the importance of self-consistent determination of 
spatiotemporal scales of equilibrium variations and of the corresponding structures, by means of nonlinear gyrokinetic theory.
Given the conceptual importance of these issues, general remarks on transport equations derived in this thesis work 
are given in the final section of this Chapter, where we also illustrate the connection of our results with earlier analysis
of collisional \cite{hinton1976theory} and fluctuation induced transport \cite{plunk2009theory,barnes2009trinity,abel2013multiscale}.
There, we also anticipate the necessity of a truly nonlinear gyrokinetic analysis of equilibrium distortions on mesoscales, which will then be presented in Chapter 4.
\section{Moments of the distribution function}
Following Hazeltine and Meiss \cite{hazeltine2003plasma} we introduce the generic tensor notation:
\begin{equation}
\label{eq:112}
M^{(N)}_{\alpha \beta \dots \tau} \equiv \left. \int d \bm v f  v_\alpha v_\beta \dots v_\tau  \right|_N 
\end{equation}
for the $Nth$ moment of the distribution function of a species constituting the plasma, where the $|_N$ notation denotes the $N$ factors of $\bm v$. $M^{(N)}$ is a tensor of rank at most $N$. Examples are (repeated subscripts are implicitly summed up):
\begin{equation*}
M^{(0)}= n \;\;\;\;\; M^{(1)}_\alpha = n V_\alpha \;\;\;\;\; M^{(2)}_{\alpha \beta} = m^{-1} P_{\alpha \beta}
\end{equation*}
\begin{equation*}
M^{(3)}_{\alpha \beta \beta}= (2/m) Q_\alpha \;\;\;\;\; M^{(4)}_{\alpha \beta \tau \tau}= (2/m) R_{\alpha \beta}
\end{equation*}
where $n, V_{\alpha}, P_{\alpha\beta}, R_{\alpha\beta} \dots$ are the components of the following tensors:
\begin{equation}
  \label{eq:18}
\nonumber n {\bm V}= \int d {\bm v} \,{\bm v}f, \quad {\bm P}= \int d {\bm v} \,{\bm v}{\bm v}f, \quad {\bm R}= \int d {\bm v} \, \frac{1}{2}m v^2 {\bm v}{\bm v}, \quad {\bm Q}= \int d {\bm v}\, \frac{1}{2}m v^2{\bm v} f.
\end{equation}
which, respectively, are the density, the average velocity, the pressure and the energy weighted stress tensor. Similarly, one introduces the $Nth$ moment of the collision operator:
\begin{equation}
\label{eq:113}
C^{(N)}_{\alpha \beta \dots \tau} \equiv \left. \int d \bm v  C v_\alpha v_\beta \dots  v_\tau  \right|_N 
\end{equation}
\begin{equation*}
C^{(0)}= 0 \;\;\;\;\; C^{(1)}_\alpha = m^{-1} F_\alpha \;\;\;\;\; C^{(2)}_{\alpha \alpha} = (2/m) ( W + F_\alpha V_\alpha)
\;\;\;\;\;  C^{(3)}_{\alpha \beta \beta}= (2/m) G_\alpha.
\end{equation*}
where:
\begin{align}
  \label{eq:18}
\nonumber {\bm F}= \int d {\bm v} \,m {\bm v}\, C, \quad {\bm G}= \int d {\bm v}\, \frac{1}{2}m v^2{\bm v}\, C, \quad W= \int d {\bm v}\, \frac{1}{2}m v^2 C.
\end{align}
Multiplying the Landau kinetic equation for the proper velocity function and integrating over the velocity space we obtain:
\begin{equation}
\label{eq:141a}
\frac{\partial}{\partial t} M^{(N)}_{\alpha \beta \dots \tau} + \frac{\partial}{\partial x_\mu}  M^{(N+1)}_{\mu \alpha \beta \dots \tau}- \frac{e}{m} \left \llbracket E_\mu M^{(N-1)}_{\alpha \beta \dots \tau} \right\rrbracket_\mu - \Omega \left \llbracket \epsilon_{\mu \kappa \lambda} M^{(N)}_{ \kappa \alpha \beta \dots \tau} \right\rrbracket_\mu b_\lambda  = C^{(N)}_{\alpha \beta \dots\tau}
\end{equation}
where the notation $\llbracket E_\mu M^{(N-1)}_{\alpha \beta \dots \tau} \rrbracket_\mu$ stands for:
\begin{equation}
\label{eq:142}
\llbracket E_\mu M^{(N-1)}_{\alpha \beta \dots \tau} \rrbracket_\mu = E_\mu \delta_{\alpha \mu} M^{(N-1)}_{\beta \dots \tau} + E_\mu \delta_{\beta \mu} M^{(N-1)}_{\alpha \gamma \dots \tau} + \dots + E_\mu \delta_{\tau \mu} M^{(N-1)}_{\alpha \gamma \dots \sigma}.
\end{equation}
This is consistent with \cite{hinton1976theory,balescu1988transport}. We note, as already stated, that the evolutive equations for each moment have always a term depending on higher order moments. Solving the infinite set of equations for the moments is equivalent to solve the kinetic equation.

Choosing $N=0$ we obtain the continuity equation:
\begin{equation}
\label{eq:143}
\partial_t n + \bm \nabla \cdot ( n \bm V ) = 0
\; ;
\end{equation}
with $N=1$ the force balance equation:
\begin{equation}
\label{eq:144a}
\partial_t \left( n m \bm V \right) + \bm \nabla \cdot \bm P - e n \left( \bm E + \bm V \times \bm B/c \right)  = \bm F
\; ;
\end{equation}
with $N=2$ the pressure equation:
\begin{equation}
\label{eq:145}
\partial_t \left( \frac{3}{2} p + \frac{m}{2} n V^2 \right)  + \bm \nabla \cdot \bm Q - e n\bm E \cdot \bm V  = W + \bm F \cdot \bm V
\; ;
\end{equation}
and with $N=3$, the energy transport equation:
\begin{equation}
\label{eq:146}
\partial_t \bm Q  + \bm \nabla \cdot \bm R - (e/m) \bm E \cdot \left[\bm P + \bm I \left( \frac{3}{2} p + \frac{m}{2} n V^2 \right) \right]  - (e/mc) \bm Q \times \bm B =  \bm G
\; ;
\end{equation}
where $p = \frac{1}{3} \mathrm{Tr} \,\wb{p}$ and:
\begin{equation}
  \label{eq:18}
\nonumber {\bm p}= \int d {\bm v}\,({\bm v}-{\bm V})({\bm v}-{\bm V})f.
\end{equation}
We note that, following the notation introduced in \cite{hazeltine2003plasma}, we have indicated with the capital letter all the moments of the distribution function in the laboratory frame of reference while we have indicated with the lower case the moments in the frame of reference locally co-moving with the plasma species.
\section{Drift ordering}
In this thesis we will deal with strongly magnetized plasmas, i.e. plasmas such that particles Larmor radii $\rho$ are much smaller than the characteristic length scale of variation of the macroscopic quantities $L$:
\begin{equation}
\label{eq:3}
\rho = \frac{v_{th}}{\Omega} \ll L.
\end{equation}
We now introduce the drift ordering between physical quantities:
\begin{equation}
\label{eq:4}
\frac{\rho}{L} =  \frac{v_{th}/L}{\Omega} \equiv \frac{\omega}{\Omega} \sim {\cal O}(\delta)
\end{equation}
where $\omega \equiv v_{th}/L$ is the characteristic frequency of the particles interacting with structures on the length scale $L$. Due to the fact that we are studying a magnetically confined plasma, we assume that the macroscopic quantities are varying on a long time scale compared with the fluctuating ones, which have characteristic frequency $\omega$:
\begin{equation}
\label{eq:1}
\omega^{-1}\frac{\partial }{\partial t}\ln{p}= \omega^{-1}p^{-1}\frac{\partial p}{\partial t}  \sim {\cal O}(\delta^{2}).
\end{equation}
The last assumption is that particles' drift due to ${\bm E} \times {\bm B}$ force is small compared with the thermal velocity:
\begin{equation}
\label{eq:5}
\frac{c E}{B v_{th}}\sim {\cal O}(\delta).
\end{equation}
This ordering is consistent with the gyrokinetic ordering, i.e. \cite{frieman1982nonlinear}. These assumptions imply the distribution function to be Maxwellian at the leading order:
\begin{equation}
\label{eq:6a}
f = f_M + {\cal O}(\delta),  \quad f_{M}= n_{0}(\pi^{1/2}v_{th})^{-3}e^{-(v/v_{th})^{2}}.
\end{equation}
We also introduce an auxiliary ordering on the derivatives operators:
\begin{equation}
\label{eq:2}
\frac{\wb{\nabla}_{\parallel}}{\wb{\nabla}_{\perp}} \sim \frac{\wb{k_{\parallel}}}{k_{\perp}} \sim \mathcal{O}(\delta).
\end{equation}
Using Eq. (\ref{eq:6a}), assuming that the parallel flow is strongly subsonic and that there is small pressure anisotropy between the directions perpendicular and parallel to ${\bm B}$ due to the collisions, as shown in \cite{chew1956boltzmann, hinton1976theory}, we obtain:
\begin{equation}
\label{eq:7}
\left\{ n \bm V, \bm F, \bm Q, [ \bm P - \bm I p ], [ \bm R - \bm I (5/2) p (T/m) ]\right\} \sim {\cal O}(\delta).
\end{equation}
Here, space and time are normalized to $|\rho|$ and $|\Omega|^{-1}$, respectively, density is expressed in units of its local equilibrium value, etc. At the lowest order of the energy flux conservation we obtain:
\begin{equation}
\label{eq:9}
\bm E=- \bm \nabla \Phi + \cal{O}(\delta);
\end{equation}
and thus a consequence of the drift ordering is that the equilibrium electric field is mainly electrostatic. Taking the lowest order of the force balance equation and of the energy flux conservation projected along the magnetic axis we obtain, respectively:
\begin{eqnarray}
\label{eq:11}
  \bm b \cdot ( \bm \nabla p + e n  \bm \nabla \Phi) & = & {\cal O}(\delta)
  \; ,\\
\nonumber \bm b \cdot ( \bm \nabla ( p T ) + e p \bm \nabla \Phi) & = & {\cal O}(\delta)
\; ,
\end{eqnarray}
which can be cast in the following form:
\begin{eqnarray}
\label{eq:12}
  \bm b \cdot \bm \nabla T & = & {\cal O}(\delta)
  \; , \\
\nonumber \bm b \cdot \bm \nabla  \left( n \exp (e \Phi /T) \right) & = & {\cal O}(\delta)
 \; . \\ \nonumber
\end{eqnarray}
These equations yield $\bm b \cdot \bm \nabla n = {\cal O}(\delta )$ and $\bm b \cdot \bm \nabla \Phi = {\cal O}(\delta)$. That is, at the lowest order, temperature and density are constant along magnetic field lines and, because of their ergodic properties \cite{hinton1976theory}, they are also constant on magnetic surfaces.

The essence of magnetic confinement stands in the expression of first order fluxes (there are no zero order fluxes in slowly rotating plasmas), which are contained in the flux surface and are given by the expressions that will be calculated in the next sections. Thus, determining cross-field (radial) transport implies computing the second order fluxes (or higher). In particular we will show that:
\begin{align}
\label{eq:114}
  & n \bm V_\perp = (n \bm V_\perp)_1 + (n \bm V_\perp)_c + (n \bm V_\perp)_{NC} + {\cal O}(\delta^3) \\
  & \bm Q_\perp = (\bm Q_\perp)_1 + (\bm Q_\perp)_c + (\bm Q_\perp)_{NC} + {\cal O}(\delta^3)
\end{align}
Note that $\mathcal{O}(\delta^{2})$ fluxes are, by the structure of momentum equations, sufficient to compute the evolution of the system on a time scale $ \mathcal{O}(1/\delta^2)\omega^{-1}$, i.e. $\mathcal{O}(1/\delta^3)\Omega^{-1}$. In ITER $T_i \simeq 20 keV$, $B = 5.3 T$, $L \sim a = 2 m$, where $a$ is the minor radius of the machine and thus $\Omega_i \simeq 5 \times 10^8 s^{-1}$, $v_{thi} \simeq 2 \times 10^6 m s^{-1}$, $\delta_i = \rho_i/a \simeq 2 \times 10^{-3}$, $\omega \simeq 1 \times 10^6 s^{-1}$, $\tau_{transp} \simeq \delta_i^{-2} \omega^{-1} \simeq 0.25 s$. Therefore this is a time scale of the order of one second. Investigating cross-field transport on longer time scales, i.e. of order 100s, would require knowledge of fluxes up to order $\mathcal{O}(\delta^{3})$ included. This topic will be explored with more details in \autoref{cap:longer}.

In the following, we present an example to illuminate the advantages of dealing with moments equations instead of solving the kinetic equation and, then, taking the moments of the solution. We will use this method in the next section in order to calculate particle and energy transport equations. Following \cite{hazeltine2003plasma} we take the cross product of Eq. (\ref{eq:144a}) with the unit equilibrium magnetic field vector ${\bm b}$ obtaining:
\begin{equation}
\label{eq:147}
n {\bm V}_{\perp}= \frac{1}{m \Omega} {\bm b} \times \left[\partial_{t} \left( n m \bm V \right) + \nabla \cdot {\bm P} - e n {\bm E} - {\bm F}\right].
\end{equation}
Applying the drift ordering and noting that there is a $1/\Omega$ factor outside the square brackets, we need to evaluate the terms inside the square brackets up to order $\delta^{n-1}$ in order to calculate the perpendicular flux up to $\delta^{n}$. For example, at the first order in $\delta$, we obtain:
\begin{equation}
\label{eq:148}
n {\bm V}_{\perp} = \frac{1}{m \Omega}{\bm b}\times \left(  \nabla p +en \nabla \Phi\right)
\end{equation}
and, therefore, only information about the zeroth-order distribution function is required in order to calculate the lowest order perpendicular flux, which, as anticipated above, is confined within the flux magnetic flux surface due to Eqs. (\ref{eq:11}) and (\ref{eq:12}). In the next sections we will systematically use this methodology in order to calculate fluxes up to the second order in $\delta$. Therefore we will need information about the first-order distribution function.

\section{Transport equations}
\label{sec-4-3}
\subsection{Flux surface average}
\label{sec-4-3-2}
In order to derive the equations describing particle and energy transport across magnetic flux surfaces, we need to introduce an averaging operation and some of its properties. In toroidal coordinates, i.e. Eq. (\ref{eq:37}), we can take the flux surface average of a physical quantity $\left\langle ... \right\rangle_\psi$ :
\begin{equation}
\label{eq:31}
\left\langle ... \right\rangle_\psi = \frac{1}{V'} \oint J d\theta d\phi (...) = \frac{1}{V' } \oint  \frac{d\theta d\phi}{\bm B \cdot \bm \nabla \theta}(...)
\end{equation}
where $J$ is the Jacobian of the change of coordinates and where:
\begin{equation}
\label{eq:34}
V' = \frac{dV}{d\psi} =\oint\frac{ d\theta d\phi}{\bm B \cdot \bm \nabla \theta}.
\end{equation}
Applying the divergence theorem on a volume $\Delta V = V'\Delta\psi$ we obtain:
\begin{equation}
\label{eq:35}
\frac{1}{\Delta V}\left\langle \bm \nabla \cdot ( ... ) \right\rangle_{\Delta V} =  \frac{1}{V'} \frac{\partial}{\partial \psi} \left[ V' \left\langle \bm \nabla \psi \cdot ( ... ) \right\rangle_\psi\right]
\; ,
\end{equation}
where the subscript $\Delta V$ to angular brackets denotes volume integral over $\Delta V$. Thus,
\begin{equation}
\label{eq:41}
\left\langle \bm \nabla \cdot ( ... ) \right\rangle_{\psi} = \frac{1}{\Delta V}\left\langle \bm \nabla \cdot ( ... ) \right\rangle_{\Delta V}.
\end{equation}

\subsection{Density transport}
\label{sec-4-3-3}
Acting with $R^2 \bm \nabla \phi \, \cdot$ on the momentum equation and taking the flux surface average, we obtain:
\begin{align*}
\label{eq:20}
\left\langle R^2 \bm \nabla \phi \cdot \partial_t (n m \bm V ) \right\rangle_\psi + \left \langle R^2 {\bm \nabla}\phi \cdot {\bm \nabla}\cdot {\bm P}\right \rangle_\psi =   \left\langle R^2 \bm \nabla \phi \cdot(en \bm E + \bm F ) \right\rangle_\psi + \left\langle R^2 \bm \nabla \phi \cdot \left(e n /c {\bm V} \times {\bm B} \right) \right\rangle_\psi .
\end{align*}
Using the drift ordering, it can be shown that the tensor ${\bm P}$ up to $\mathcal{O}(\delta)$ is symmetric. Furthermore using the explicit form:
\begin{equation}
\label{eq:90}
\bm \nabla ( R^2 \bm \nabla \phi )= (\bm \nabla R) (R \bm \nabla \phi) + R \bm \nabla \phi \frac{\partial}{\partial \phi} (R \bm \nabla \phi) = (\bm \nabla R) (R \bm \nabla \phi) -
(R \bm \nabla \phi) (\bm \nabla R)\; ,
\end{equation}
we can see that $\bm \nabla ( R^2 \bm \nabla \phi )$ is anti-symmetric and, thus, vanishes when contracted on any symmetric tensor. Using this result and Eq. (\ref{eq:90}), we can manipulate the second term of the LHS obtaining:
\begin{eqnarray}
\label{eq:93}
\left\langle R^2 \bm \nabla \phi \cdot \partial_t (n m \bm V ) \right\rangle_\psi & + & \frac{1}{V'} \frac{\partial}{\partial \psi} \left\langle V' \bm \nabla \psi \cdot \bm P \cdot R^2 \bm \nabla \phi\right\rangle_\psi  = \left\langle R^2 \bm \nabla \phi \cdot(en \bm E + \bm F ) \right\rangle_\psi+ \\
\nonumber & & + \left\langle R^2 \bm \nabla \phi \cdot \left(e n /c {\bm V} \times {\bm B} \right) \right\rangle_\psi .
\end{eqnarray}
In the derivation, we will use the following identity which can be verified substituting the general expression for the equilibrium magnetic field in flux coordinates, i.e. Eq. (\ref{eq:37}):
\begin{equation}
\label{eq:94}
\bm B_0 \times R^2 \bm \nabla \phi = \bm \nabla \psi.
\end{equation}
With some algebra, we can also obtain the more general expression:
\begin{equation}
\label{eq:100}
{\cal V} \cdot \bm \nabla \psi =  {\cal V} \times \bm B_0 \cdot R^2 \bm \nabla \phi
\end{equation}
which holds for any generic vector $\mathcal{V}$. We now separate the equilibrium fields from the fluctuating ones and we use Eq. (\ref{eq:94}) in order to obtain:
\begin{eqnarray}
\label{eq:95}
\left\langle R^2 \bm \nabla \phi \cdot \partial_t (n m \bm V ) \right\rangle_\psi & + & \frac{1}{V'} \frac{\partial}{\partial \psi} \left\langle V' \bm \nabla \psi \cdot \bm P \cdot R^2 \bm \nabla \phi\right\rangle_\psi  = \left\langle (en \bm E + \bm F ) \cdot R^2 \bm \nabla \phi \right\rangle_\psi \\
\nonumber & & + \left\langle (en/c) \bm V \cdot \nabla \psi \right\rangle_\psi +
\left\langle (en/c) \bm V \cdot (\bm B - \bm B_0) \times R^2 \bm \nabla \phi \right\rangle_\psi.
\end{eqnarray}
Applying the drift ordering, at ${\cal O}(\delta \omega n_{0}m v_{th}L)$, we obtain:
\begin{eqnarray}
\label{eq:96}
\left\langle (en/c) \bm V \cdot \nabla \psi \right\rangle_\psi  & = & - \left\langle (en \bm E + \bm F ) \cdot R^2 \bm \nabla \phi \right\rangle_\psi +
\\\nonumber  & & + \left\langle (en/c) \bm V \cdot (\bm B_0 - \bm B) \times R^2 \bm \nabla \phi \right\rangle_\psi
\end{eqnarray}
which is the analogue of Eq. (2.93) in Ref. \cite{hinton1976theory} where  also the contribution of the fluctuating fields have been considered. In general the term $\left\langle R^2 \bm \nabla \phi \cdot \partial_t (n m \bm V ) \right\rangle_\psi$ is non negligible at this order. Since $\sav{\partial_t \left(n \bm V\right)} \sim \bm V \sav{\partial_t n}$ we can estimate its magnitude from the surface averaged continuity equation:
\begin{equation}
\label{eq:159}
\sav{\partial_{t} n} = -  \sav{\wb{\nabla} \cdot\left( n \wb{V} \right)}.
\end{equation}
Decomposing the right hand side of this expression in Fourier components, we note that only the terms generated by a coupling between fluctuations with both poloidal and toroidal mode numbers equal in absolute value and opposite sign are not annihilated by the surface average. In principle these terms may have a generic characteristic length-scale in between that of the turbulent fluctuation spectrum and equilibrium itself; that is, mesoscale structures. Thus, an approach that postulates a systematic scale separation between fluctuating quantities and equilibrium profiles is questionable. Labeling the characteristic length-scale $ k_{z}^{-1}$, we obtain $\sav{\partial_{t} n} \sim \delta^{2}k_{z}n_{0}v_{th}$. In the transport equations derivation of this chapter we adopt the drift ordering and, therefore, we  postulate that equilibrium modification due to transport processes can occur on sufficiently long (radial) scales only. Consistently with this assumption we require that $k_{z}^{-1}\sim L$ and, therefore,  Eq.(\ref{eq:96}) follows. Starting from an equation for the first moment of the distribution function and applying the drift ordering, we have found an expression for the particle flux, which can be used to compute the evolution of the density profile in the continuity equation. Using this method, we can describe the fluxes up to second order in the drift parameter using the information on the distribution function accurate up to first order. This is crucial for the description of the fluctuation induced transport, which requires gyrokinetic theory in order to express the distribution function up to a certain order in $\delta$. This theory is completely general \cite{frieman1982nonlinear,brizard2007foundations}; however, general expressions of the particle response (distribution function) up to order ${\cal O}(\delta^2)$ for generic fluctuations in nonuniform toroidal equilibria are not available to date althought a recent work, i.e. \cite{tronko2016second}, is addressing this problem. The moments equation approach allows us to compute the fluctuation induced fluxes with the required precision (${\cal O}(\delta^2)$) by means of the fluctuation induced particle response at ${\cal O}(\delta)$.

Expression Eq. (\ref{eq:96}) include classical, neoclassical and fluctuation-induced transport and, therefore, generalize the result derived in \cite{hinton1976theory}. 
Using the following relation:
\begin{equation}
\label{eq:99}
{\bm b} \times {\bm\nabla} \psi = F {\bm b} - B R^2 {\bm \nabla}\phi,
\end{equation}
starting from the expression for the fluxes derived in \cite{hinton1976theory}, we can identify the classical and neoclassical contributions in Eq(\ref{eq:96}). The distinction of classical and neoclassical fluxes is somewhat conventional \cite{hinton1976theory,wimmel1970energy}, since it ultimately resorts to the effect of Coulomb collisions. In particular, for the classical particle flux we obtain:
\begin{equation}
\label{eq:97}
(n \bm V_\perp)_c = - \frac{\bm b}{m\Omega} \times \bm F \;\;\;\;\; \Rightarrow \;\;\;\;\;  \left\langle (en/c) \bm V \cdot \nabla \psi \right\rangle_{\psi c} = - \left\langle \bm F_\perp \cdot R^2 \bm \nabla \phi \right\rangle_\psi
\end{equation}
while, analogously, for the neoclassical one:
\begin{equation}
\label{eq:98}
\left\langle (en/c) \bm V \cdot \nabla \psi \right\rangle_{\psi NC} = - \left\langle ( en \bm E_0 + \bm F_\parallel ) \cdot R^2 \bm \nabla \phi \right\rangle_\psi
\end{equation}
where $\bm F_\parallel \cdot R^2 \bm \nabla \phi = (\bm b \cdot \bm F) \bm b \cdot R^2 \bm \nabla \phi = (\bm b \cdot \bm F) (F/B)$, $\bm F_\perp = \bm F - \bm F_\parallel$.
The remaining terms of Eq. (\ref{eq:96}), which can be attributed to fluctuations, as they vanish in the absence of them, read:
\begin{eqnarray}
\label{eq:101}
\left\langle (en/c) \bm V \cdot \nabla \psi \right\rangle_{\psi gk}  = - \left\langle (en \bm E - en \bm E_0 ) \cdot R^2 \bm \nabla \phi \right\rangle_\psi + \left\langle (en /c) \bm V \cdot (\bm B_0 - \bm B) \times R^2 \bm \nabla \phi \right\rangle_\psi 
\end{eqnarray}
Collecting the various contributions derived above, the density transport
equation can be written as:
\begin{equation}
\label{eq:116}
\left\langle\left\langle \partial_t f\right\rangle_v\right\rangle_\psi = - \frac{1}{V'} \frac{\partial}{\partial \psi} \left[ V' \left\langle n \bm V \cdot \nabla \psi \right\rangle_{\psi c} + V' \left\langle n \bm V \cdot \nabla \psi \right\rangle_{\psi NC} + V' \left\langle n \bm V \cdot \nabla \psi \right\rangle_{\psi gk} \right]
\end{equation}
This additive form does not imply that transport processes are independent
of each other. It is readily recognized that, e.g., the neoclassical flux in Eq. (\ref{eq:98}) could also depend on fluctuations intensity, although at higher (negligible) order. Exploring transport processes more in depth, fluctuations may enhance the deviation of system from local thermodynamic equilibrium and cause structures in the phase space \cite{chen2007theory}, \cite{zonca2006physics}, \cite{zonca2015nonlinear} which are eventually damped by collisions (enhanced collisional damping). Furthermore collisions may damp long lived structures formed by saturated instabilities, such as zonal flows \cite{hasegawa1979nonlinear,lin1998turbulent,rosenbluth1998poloidal,hinton1999dynamics,diamond2005zonal,itoh2006physics} or more generally zonal structures \cite{zonca2006physics}, which, in turn, regulate turbulent transport itself.
\subsection{Energy transport}
The simplest way to compute flux surface average energy transport (energy
conservation) is proceeding as for the density transport. Taking the dot product of $R^{2}\wb{\nabla}\phi$ with  the energy transport equation, i.e. Eq. (\ref{eq:146}), and taking the flux surface average yields:
\begin{align*}
\label{eq:108}
  \left \langle \frac{\partial {\bm Q}}{\partial t}\cdot R^2 \wb{\nabla}\phi \right \rangle_\psi + \left \langle{\bm \nabla} \cdot {\bm R}\cdot R^2 \wb{\nabla} \phi\right \rangle_\psi- \left \langle \frac{e}{m}{\bm E}\cdot {\bm P}\cdot R^2{\bm \nabla}\phi \right \rangle_\psi &- \left \langle{\bm I}\left( \frac{3}{2}p + \frac{m}{2} n V^2\right) \cdot R^2 {\bm \nabla}\phi \right \rangle_\psi +\\ \nonumber
  &- \left \langle \frac{e}{mc} {\bm Q} \times {\bm B} \cdot R^2\cdot {\bm \nabla}\phi\right \rangle = \left \langle {\bm G} \cdot R^2 {\bm \nabla}\phi\right \rangle.
\end{align*}
Applying the drift ordering we obtain, up to order $\mathcal{O}(\delta^{2})$:
\begin{equation}
\label{eq:129}
\left\langle \bm Q \cdot \nabla \psi \right\rangle_\psi = \left\langle \bm Q \cdot \nabla \psi \right\rangle_{\psi c} + \left\langle \bm Q \cdot \nabla \psi \right\rangle_{\psi NC} + \left\langle \bm Q \cdot \nabla \psi \right\rangle_{\psi gk}
\end{equation}
where:
\begin{align}
\label{eq:144}
  \left\langle \bm Q \cdot \nabla \psi \right\rangle_{\psi c} &=  - (mc/e)\left\langle \bm G_\perp \cdot R^2 \bm \nabla \phi \right\rangle_\psi \\
  \left\langle \bm Q \cdot \nabla \psi \right\rangle_{\psi NC} &=  - \left\langle (c \bm E_0 (2 p_\perp + p_\parallel/2) + (mc/e) \bm G_\parallel ) \cdot R^2 \bm \nabla \phi \right\rangle_\psi\\
  \left\langle \bm Q \cdot \nabla \psi \right\rangle_{\psi gk}  = & - \left\langle c (\bm E - \bm E_0 ) (2 p_\perp + p_\parallel/2)  \cdot R^2 \bm \nabla \phi \right\rangle_\psi
  - \left\langle  \bm Q \cdot (\bm B - \bm B_0) \times R^2 \bm \nabla \phi \right\rangle_\psi.
\end{align}
In order to compare these results with the particle fluxes calculated in the previous section, we note that Eq. (\ref{eq:145}) can be cast in the following form:
\begin{equation}
\label{eq:149}
\partial_t \left( p_\perp  + p_\parallel/2  \right)  + \bm \nabla \cdot \bm Q   = W + \left( e n\bm E + \bm F \right) \cdot \bm V - \partial_t \left( \frac{m}{2} n V^2 \right).
\end{equation}
We now want to rewrite the last two term of the RHS and therefore we take the dot product of Eq. (\ref{eq:144a}) with $\wb{V}$:
\begin{equation}
\label{eq:150}
\left( e n\bm E + \bm F \right) \cdot \bm V
- \partial_t \left( \frac{m}{2} n V^2 \right) = \frac{m}{2} V^2 \partial_t n + \left( \bm \nabla \cdot \bm P \right) \cdot \bm V.
\end{equation}
Applying the drift ordering up to the leading order $\mathcal{O}(\delta^{3})$ we obtain:
\begin{equation}
\label{eq:151}
\partial_t \left( p_\perp  + p_\parallel/2  \right)  + \bm \nabla \cdot \bm Q   - \left( \bm \nabla \cdot \bm P \right) \cdot \bm V   = W + {\cal O}(\delta^5)
\end{equation}
Using the leading order expression of $\wb{V}$:
\begin{equation}
\label{eq:152}
\bm V = \bm b V_\parallel + \frac{\bm b}{m\Omega} \times \left( \frac{1}{n} \bm \nabla \cdot \bm P - e \bm E\right)
\; ,
\end{equation}
we can write:
\begin{align}
\label{eq:153}
\left( \bm \nabla \cdot \bm P \right) \cdot \bm V   & \simeq    \left( \bm \nabla \cdot \bm P \right) \cdot  \frac{\bm b}{m\Omega} \times \left( - e \bm E\right)
  \simeq \\
\nonumber   & \simeq  \frac{e \bm b}{m\Omega} \cdot \left( \bm \nabla \cdot \bm P \right)  \times  \bm E  \simeq
  \\\nonumber  & \simeq \bm \nabla \cdot \left( \frac{e}{m\Omega} P_\perp  \bm E \times \bm b \right)
\; ,
\end{align}
which, at the lowest order, describes the advection of $P_{\perp}\approx p_{\perp}$ at the $\wb{E}\times\wb{B}$ speed.

Considering the flux surface average of the energy evolution equation, we
have, at the leading order:
\begin{equation}
\label{eq:154}
\partial_t \left\langle p_\perp  + p_\parallel/2  \right\rangle_\psi  + \frac{1}{V'}
 \frac{\partial}{\partial \psi} \left[ V' \left(  \left\langle \bm Q \cdot  \bm \nabla \psi \right\rangle_\psi  + \left\langle c p_\perp \bm E \cdot R^2 \bm \nabla \phi \right\rangle_\psi  \right) \right]   = \left\langle W \right\rangle_\psi
 \; ;
\end{equation}
and, thus, we have demonstrated that, at the relevant order in our asymptotic expansion in the drift parameter, the evolution equation for $(p_{\perp}+p_{\parallel}/2)$ is a transport equation with a collisional heating source $\left\langle W \right\rangle_\psi$ and with an effective radial flux:
\begin{equation}
\label{eq:155}
\left\langle \bm Q_{\mathrm eff} \cdot  \bm \nabla \psi \right\rangle_\psi \equiv \left\langle \bm Q \cdot  \bm \nabla \psi \right\rangle_\psi  + \left\langle c p_\perp \bm E \cdot R^2 \bm \nabla \phi \right\rangle_\psi
\; .
\end{equation}
Using this result we can write the expressions for the effective fluxes to be used in the energy evolution equation:
\begin{equation}
\label{eq:156}
\left\langle \bm Q \cdot \nabla \psi \right\rangle_\psi = \left\langle \bm Q \cdot \nabla \psi \right\rangle_{\psi c} + \left\langle \bm Q \cdot \nabla \psi \right\rangle_{\psi NC} + \left\langle \bm Q \cdot \nabla \psi \right\rangle_{\psi gk}
\; ,
\end{equation}
which are the following:
\begin{align}
\label{eq:157}
\left\langle \bm Q_{\mathrm eff} \cdot \nabla \psi \right\rangle_{\psi c} & = - (mc/e)\left\langle \bm G_\perp \cdot R^2 \bm \nabla \phi \right\rangle_\psi \\  \left\langle \bm Q_{\mathrm eff} \cdot \nabla \psi \right\rangle_{\psi NC} & = - \left\langle (c \bm E_0 (p_\perp + p_\parallel/2) + (mc/e) \bm G_\parallel ) \cdot R^2 \bm \nabla \phi \right\rangle_\psi \\  \left\langle \bm Q_{\mathrm eff} \cdot \nabla \psi \right\rangle_{\psi gk} & = - \left\langle c (\bm E - \bm E_0 ) (p_\perp + p_\parallel/2)  \cdot R^2 \bm \nabla \phi \right\rangle_\psi
  - \left\langle  \bm Q \cdot (\bm B - \bm B_0) \times R^2 \bm \nabla \phi \right\rangle_\psi
\end{align}
By direct comparison with the collisional and gyrokinetic particle fluxes, we readily see that the expression are \emph{formally} the same, with energy fluxes weighted by $m v^{2}/2$, consistently with the evolution equation of phase space zonal structures which will be introduced in the next chapter.

\subsection{Gyrokinetic description of particle distribution function}
As it is shown in \autoref{cap:appA}, the particles distribution function can be expressed in terms of the guiding-center distribution function $F$ which, in turn, can be written in terms of the gyrocenter distribution function $\bar{F}$ \cite{brizard2007foundations}:
\begin{eqnarray}
\label{eq:102}
f= e^{- {\bm \rho}\cdot {\bm \nabla}}F  =&  e^{- {\bm \rho}\cdot {\bm \nabla}} \bar F  - \frac{e}{m} e^{- {\bm \rho}\cdot {\bm \nabla}}\left \langle \delta \psi_{gc} \right \rangle \left( \frac{\partial \bar F}{\partial \cal E} + \frac{1}{B_0} \frac{\partial \bar F}{\partial \mu} \right) +
\left[ \frac{e}{m} \delta \phi \frac{\partial \bar F}{\partial \cal E}  \right] + \\  \nonumber
 &+  \left[  \frac{e}{m} \left( \delta \phi - \frac{v_\parallel}{c} \delta A_\parallel \right)  \frac{1}{B_0} \frac{\partial \bar F}{\partial \mu} + \delta \bm A_\perp \times \frac{\bm b}{B_0} \cdot \bm \nabla \bar F \right]
 \; ,
\end{eqnarray}
where  ${\cal E} = v^2/2$ is the energy per unit mass, $\mu$ is the magnetic moment adiabatic invariant $\mu = v_\bot^2/(2 B_0) + \ldots$ and:
\begin{equation}
\label{eq:103}
\delta \psi_{gc} = \delta \phi_{gc} - \frac{\bm v}{c} \cdot \delta \bm A_{gc} = e^{\bm \rho \cdot \bm \nabla} \left( \delta \phi - \frac{\bm v}{c} \cdot \delta \bm A \right) \equiv e^{\bm \rho \cdot \bm \nabla} \delta \psi.
\end{equation}
In Eq. (\ref{eq:102}),  all terms that are acted upon by $e^{\bm \rho \cdot \bm \nabla}$ are the adiabatic response of the particle distribution function, while other terms represent the non-adiabatic response of the guiding-center distribution. This subdivision, introduced for convenience of formal treatments in the early works on gyrokinetic theory \cite{antonsen1980kinetic,catto1981generalized,frieman1982nonlinear}, can be rigorously motivated ``in the context of the action of the pull-back operators used in the derivation of the nonlinear gyrokinetic
Vlasov equation''. The ``physical interpretation of the pull-back operator is that it performs a partial solution of the Vlasov equation associated with fast-time gyromotion dynamics'' \cite{brizard2007foundations}. The gyrophase average $\left\langle \delta \psi_{gc}\right\rangle$ involves the introduction of Bessel functions as integral operators:
\begin{equation}
\label{eq:104}
\left\langle \delta \psi_{gc}\right\rangle  =  \hat I_0  \left( \delta \phi - \frac{v_\parallel}{c} \delta A_\parallel \right) + \frac{m}{e} \mu \hat I_1  \delta B_\parallel \;\; .
\end{equation}
where $\hat I_n (x) \equiv (2/x)^n J_n(x)$ \cite{brizard1992nonlinear} , $J_n(x)$ are the Bessel functions, $\lambda^2 \equiv 2 (\mu B_0/\Omega^2) k_\perp^2$ and the definition of $\hat I_n$ acting on a generic function $g(\bm r) = \int \hat g (\bm k) \exp (i \bm k \cdot \bm r) d \bm k$ is  the following:
\begin{equation}
\label{eq:105}
\hat I_n g (\bm r) \equiv \int d \bm k e^{i \bm k \cdot \bm r} \hat I_n(\lambda) \hat g (\bm k) .
\end{equation}
At the leading order we can show that:
\begin{equation}
\label{eq:106}
\left\langle e^{-\bm \rho \cdot \bm \nabla} (...) \right\rangle = \hat I_0 (...) \; ; \;\;\;\; \left\langle e^{-\bm \rho \cdot \bm \nabla} \bm v (...) \right\rangle = \hat I_0  v_\parallel \bm b (...) + \frac{mc}{e} \mu  \hat I_1  \bm b \times \bm \nabla (...).
\end{equation}
Introducing the velocity space integration $\langle \dots \rangle_{v}$, using the previous relations, we can show that the following relations hold:
\begin{eqnarray}
\label{eq:107}
\left\langle f \right\rangle_v  & = &
\left \langle \hat I_0 \left[ \bar F - \frac{e}{m} \left(  \frac{\partial \bar F}{\partial \cal E} + \frac{1}{B_0} \frac{\partial \bar F}{\partial \mu} \right) \left\langle \delta \psi_{gc} \right\rangle \right] \right\rangle_v + \\
 \nonumber &  +& \frac{e}{m} \left\langle \frac{\partial \bar F}{\partial \cal E} \right\rangle_v \delta \phi + \frac{e}{m} \left\langle \frac{1}{B_0} \frac{\partial \bar F}{\partial \mu} \left( \delta \phi - \frac{v_\parallel}{c} \delta A_\parallel \right) \right\rangle_v  +  \delta \bm A_\perp \times \frac{\bm b}{B_0} \cdot \bm \nabla \left\langle \bar F \right \rangle_v  \;\;
\end{eqnarray}
\begin{eqnarray}
\label{eq:136a}
\hspace{-7em}\left\langle \bm v_\perp f \right\rangle_v  =  \frac{mc}{e} \bm b \times \left\langle \mu \hat I_1 \bm \nabla \left[  \bar F - \frac{e}{m} \left(  \frac{\partial \bar F}{\partial \cal E} + \frac{1}{B_0} \frac{\partial \bar F}{\partial \mu} \right) \left\langle \delta \psi_{gc} \right\rangle \right]\right\rangle_v \;\;.
\end{eqnarray}
The previous relations are derived in \autoref{cap:appB}. Using this result, we can compute the leading order of Eq. (\ref{eq:101}). In particular we obtain the following expressions:
\begin{eqnarray}
\label{eq:109}
- \left\langle (en \bm E - en \bm E_0 ) \cdot R^2 \bm \nabla \phi \right\rangle_\psi & = & e \left \langle\left( \frac{1}{c} \frac{\partial}{\partial t} \delta \bm A + \bm \nabla \delta \phi \right) \cdot R^2 \bm \nabla \phi \left\langle f \right\rangle_v \right\rangle_\psi
\nonumber \\ & = & e \left \langle R^2 \bm \nabla \phi \cdot \bm \nabla \delta \phi   \left\langle \hat I_0 (\lambda) \delta \bar G \right\rangle_v \right\rangle_\psi
\; ,
\end{eqnarray}
where we have introduced the function $\bar G$ \cite{frieman1982nonlinear}:
$$\bar G = \bar F - \frac{e}{m} \frac{\partial \bar F}{\partial \cal E} \left\langle \delta \psi_{gc} \right\rangle $$
that satisfies the Frieman-Chen nonlinear gyrokinetic equation \cite{chen2016physics} up to ${\cal O}(\epsilon)$. After some calculations, shown in \autoref{cap:appC}, we obtain:
\begin{eqnarray}
\label{eq:110}
& &\left\langle (en /c) \bm V \cdot (\bm B_0 - \bm B) \times R^2 \bm \nabla \phi \right\rangle_\psi = \left\langle (e/c) \left\langle \bm v f \right\rangle_v \cdot R^2 \bm \nabla \phi \times  (\bm \nabla \times \delta \bm A)  \right\rangle_\psi \\
\nonumber & & \hspace*{1cm} = e \left\langle\left\langle \frac{v_\parallel}{c} \frac{\bm \nabla \psi \cdot \delta \bm B_\perp}{B_0} \hat I_0 (\lambda) \delta \bar G \right\rangle_v \right\rangle_\psi
\nonumber\\ & & \hspace*{1.2cm} - m \left\langle\left\langle   \left( \delta B_\parallel R^2 \bm \nabla \phi - \frac{F}{B_0}
\delta \bm B_\perp \right) \cdot  \mu \hat I_1 (\lambda)  \bm \nabla_\perp  \delta \bar G \right\rangle_v \right\rangle_\psi
\nonumber\\ & & \hspace*{1cm} =
\nonumber e \left\langle R^2 \bm \nabla \phi \cdot \left\langle \bm \nabla  \left( - \frac{v_\parallel}{c} \delta A_\parallel \right) \hat I_0 (\lambda) \delta \bar G
+ \bm \nabla  \left( \frac{m}{e} \mu \delta B_\parallel \right) \hat I_1 (\lambda) \delta \bar G \right\rangle_v \right\rangle_\psi.
\end{eqnarray}
From this expression, recalling that $\wb{\nabla}\phi \cdot \wb{\nabla} \sim \partial_{\phi}$, we can see that the fluctuation induced transport is due only to toroidally symmetry breaking perturbations. The push forward expression for the energy fluxes are identical to the density fluxes except for the weight $m v^{2}/2$ which multiplies every term.
\section{General remarks on transport equations}
Note that the expressions for fluctuation induced fluxes and ensuing transport are valid for generic short-wavelength turbulence;
that is for drift wave fluctuations at frequencies much lower than the cyclotron frequency but wavelength as short as the particle 
Larmor radius. Nonetheless, our moment approach is based on a small drift-parameter asymptotic expansion, which assumes that the 
effect of fluctuation induced transport is given for structures that are sufficiently longer scale than the Larmor radius. In other words,
although drift-wave turbulence is described by nonlinear gyrokinetic theory, its effect on transport is accounted for on the
length scale typical of the plasma equilibrium. This assumption has been used in the derivation of Eq. (\ref{eq:151}) and Eq. (\ref{eq:129}) several times, e.g. neglecting terms with the partial derivative of the density of momentum/energy.

As anticipated in the introduction to Chapter 3, our moment equation approach to transport equations \cite{hinton1976theory}, combined with 
the conceptual framework of nonlinear gyrokinetic theory and the push-forward representation of particle moments \cite{frieman1982nonlinear,brizard2007foundations}, allows a compact and physically transparent derivation of cross-field particle
and energy fluxes, which include collisional and fluctuation induced transport processes on the same footing. These are original
results of this thesis work, and generalize and extend the analysis of Ref. \cite{hinton1976theory}. At the same time, our results recover
those originally proposed in Refs. \cite{plunk2009theory,barnes2009trinity,abel2013multiscale} derived assuming a systematic
spatiotemporal scale separation between dynamically evolving plasma equilibrium and turbulent fluctuation spectrum. In fact, by
introduction of suitable radial and time averages, Refs. \cite{plunk2009theory,barnes2009trinity,abel2013multiscale} compute
the slow evolution of smoothed equilibrium density and pressure profiles. Our approach instead, based on moment equations and 
nonlinear gyrokinetic theory, follows a different theoretical framework, which assumes that equilibrium modification due to
transport processes can occur on sufficiently long (radial) scales only without introducing any averaging operation. For this reason the expressions that we derive hold point-wise in time and space contrarily to the results obtained in \cite{plunk2009theory,barnes2009trinity,abel2013multiscale}. It is to be expected that the spatiotemporal 
average description of Refs. \cite{plunk2009theory,barnes2009trinity,abel2013multiscale} and our novel approach are 
consistent. We can verify this by substituting the pull-back representation of the distribution function, i.e. Eq. (\ref{eq:102}), into Eq. (A.20) of \cite{plunk2009theory}, which describes the transport of particles analogously to Eq. (\ref{eq:116}), and into Eq. (A.25), which describes the transport of heat analogously to Eq. (\ref{eq:154}), obtaining, up to the required order, the averaged version of the equations already derived by means of the moment method. The originality of our present results does not only consist in the different theoretical framework adopted, based on moment equations and 
nonlinear gyrokinetic theory, yielding a significantly more compact formulation analogous to that of Ref. \cite{hinton1976theory}. It also 
naturally introduces the notion of spatiotemporal scales of equilibrium variations and of the corresponding structures, which must be
self-consistently determined by nonlinear gyrokinetic theory. The implications of this is further elaborated in the next chapter, where we
discuss the importance of ``zonal structures'' and their counterpart in the particle phase space as crucial elements for the nonlinear
evolution of magnetized plasmas and for the understanding of underlying transport processes. The results obtained in \autoref{cap:zonal} can be applied in order to put a constraint on the characteristic length-scale of the average operation introduced in \cite{plunk2009theory}: $l_{h}$. By direct evaluation of the order of magnitude of the terms of Eq. (\ref{eq:999}) we can show that the evolution equations for the equilibrium profiles are consistent with the results obtained by means of phase space zonal structure theory if all the zonal structures characterized by $k_{z}L > \delta^{-1/2}$ are annihilated by spatial averaging and, therefore, we find $l_{h}\gtrsim \delta^{1/2}L$ which is in agreement with \cite{plunk2009theory}. In the next Chapter, the crucial role of nonlinear gyrokinetic theory is made evident. Here, we further emphasize that 
the results discussed above, based on moment equations, and the nonlinear gyrokinetic theory of transport, discussed below,
are both essential elements of this thesis work. The moment equation approach, in fact, illuminates the possibility of 
providing a unified theory of collisional and fluctuation induced transport by means of a compact and intuitive formulation.
It however fails where mesoscale structures become increasingly more important, as discussed in the next Chapter. On the other hand, 
nonlinear gyrokinetic theory, despite its generality, is based on spatiotemporal scale separation between plasma equilibrium 
and fluctuation spectra \cite{frieman1982nonlinear,brizard2007foundations}. 
A global transport analysis based on nonlinear gyrokinetic theory remains, thus, a challenge that is 
one of the main topics of interest of present research in magnetic confinement physics. Therefore, the possibility of adopting both
approaches and showing, as original result of this thesis work, that they coincide in the appropriate limit provides confidence that 
a global theory of collisional and fluctuation induced particle transport treated on the same footing can be formulated 
as proposed here.

\chapter{Zonal structures}
\label{cap:zonal}
In the previous Chapter we have derived the equations governing transport of particles and energy on a time scale $\mathcal{O}(\delta^{-2}\omega^{-1})$. The result is consistent with \cite{plunk2009theory,barnes2009trinity,abel2013multiscale} and these equations reduce to the ones found by G. Plunk if we introduce radial (patch) averages \cite{plunk2009theory}. We also noted that our formulation, which is valid point wise in space and time, naturally introduces the notion of spatiotemporal scales of equilibrium variations and of the corresponding structures. In fact, starting from given plasma profiles, the spatiotemporal features of the corresponding dynamic evolution is
given by collisional and fluctuation induced fluxes, self-consistently. While collisions generally tend to ``smooth out'' distortions in the phase space, fluctuations are to be considered as ``sources'' of those distortions. And, in general, we cannot conclude that the spatiotemporal scales of the considered plasma equilibrium
will be preserved by the nonlinear evolution. What we generally know, from the analysis of Chapter 3, is that our moment based macroscopic transport equations are valid as long as the asymptotic expansion in the small drift parameter is consistent. This intuitively suggests that the length scale of a few magnetic drift orbit widths is the natural scale over which mesoscale structures may appear in the nonlinear equilibrium evolution.

In this Chapter, using  nonlinear gyrokinetic theory \cite{frieman1982nonlinear,brizard2007foundations}, we introduce governing equations for structures that nonlinearly modify the plasma equilibrium and,
in general, can be characterized by length scales of the order of the particle Larmor radius. In order to be free of strong and rapid collisionless dissipation
\cite{rosenbluth1998poloidal,hinton1999dynamics}, such zonal structures are poloidally and toroidally symmetric. In general, they also have a phase space counterpart, the phase space zonal structures \cite{zonca2015nonlinear,chen2016physics}, which may be of particular relevance when resonant wave-particle interactions are origin of fluctuation induced transport.

Here, as original novel result of this thesis work, we derive transport equations based on the description of phase space zonal structures, and 
demonstrate that they reduce, as expected, to the equations obtained in Chapter 3, when long spatial scale corrugations to the nonlinear evolving equilibrium are considered. As collisional transport manifests itself on long length scales only, and gyrokinetic theory is based on spatiotemporal scale separation between plasma equilibrium and fluctuation spectra \cite{frieman1982nonlinear,brizard2007foundations}, we discuss gyrokinetic transport equations
in the collisionless, short wavelength limit, showing that they reduce to transport equations derived earlier in the proper parameter range.
By doing so, we are able to isolate the linear polarization response \cite{rosenbluth1998poloidal,hinton1999dynamics}, which can be considered of higher order in the usual macroscopic plasma transport analysis, and the fluctuation induced nonlinear fluxes, suitably modified at short scales.
In general, we show that fast radial oscillations of the equilibrium (slowly varying) profiles are of crucial importance in the self-consistent description of the transport processes in a magnetically confined plasma.
\section{Phase space zonal structures}
\label{sec-1-2}
Mode mode coupling processes between fluctuating fields in toroidal fusion plasmas can generate toroidal symmetric structures, usually linearly stable, in the density and temperature profiles with slow time variation which can be considered as modifications to the slow evolving, i.e. equilibrium, profiles \cite{diamond2005zonal}. The poloidally symmetric response of these structures is unaffected by rapid collisionless dissipation \cite{rosenbluth1998poloidal,hinton1999dynamics}, and may be regarded as radial corrugations of the "smooth" equilibrium parameters. Therefore the dynamics of the plasma must be described by new self-consistent neighboring (nonlinear) equilibria \cite{chen2007theory} (we call equilibrium the slow evolving part of the fields). These modifications are called zonal structures and, due to their slow temporal dynamics, it is required that they are unaffected by collisionless dissipation processes, i.e. Landau damping \cite{rosenbluth1998poloidal} as anticipated above. In summary, zonal structures must satisfy $k_{\parallel}\equiv 0$ everywhere, e.g. $\delta \bm E_r = - \bm \nabla \psi \partial_\psi \delta \phi (\psi)$. Thus, in magnetically confined fusion plasmas, zonal structures correspond to long-lived or oscillating electromagnetic perturbations with predominant variations in the radial direction and, tipically, characterized by mainly electrostsatic component.

As the zonal structures are nonlinearly excited (linearly stable), they will scatter the primary driving instabilities to shorter-wavelength stable regime and stabilize the driving instabilities. For this reason they can importantly regulate turbulence saturation level \cite{hasegawa1979nonlinear,lin1998turbulent,rosenbluth1998poloidal,hinton1999dynamics,itoh2006physics,diamond2005zonal} and, eventually, turbulent transport; and thus they must be properly accounted for a self-consistent description of gyrokinetic transport. In addition to zonal structures, more general phase space zonal structures \cite{zonca2015nonlinear,chen2016physics} can exist. They represent a deviation of the plasma from the local thermodynamic equilibrium and are crucial to determine the statistical properties of transport events such as intermittency, avalanches, bursting and/or non-local behaviors. They are particularly important when resonant wave-particle interactions are crucial in the instability and transport processes \cite{zonca2006physics,zonca2015nonlinear,chen2016physics}. In this context and theoretical framework, zonal structures and phase space zonal structures are self-consistent counterparts of collisionless undamped (long-lived) nonlinear deviation of the slowly varying plasma equilibrium from the reference local thermodynamic equilibrium state. For this reason they can increase the transport induced by collisions with a term, in principle, of the same order of the neoclassical flux or even larger. The existence of phase space zonal structures is the natural consequence of the collisionless nature of high temperature plasmas and of the important role played by resonant wave-particle interactions. They are eventually damped by collisions; but a realistic description of transport in collisionless plasmas must self-consistently take them into account.
\section{Evolutive equations}
\label{sec-1-3}
Assuming that plasma turbulence is characterized by low frequencies with respect to the gyration frequency, i.e. $\omega \sim \delta \Omega$, the leading order plasma response to zonal structures can be described \cite{chen2016physics}, using the same notation of \autoref{cap:gyrotransp}, as:
\begin{equation}
\label{zonaleq:1}
\delta f_z  =  e^{-\bm \rho \cdot \bm \nabla}  \delta \bar G_z  + \frac{e}{m} \delta \phi_{0,0} \frac{\partial \bar F_0}{\partial \cal \mathcal{E}}
\end{equation}
where, $0,0$ subscript to $\delta \phi$ indicates the $m=n=0$ component with $m$ and $n$ being respectively the poloidal and toroidal mode numbers of the fluctuation. We have also assumed that the equilibrium guiding center distribution is isotropic, that $\partial_\mu \bar F_0 = 0$, and that the usual low-$\beta$ tokamak ordering applies. The non-adiabatic gyrocenter plasma response $\delta \bar G_z$,  is obtained from the solution of the Frieman-Chen nonlinear gyrokinetic equation \cite{frieman1982nonlinear}:
\begin{equation}
\label{zonaleq:2}
\left(\partial_t + v_\parallel \nabla_\parallel + \bm v_d \cdot \bm \nabla \right) \delta \bar G_z = - \frac{e}{m} \frac{\partial \bar F_0}{\partial \cal E} \frac{\partial}{\partial t} \left\langle \delta \psi_{gc}\right\rangle_z  - \frac{c}{B_0} \bm b \times \bm \nabla \left\langle \delta \psi_{gc}\right\rangle \cdot \bm \nabla \delta \bar G
\; ,
\end{equation}
where:
\begin{equation}
\label{zonaleq:3}
\left\langle \delta \psi_{gc}\right\rangle_z  =  \hat I_0  \left( \delta \phi_{0,0} - \frac{v_\parallel}{c} \delta A_{\parallel 0,0} \right) + \frac{m}{e} \mu \hat I_1 \delta B_{\parallel 0,0}.
\end{equation}
Note that the gyro-center zonal structure response $\delta \bar{G}_{z}$ must be axisymmetric in order to avoid Landau damping. This equation states that zonal structures are driven by the zonal fields, i.e. fields with $n=m=0$, and by nonlinear coupling between the gyro-center response and the perpendicular gradient of the fluctuating fields that generate terms with the same property. Particle drift velocity due to the equilibrium fields can be written in the following form:
\begin{equation}
\label{zonaleq:4}
\bm v_d = \frac{v_\parallel}{\Omega} \bm \nabla \times ( \bm b v_\parallel ).
\end{equation}
Using toroidal coordinates we can write the operator $\wb{v}_{d}\cdot \wb{\nabla}$ acting on axisymmetric perturbations as:
\begin{equation}
\label{zonaleq:6}
\bm v_d \cdot \bm \nabla = \frac{v_\parallel}{J \Omega} \left[ \frac{\partial}{\partial \theta} \left( \frac{F v_\parallel}{B_0}\right) \frac{\partial}{\partial \psi} -
\frac{\partial}{\partial \psi} \left( \frac{F v_\parallel}{B_0}\right) \frac{\partial}{\partial \theta} \right].
\end{equation}
Furthermore, we can write the following relation between derivative operators acting on an axisymmetric perturbation:
\begin{equation}
\label{zonaleq:7}
\wb{\nabla}_{\parallel}= \frac{1}{J B} \frac{\partial}{\partial \theta}
\; ;
\end{equation}
and, therefore, rewrite the free streaming operator as:
\begin{equation}
\label{zonaleq:8}
\partial_t + v_\parallel \left[ 1 - \frac{\partial}{\partial \psi} \left( \frac{F v_\parallel}{\Omega}\right) \right] \nabla_\parallel + v_\parallel \nabla_\parallel \left( \frac{F v_\parallel}{\Omega}\right) \frac{\partial}{\partial \psi}.
\end{equation}
This operator carries information about the free streaming of particles belonging to the phase space zonal structure. Introducing the toroidal angular momentum $P_\phi = (e/c) (F v_\parallel/\Omega - \psi)$, this can be rewritten as:
\begin{equation}
\label{zonaleq:9}
\partial_t - \frac{v_\parallel c}{e} \left[ \frac{\partial P_\phi}{\partial \psi} \nabla_\parallel - \nabla_\parallel P_\phi \frac{\partial}{\partial \psi} \right]
= \partial_t - \frac{v_\parallel c}{e} \left[ \frac{\partial P_\phi}{\partial \psi} \right]_\theta \left. \nabla_\parallel \right|_{P_\phi}
\end{equation}
therefore, we have verified that particles belonging to the phase space zonal structure move along surfaces of constant $P_{\phi}$. Up to the leading order in $\delta$, the particles free streaming operator can be  written as:
\begin{equation}
\label{zonaleq:12}
\partial_t + v_\parallel  \nabla_\parallel + v_\parallel \nabla_\parallel \left( \frac{F v_\parallel}{\Omega}\right) \frac{\partial}{\partial \psi}
\; ,
\end{equation}
showing the two different components of the velocity respectively parallel and perpendicular to the magnetic surface. Following \cite{chen2007nonlinear1}, we further decompose $\delta \bar G_z = e^{- i Q_z} \delta \bar g_z$ obtaining the following equation for $Q_{z}$ in order to simplify the Frieman-Chen nonlinear gyrokinetic equation:
\begin{equation}
\label{zonaleq:13}
i \nabla_\parallel Q_z = i \nabla_\parallel \left( \frac{F v_\parallel}{\Omega}\right) \frac{k_z}{d\psi/dr}
\; ;
\end{equation}
where $k_{z} \equiv (-i \partial_{r})$. This can be integrated obtaining:
\begin{equation}
\label{zonaleq:14}
Q_z =   F(\psi) \left[ \frac{v_\parallel}{\Omega} - \overline{\left( \frac{v_\parallel}{\Omega} \right)}\right] \frac{k_z}{d\psi/dr}
\; ,
\end{equation}
where we have introduced the average along unperturbed particle orbits:
\begin{equation}
\label{zonaleq:16}
\overline{\left[ \ldots \right]} \equiv \tau_b^{-1} \oint \frac{d \ell}{v_\parallel} \left[ \ldots \right]
\; ;
\end{equation}
and $\tau_{b}$ is the time required for particles to complete an (integrable) close poloidal orbit in the equilibrium magnetic field. We can rewrite Eq. (\ref{zonaleq:2}) in the following form:
\begin{equation}
\label{zonaleq:15}
\left(\partial_t + v_\parallel \nabla_\parallel \right) \delta \bar g_z = e^{i Q_z} \left(
- \frac{e}{m} \frac{\partial \bar F_0}{\partial \cal E} \frac{\partial}{\partial t} \left\langle \delta \psi_{gc}\right\rangle_z  -
\frac{c}{B_0} \bm b \times \bm \nabla \left\langle \delta \psi_{gc}\right\rangle \cdot \bm \nabla \delta \bar G \right)
\; ,
\end{equation}
where the $e^{- i Q_z}$ operator is the pull-back of the drift/banana center zonal structure response $\delta \bar g_z$ (the phase space zonal structure) to the gyro-center response $\delta \bar G_z$. If we define:
\begin{equation}
\label{zonaleq:18}
\rho_{drift} \equiv \frac{F(\psi)}{d \psi /d r} \left[ \frac{v_\parallel}{\Omega} - \overline{\left( \frac{v_\parallel}{\Omega} \right)}\right]
\end{equation}
we can write the pullback operator as $e^{\rho_{drift}k_{z}}$, which is the same formal expression used for the guiding center pullback operator. Also the physical meaning is the same: it allows a simplified description of the plasma in terms of ``moving drifting orbits''. The pullback operator does not depend on the $\phi$ coordinate and, therefore, the requirement for the phase space zonal structure to be long lived, i.e. that is undamped by collisionless processes, is that $\nabla_\parallel \delta \bar g_z = 0$. Thus, the only term contributing to the phase space zonal structure is:
\begin{equation}
\label{zonaleq:17}
\partial_t \delta \bar g_z = \overline{\left[e^{i Q_z} \left(
- \frac{e}{m} \frac{\partial \bar F_0}{\partial \cal E} \frac{\partial}{\partial t} \left\langle \delta \psi_{gc}\right\rangle_z  -
\frac{c}{B_0} \bm b \times \bm \nabla \left\langle \delta \psi_{gc}\right\rangle \cdot \bm \nabla \delta \bar G \right)\right]}.
\end{equation}
In order to derive the evolutive equation for moments of $\delta f_{z}$, we show an important relationship between lowest order bounce averaging and the flux surface average of a velocity space integral. Recalling the definitions of flux surface average and velocity average:
\begin{align}
\label{zonaleq:19}
  &\nonumber \left\langle \ldots \right\rangle_v = 2\pi B_0 \sum_{v_\parallel/|v_\parallel|=\pm}\int \frac{d\mu d{\cal E}}{|v_\parallel|} (\ldots)
\; ,  \\
&\nonumber \left\langle ... \right\rangle_\psi = \frac{1}{V'} \oint J d\theta d\phi (...) = \frac{1}{V' } \oint d\theta d\phi (...)/\bm B \cdot \bm \nabla \theta
\end{align}
where $V' = dV(\psi)/d\psi$, noting that the flux surface volume element $V' = \oint d\theta d\phi/\bm B \cdot \bm \nabla \theta$ and noting that particle orbit is along $\bm B_0$ at the lowest order in the drift parameter expansion, we can derive the following relation:
\begin{equation}
\label{eq:158}
\int d \theta/\bm B \cdot \bm \nabla \theta \simeq \int d\ell/B_0.
\end{equation}
Using this identity, we can show that, for any velocity space function $f$, the following expression holds
\begin{equation}
\label{zonaleq:20}
\left\langle \left\langle f \right\rangle_v\right\rangle_\psi = \frac{4\pi^2}{V' } \sum_{v_\parallel/|v_\parallel|=\pm}\int \tau_b \overline{f_{n=0}} d\mu d{\cal E}.
\end{equation}
This result shows that, at the leading order in the asymptotic expansion, the flux surface average of a velocity integral depends only on the
bounce averaged response of the $n=0$ toroidal Fourier harmonic. This is clearly connected with phase space zonal structures. In fact, in the presence of fluctuations in the gyro-center particle distribution, the drift/banana-center non-adiabatic particle response yields the following form of the phase space zonal structure \cite{zonca2015nonlinear,chen2016physics}:
\begin{equation}
\label{zonaleq:21}
\overline{\left\langle \delta f_z \right\rangle} = \overline{\left(e^{-iQ_z}\hat I_0\right)} \delta \bar g_z  + \frac{e}{m} \delta \phi_{0,0} \frac{\partial \bar F_0}{\partial \cal E}  \;,
\end{equation}
where we recall that the gyrophase average is indicated with $\gav{\dots}$. Acting on this expression by $\partial_t$ and integrating in velocity space we obtain:
\begin{eqnarray}
\label{eq:999}
\partial_t \left\langle \left\langle \delta f_z \right\rangle_v\right\rangle_\psi & = & \frac{e}{m} \partial_t \delta \phi_{0,0} \left\langle \frac{\partial \bar F_0}{\partial \cal E} \right\rangle_v + \frac{4\pi^2}{V' } \sum_{v_\parallel/|v_\parallel|=\pm}\int \tau_b d\mu d{\cal E} \nonumber \\
& & \times \overline{\left(e^{-iQ_z}\hat I_0\right)} \left[ - \frac{e}{m} \frac{\partial \bar F_0}{\partial \cal E} \overline{\left(e^{i Q_z} \hat I_0\right)} \partial_t \delta \phi_{0,0}  \right. \nonumber \\
& & \hspace*{1cm} \left. -  \overline{ e^{i Q_z}\left( \frac{c}{B_0} \bm b \times \bm \nabla \left\langle \delta \psi_{gc}\right\rangle \cdot \bm \nabla \delta \bar G \right) } \right] \;,
\end{eqnarray}
where the gyrophase average has been included in the velocity space average and we have used Eq. (\ref{zonaleq:20}) in order to remove the bounce average from the LHS.

Equation (\ref{eq:999}) is the gyrokinetic extension of Eq. (\ref{eq:116}), derived in Chapter 3, and valid for equilibrium distortions on the
particle Larmor radius scale. As anticipated above, collisional transport is suppressed here but could be readily restored. If weighted 
over $mv^2/2$, Eq. (\ref{eq:999}) would give the gyrokinetic extension of Eq. (\ref{eq:154}) and, as noted in Chapter \ref{cap:gyrotransp}, show that 
fluctuation induced particle and energy transport are obtained from the same ``formal'' expressions. These points are further analyzed and articulated in the remaining part of this Chapter.
\section{Connection to transport}
\label{sec-1-4}
In this section we show that, considering only the contribution of zonal structures with long wave length, i.e. $k_{z}L < \delta^{-1/2}$, we obtain a transport equation for the density which is identical to the fluctuation induced part of Eq. (\ref{eq:116}) derived by means of the moments approach.

\begin{figure}
  \centering
\includegraphics[width=1\linewidth]{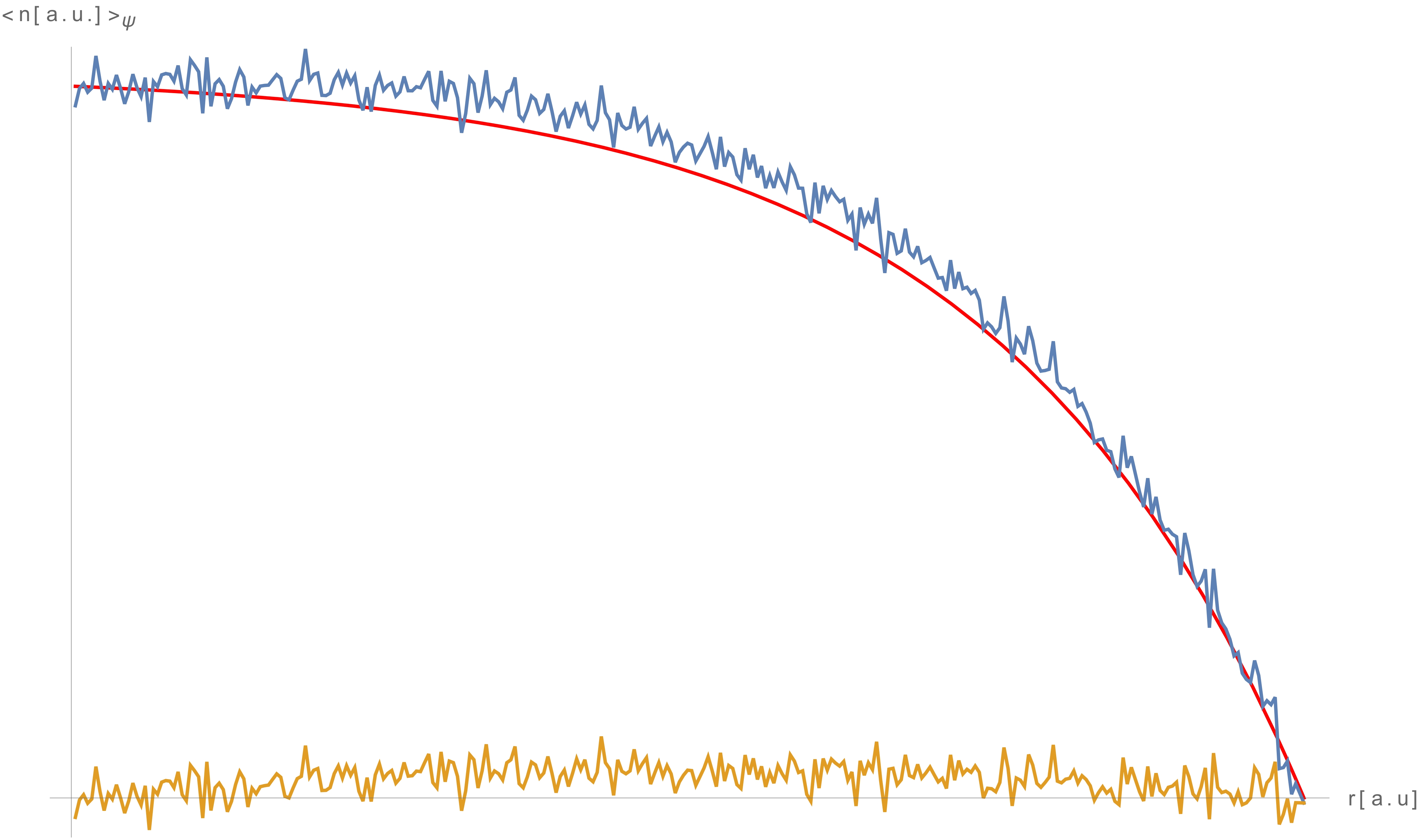}
\caption{Cartoon with a schematic plot of the surface averaged density, the equilibrium profile and the surface averaged fluctuations respectively in blue, red and orange.}
 \label{fig:nwithfluctuations}
\end{figure}

\begin{figure}
  \centering
\includegraphics[width=1\linewidth]{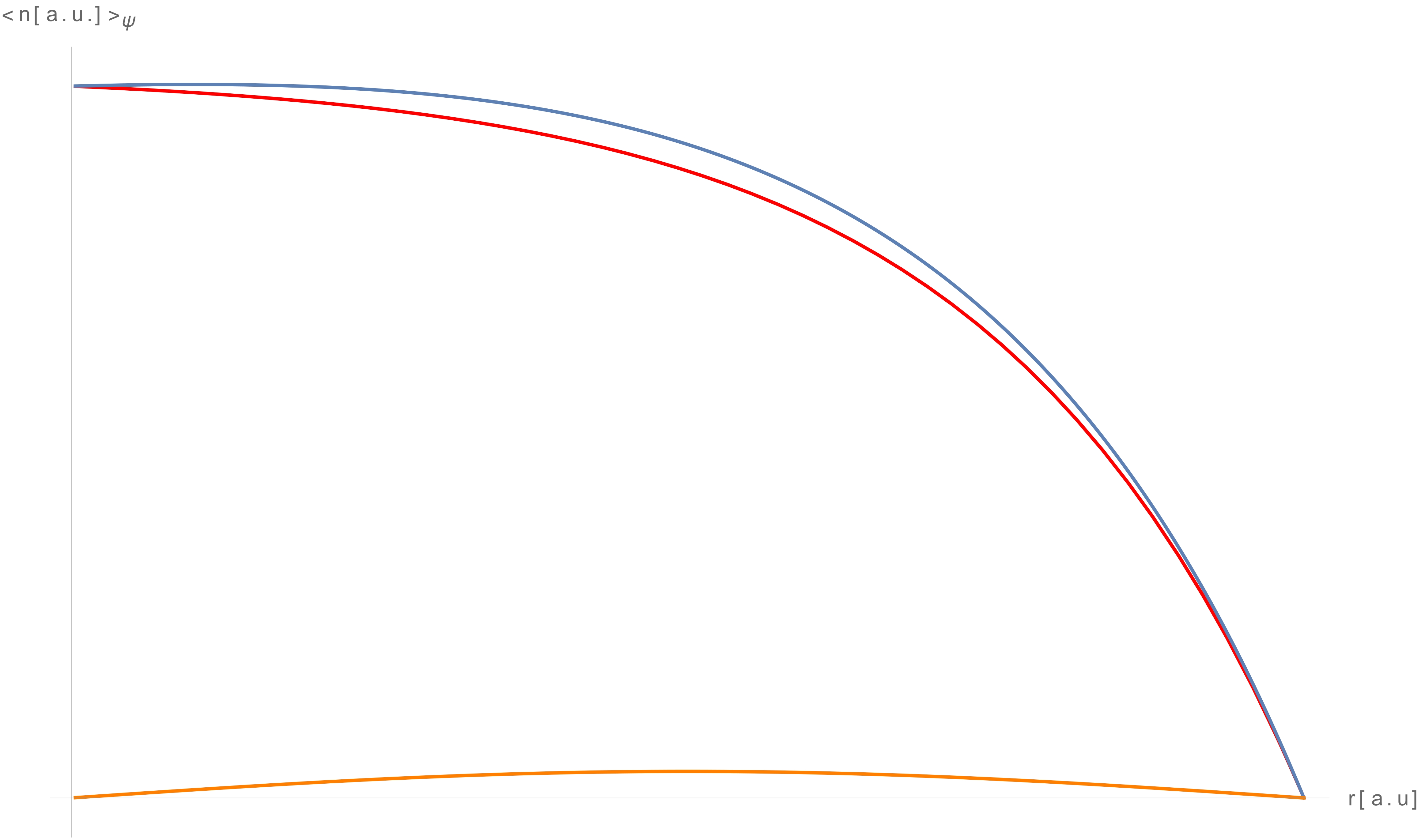}
\caption{Cartoon with a schematic plot of the long wave limit of the surface averaged density, the equilibrium profile and the surface averaged fluctuations respectively in blue, red and orange.}
 \label{fig:nwithoutfluctuations}
\end{figure}

\begin{figure}
  \centering
\includegraphics[width=1\linewidth]{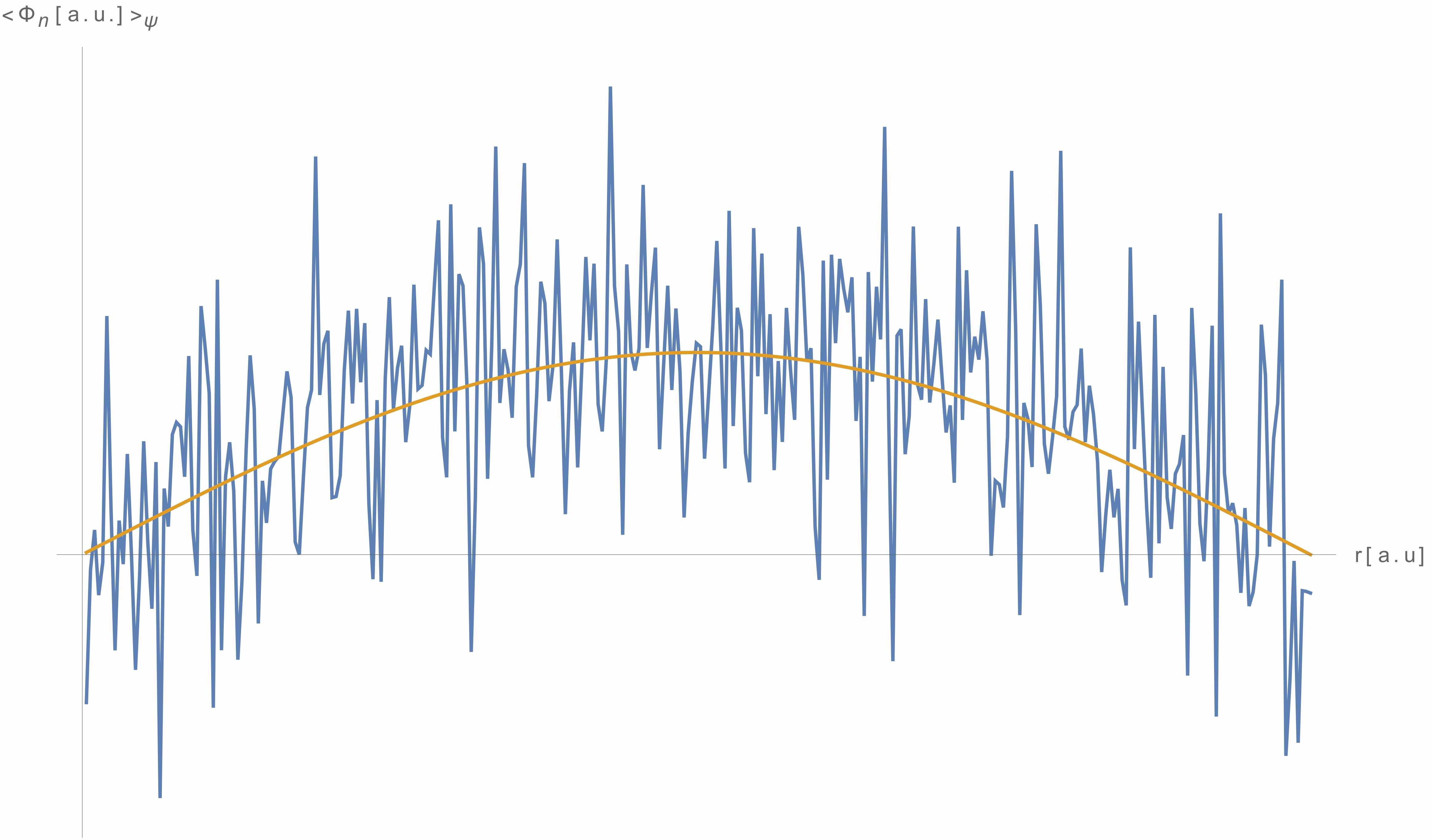}
\caption{Cartoon with a schematic plot of the surface averaged fluctuation induced particle flux and its long wave limit respectively in blue and Orange.}
 \label{fig:turbulentflux}
\end{figure}

We can re-write the last term of Eq. (\ref{eq:999}):
\begin{align}
\label{zonaleq:28}
& \bm b \times \bm \nabla \left\langle \delta \psi_{gc}\right\rangle \cdot \bm \nabla \delta \bar G =
\bm b \cdot \left( \bm \nabla r \times \bm \nabla \theta \right) \left( \frac{\partial \left\langle \delta \psi_{gc}\right\rangle}{\partial r}
\frac{\partial \delta \bar G }{\partial \theta} - \frac{\partial \left\langle \delta \psi_{gc}\right\rangle}{\partial\theta}
\frac{\partial \delta \bar G }{\partial r} \right) +\\
\nonumber &\hspace*{2em} + \bm b \cdot \left( \bm \nabla r \times \bm \nabla \phi \right) \left( \frac{\partial \left\langle \delta \psi_{gc}\right\rangle}{\partial r}
\frac{\partial \delta \bar G }{\partial \phi} - \frac{\partial \left\langle \delta \psi_{gc}\right\rangle}{\partial\phi}
\frac{\partial \delta \bar G }{\partial r} \right) + \\
\nonumber&\hspace*{2em} + \bm b \cdot \left( \bm \nabla \phi \times \bm \nabla \theta \right) \left( \frac{\partial \left\langle \delta \psi_{gc}\right\rangle}{\partial \phi}
\frac{\partial \delta \bar G }{\partial \theta} - \frac{\partial \left\langle \delta \psi_{gc}\right\rangle}{\partial\theta}
\frac{\partial \delta \bar G }{\partial \phi} \right).
\end{align}
Using toroidal coordinates for the magnetic equilibrium we can show that:
\begin{equation}
\label{zonaleq:29}
\bm b \cdot \left( \bm \nabla r \times \bm \nabla \theta \right) = \frac{F}{(d\psi/dr)} \frac{1}{JB} \; ; \;\;\;\;\; \bm b \cdot \left( \bm \nabla r \times \bm \nabla \phi \right) = \frac{F}{(d\psi/dr)} \left( \frac{q}{JB} - \frac{B}{F} \right).
\end{equation}
Substituting this expression into Eq. (\ref{zonaleq:28}) and noting that:
\begin{equation}
\label{eq:8}
{\cal O}\left( \frac{1}{qR_0} \right) \sim  k_\parallel = - \frac{i}{JB} \left( \frac{\partial}{\partial \theta} + q(r) \frac{\partial}{\partial \phi} \right),
\end{equation}
that $(J B)^{-1}\sim 1/q R_{0}$ and that $\partial_{\theta}\sim L k_{\perp}\sim \delta^{-1}k_{\parallel}$ when acting on plasma turbulence thus:
\begin{equation}
\label{zonaleq:27}
\frac{\partial}{\partial \theta} \simeq - q (r)  \frac{\partial}{\partial \phi} + {\cal O}(\delta)
\end{equation}
we obtain, at the leading order, the following expression:
\begin{equation}
\label{zonaleq:30}
\left. \hspace*{-1cm} \frac{c}{B} \bm b \times \bm \nabla \left\langle \delta \psi_{gc}\right\rangle \cdot \bm \nabla \delta \bar G \right|_z= \frac{c}{(d\psi/dr)}
 \left( \frac{\partial \left\langle \delta \psi_{gc}\right\rangle}{\partial \phi}
\frac{\partial \delta \bar G }{\partial r} - \frac{\partial \left\langle \delta \psi_{gc}\right\rangle}{\partial r}
\frac{\partial \delta \bar G }{\partial \phi} \right)_z + {\cal O}(\delta).
\end{equation}
Specializing the expression above for phase space zonal structures, which have an overall behavior independent of $\phi$ we obtain:
\begin{align}
\label{zonaleq:31}
&\hspace*{-1cm} \frac{c}{B} \bm b \times \bm \nabla \left\langle \delta \psi_{gc}\right\rangle \cdot \bm \nabla \delta \bar G = \frac{c}{(d\psi/dr)}
 \left( \frac{\partial \left\langle \delta \psi_{gc}\right\rangle}{\partial \phi}
\frac{\partial \delta \bar G }{\partial r} - \frac{\partial \left\langle \delta \psi_{gc}\right\rangle}{\partial r}
  \frac{\partial \delta \bar G }{\partial \phi} \right) + {\cal O}(\delta) \simeq \\
& \nonumber \hspace*{1cm} \simeq \frac{c}{(d\psi/dr)}
 \left( i n \left\langle \delta \psi_{gc}\right\rangle
\frac{\partial \delta \bar G }{\partial r} + i n \frac{\partial \left\langle \delta \psi_{gc}\right\rangle}{\partial r}
  \delta \bar G  \right) + {\cal O}(\delta) \simeq\\
&\nonumber \hspace*{1cm} \simeq \frac{i n c}{(d\psi/dr)}
  \frac{\partial}{\partial r}\left ( \delta \bar G \left\langle \delta \psi_{gc}\right\rangle  \right ) + {\cal O}(\delta) \simeq\\
&\nonumber \hspace*{1cm} \simeq c \frac{\partial}{\partial \psi}
 \left( R^2 \bm \nabla \phi \cdot \bm \nabla \left\langle \delta \psi_{gc}\right\rangle \delta \bar G \right) + {\cal O}(\delta)
\end{align}
where we have noted that zonal structures must have $n=0$ which is obtained only if the toroidal mode number of $\left\langle \delta \psi_{gc}\right\rangle$ is equal to the mode number of $\delta \bar G$ with the sign changed. Therefore we can re-write the evolutive equation for the zonal structures:
\begin{align}
\label{zonaleq:26}
\partial_t \left\langle \left\langle \delta f_z \right\rangle_v\right\rangle_\psi  & =  \frac{e}{m} \partial_t \delta \phi_{0,0} \left\langle
\left[ 1 - \left(e^{-iQ_z}\hat I_0\right) \overline{\left(e^{i Q_z} \hat I_0\right)} \right]\frac{\partial \bar F_0}{\partial \cal E} \right\rangle_v \nonumber \\
&    - \frac{1}{V'} \frac{\partial}{\partial \psi} \left\langle\left\langle V' \left(e^{-iQ_z}\hat I_0\right)
\overline{\left[ c e^{i Q_z} R^2 \bm \nabla \phi \cdot \bm \nabla \left\langle \delta \psi_{gc}\right\rangle \delta \bar G \right]} \right\rangle_v\right\rangle_\psi
\end{align}
where the second term on the RHS is the long-lived effect (not damped by collisionless processes) of turbulent transport. Thus, phase space zonal structures bear fundamental information on the nonlinear evolution of plasma equilibria and related transport and give back expressions of turbulent transport in the long wavelength limit $\left(e^{i Q_z} \hat I_0\right) \rightarrow 1$. In order to show this result we note that the first term on the RHS of Eq:(\ref{zonaleq:26}) reads:
\begin{equation}
\label{eq:998}
\frac{e}{m} \partial_t \delta \phi_{0,0} \left\langle
\left[ 1 - \left(e^{-iQ_z}\hat I_0\right) \overline{\left(e^{i Q_z} \hat I_0\right)} \right]\frac{\partial \bar F_0}{\partial \cal E} \right\rangle_v \sim (Q_{z}^{2} + \lambda^{2})\frac{e \delta \phi}{T}n_{0} \sim (k_z \rho_{drift})^2 \delta \omega n_{0}
\end{equation}
and, therefore, we can neglect this term in the study of the effect of zonal structures with long wave length such that $k_{z}L < \delta^{-1/2}$ on transport up to order $\delta^{2}\omega$. With this assumption the second term on the RHS reads:
\begin{align}
\label{eqappzonal:1}
&\nonumber - \frac{1}{V'} \frac{\partial}{\partial \psi} \left\langle\left\langle V' \left(e^{-iQ_z}\hat I_0\right)
\overline{\left[ c e^{i Q_z} R^2 \bm \nabla \phi \cdot \bm \nabla \left\langle \delta \psi_{gc}\right\rangle \delta \bar G \right]} \right\rangle_v\right\rangle_\psi \sim \\
& -\frac{1}{V'} \frac{\partial}{\partial \psi} \left\langle\left\langle V'
\overline{\left[ c  R^2 \bm \nabla \phi \cdot \bm \nabla \left\langle \delta \psi_{gc}\right\rangle \delta \bar G \right]} \right\rangle_v\right\rangle_\psi.
\end{align}
The transport equations derived using the moment approach automatically satisfy the condition on $k_{z}$ because we have assumed that $k_{z}L \sim 1$. In Fig:\ref{fig:nwithfluctuations} and Fig:\ref{fig:nwithoutfluctuations} we show the effect of the long wave limit on the surface averaged density $\sav{n}$. Furthermore in Fig:\ref{fig:turbulentflux} we show the fluctuation induced particle flux.

In order to prove that Eq. (\ref{zonaleq:26}) is equivalent to the sum of Eq. (\ref{eq:109}) and Eq. (\ref{eq:110}) it remains to show that, given a scalar field $F$:
\begin{equation}
\label{eqappzonal:1}
\sav{F \bs{0}\delta \bar{G}}= \sav{\delta \bar{G} \bs{0} F}.
\end{equation}
Given two functions $f_{1}(r)$ and $f_{2}(r)$, if we are interested in their zonal component we can calculate:
\begin{align}
\label{eqappzonal:2}
&\widehat{f_{1}f_{2}}(k_{z})= \frac{1}{2 \pi}\int \mathit{d} r \, e^{-i k_{z} r} \int \mathit{d} k_{1r} \mathit{d} k_{2r}\, e^{i k_{1r}r}e^{i k_{2r}r}\hat{f}_{1}\hat{f}_{2}= \\
&\nonumber =\int \mathit{d} k_{1r} \mathit{d} k_{2r}\, e^{i k_{1r}r}e^{i k_{2r}r}\hat{f}_{1}\hat{f}_{2}\delta(k_{1}+k_{2}-k_{z})
\end{align}
and, therefore:
\begin{equation}
\label{eqappzonal:3}
k_{1r}+k_{2r}=k_{z} \Rightarrow k_{1r}\sim - k_{2r} + \mathcal{O}(\delta)
\end{equation}
where the implication sign is due to the fact that $k_{z}\rho_{d}\sim k_{z}\rho \sim \delta k_{r} \rho$. On the other hand:
\begin{align}
\label{eqappzonal:4}
&\sav{f_{1}f_{2}}= \int \mathit{d}\phi \, \int \mathit{d}\theta \, J \int \mathit{d} k_{1\theta} \mathit{d} k_{2\theta}\, e^{i k_{1\theta}\theta}e^{i k_{2\theta}\theta}\hat{f}_{1}\hat{f}_{2} = \\
& = \int \mathit{d}\phi \, \int \mathit{d}\theta \, \sum_{m}e^{i m \theta}J_{m}\int \mathit{d} k_{1\theta} \mathit{d} k_{2\theta}\, e^{i k_{1\theta}\theta}e^{i k_{2\theta}\theta}\hat{f}_{1}\hat{f}_{2}.
\end{align}
The only terms surviving the flux surface average must satisfy the following selection rule:
\begin{equation}
\label{eqappzonal:5}
k_{1 \theta} + k_{2\theta} + m = 0
\end{equation}
where due to the dependence of the Jacobian on the equilibrium magnetic field when $m \gg 1$ it must hold that $J_{m} \ll 1$ and, using the  gyrokinetic ordering, i.e. $m/k_{1 \theta}\sim \delta$, we can write:
\begin{equation}
\label{eqappzonal:6}
k_{1 \theta}= - k_{2 \theta} + \mathcal{O}(\delta).
\end{equation}
In the same way we can show that $k_{1\phi}=k_{2\phi}$ and, therefore, $\wb{k_{1}} \sim - \wb{k_{2}}$. Choosing $f_{1}=F$ and $f_{2}= \delta \bar{G}$ We can write at the leading order:
\begin{align*}
&\sav{F \bs{0}\delta \bar{G}}= \int \mathit{d} \wb{k}_{1} e^{i \wb{k}_{1}\cdot \wb{r}}\hat{F}(\wb{k_{1}})  \int \mathit{d} \wb{k}_{2} e^{i \wb{k}_{2}\cdot \wb{r}}I_{0}(\wb{k_{2}}) \widehat{\delta \bar{G}}(\wb{k_{2}}) = \\
&=\int \mathit{d} \wb{k}_{1} e^{i \wb{k}_{1}\cdot \wb{r}}I_{0}(-\wb{k}_{1})\hat{F}(\wb{k_{1}})  \int \mathit{d} \wb{k}_{2} e^{i\wb{k}_{2}\cdot \wb{r}} \widehat{\delta \bar{G}}(\wb{k}_{2})= \sav{\delta \bar{G} \bs{0} F}
\end{align*}
where we have used the analyticity and the parity of $I_{0}(\lambda)$. This result is expected and suggests, as it is shown below, that particle and energy fluctuation induced fluxes should be obtained from the same formal expression, with a different weighting in the velocity space.

In summary  phase space zonal structures are deeply connected with plasma transport processes and must be taken into account in order have a realistic and self-consistent description of a thermonuclear plasma.

\chapter{Applications: collisional fluxes}
\label{cap:coll}
In this chapter we calculate the collisional particle flux using neoclassical transport theory \cite{hinton1976theory} in a realistic geometry. In particular we study the particle flux in the DTT \cite{pizzuto2016dtt}, which is the Italian project proposal for a Tokamak capable of eventually integrating all relevant physics and technologic issues concerning alternative power exhaust solutions for ITER and DEMO \cite{romanelli2013roadmap,ihli2007recent} currently under evaluation. We recall that a complete analysis of collisional fluxes requires the study of the effect of zonal structures. In principle, all moments of the collisional operator are modified by the presence of zonal structures.
The calculation of these modifications requires the knowledge of the phase space zonal structure, which can be extracted from a gyrokinetic turbulent code (cf. Ref. \cite{garbet2010gyrokinetic} for a recent review)  and will be analyzed in future works. In this chapter, as illustrative application of the theoretical framework
presented in Chapters 3 and 4, we have not calculated this contribution and, instead, we have used neoclassical closure relations in order to calculate collisional fluxes in a realistic case of practical interest for plasma operations foreseen in the DTT.
\section{Collisional fluxes}
\label{sec-5-2}
The radial flux of particles is described by the following term of Eq. (\ref{eq:96}):
\begin{equation}
\label{eq:52}
\left\langle (en/c) \bm V \cdot \nabla \psi \right\rangle_\psi   =  - \left\langle (en \bm E + \bm F ) \cdot R^2 \bm \nabla \phi \right\rangle_\psi.
\end{equation}
In the previous chapters, and in the absence of fluctuations, we have shown how to write the right hand side of this equation as the sum of two contributions: the classical particle flux and the neoclassical one. In the following, we will calculate both terms with the neoclassical particle flux evaluated  in different collisionality regimes, i.e. Banana and Pfirsch Schl\"uter \cite{hinton1976theory}. It is well known \cite{rutherford1970collisional} that particle fluxes are ambipolar up to $\mathcal{O}(\delta^{2})$, i.e.:
\begin{equation}
\label{eq:88}
\sum_{\alpha} q_{\alpha} \langle n_{\alpha} {\bm V} \cdot {\bm \nabla} \psi \rangle_{\psi}=0.
\end{equation}
In this work, for simplicity, we will deal only with a two species plasma with $Z=1$. Restricting ourselves to this case and applying Eq. (\ref{eq:88}), we obtain that the ionic flux of particles is identical to the electronic one. Therefore, in the next sections we will calculate only the electronic particle flux. For the sake of simplicity, in the next pages it will be called simply particle flux.
\subsection{Classical particle flux}
\label{sec-5-2-2}
Following the derivation described in \cite{hinton1976theory} and using the expressions introduced in the previous chapters, we obtain the following expression for the radial particle flux averaged on a flux surface:
\begin{equation}
\label{eq:53}
\left\langle {\bm F}_{\perp}\cdot R^{2} {\bm \nabla}\phi \right\rangle_{\psi}= - \left \langle  \frac{1}{\tau_{e} \left| \Omega_{e} \right|}  {\bm b}\times \left[ {\bm \nabla}P- \frac{3}{2}{\bm \nabla}T_{e} \right] \cdot R^{2} {\bm \nabla}\phi  \right \rangle_{\psi}
\; ,
\end{equation}
where we have introduced the electron-ion momentum exchange time:
\begin{equation}
\label{eq:54}
\tau_{e}\equiv \frac{3}{16 \pi^{1/2}}\frac{m_{e} v_{the}^3}{Z^{2}e^{4} \ln{\Lambda}},
\end{equation}
the Coulomb logarithm as $\ln{\Lambda} \equiv \ln{9 N}$, where $N$ is the number of particles inside the Debye sphere, $v_{the}= \sqrt{T_{e}/m_{e}}$ is the electron thermal velocity and $P=P_{e}+P_{i}$ is the total plasma pressure. We need to calculate the surface average of the following quantities:
\begin{eqnarray}
\label{eq:55}
&\left(  {\bm b}\times {\bm \nabla}P\right)\cdot {\bm \nabla\phi}\\
\nonumber &\left(  {\bm b}\times {\bm \nabla}T_{e}\right)\cdot {\bm \nabla\phi}.
\end{eqnarray}
In this work, we are dealing with axi-symmetric equilibria and, using this assumption we obtain the following relation valid in arbitrary flux coordinates:
\begin{equation}
\label{eq:56}
\left(  {\bm b}\times {\bm \nabla}P\right)\cdot {\bm \nabla\phi} = - \frac{1}{B R^{2}}{\bm \nabla}\psi\cdot{\bm \nabla}P
\; ;
\end{equation}
and the analogous one for the electronic temperature $T_{e}$. We now choose to use Hamada coordinates, i.e. such that Eq. (\ref{eq:46}) is valid, in order to express the surface average as an integral over the $\theta$ coordinate:
\begin{equation}
\label{eq:57}
\left\langle {\bm F}_{\perp} \cdot R^{2} {\bm \nabla}\phi \right\rangle_{\psi}= \frac{1}{2 \pi}\left( \frac{3}{2}n_{e}\frac{\partial T_{e}}{\partial \psi}- \frac{\partial P}{\partial \psi} \right) \frac{1}{\tau_{e}} \int \mathit{d} \theta \frac{\left| {\bm \nabla}\psi \right|^{2}}{B \left| \Omega_{e} \right|}.
\end{equation}
From this expression, we note that Hamada coordinates are particularly useful for this calculation because the dependence on the Jacobian is simplified. The integral in Eq. (\ref{eq:57}) retains the information about the geometry of the magnetic surfaces while the term outside the integral describe  the physical quantities responsible for the transport process. An exact, analytic expression for the geometrical factor $\left| \bm{\nabla} \psi \right|^{2}/B^{2}$ can be obtained in the case of circular flux surfaces and large aspect ratio up to the second order in the $r/R_{0} \ll 1$ limit, yielding:
\begin{equation}
\label{eq:72}
\left \langle \frac{\left| {\bm \nabla}\psi \right|^2}{B^{2}}\right \rangle_\psi= \frac{r^2}{q^{2}}\left[ 1 + \left( \frac{r}{R_{0}} \right)^{2} \left( \frac{3}{2}- q^{-2} \right) \right]
\; ,
\end{equation}
where the distance between the symmetry axis of the torus and the  magnetic axis is $R_{0}$. 
\subsection{Pfirsch Schl\"uter particle flux}
\label{sec-5-2-3}
Following \cite{hinton1976theory}, we obtain this expression for the neoclassical Pfirsch Schl\"uter particle flux:
\begin{eqnarray}
\label{eq:69}
&\left\langle ( en \bm E_0 + \bm F_\parallel ) \cdot R^2 \bm \nabla \phi \right\rangle_\psi=- \frac{c m_{e}}{e \tau_{e}} F^{2}\left[ \left \langle \frac{1}{B^{2}} \right \rangle_{\psi}- \frac{1}{\left \langle B^{2}\right \rangle_{\psi}} \right]\cdot \\
\nonumber &\cdot \left[ \frac{k_{22}}{k} \frac{\partial P}{\partial \psi}- \frac{5}{2}\frac{k_{12}}{k} n_{e} \frac{\partial T_e}{\partial \psi}\right]
\end{eqnarray}
where $k= k_{11}k_{22}-k_{12}^{2}$. These coefficients connect the flux of particles, on the LHS of Eq.(\ref{eq:69}), with the thermodynamic forces which are driving it and are calculated by means of neoclassical transport theory. In particular the values used in this work have been estimated in \cite{spitzer1953transport} for the special case of $Z=1$ and read:
\begin{equation}
\label{eq:86}
k_{11}= 1.975 \quad k_{12}=1.381, \quad k_{22}=4.174.
\end{equation}
The Pfirsch Schl\"uter  geometrical factor can be expressed in terms of an average over the Hamada angular coordinate:
\begin{equation}
\label{eq:70}
\left \langle \frac{1}{B^{2}} \right \rangle_{\psi}- \frac{1}{\left \langle B^{2}\right \rangle_{\psi}}= \frac{1}{2 \pi}\left( \int \mathit{d} \theta \frac{1}{B^{2}} - \frac{4 \pi^{2}}{\int \mathit{d} \theta B^{2}} \right).
\end{equation}
These integrals can be calculated analytically in the particular case of circular flux surfaces and large aspect ratio up to the second order in $r/R_{0}$, obtaining the following expression:
\begin{equation}
\label{eq:71}
\left \langle \frac{1}{B^{2}} \right \rangle_{\psi}- \frac{1}{\left \langle B^{2}\right \rangle_{\psi}}= \frac{2 r^{2}}{B_{0}^{2} R_{0}^{2}}.
\end{equation}
In the next section, both the geometrical factors will be evaluated numerically using the data of the DTT reference scenario.

The calculation of the low-collisional particle flux requires a different approach, which will be introduced in the next section.
\subsection{Banana particle flux}
\label{sec-5-2-4}
In \cite{angioni2000neoclassical}, using neoclassical transport theory, a set of kinetic equations have been derived  and solved by the code CQL3D \cite{harvey1992cql3d}. The transport coefficients in the low collisionality regime can be expressed as integrals of the resulting distribution functions. The authors have applied this method on a wide range of equilibrium parameters and then fitted the results with functions of the effective trapped particle fraction in order to obtain simple formulas for the neoclassical transport coefficients for an arbitrary geometry of the equilibrium. In particular in this work we will use these results in order to evaluate the diffusion coefficient in the low-collisional case for the DTT reference scenario.

\begin{figure}[htb!]
\centering
\includegraphics[width=.9\linewidth]{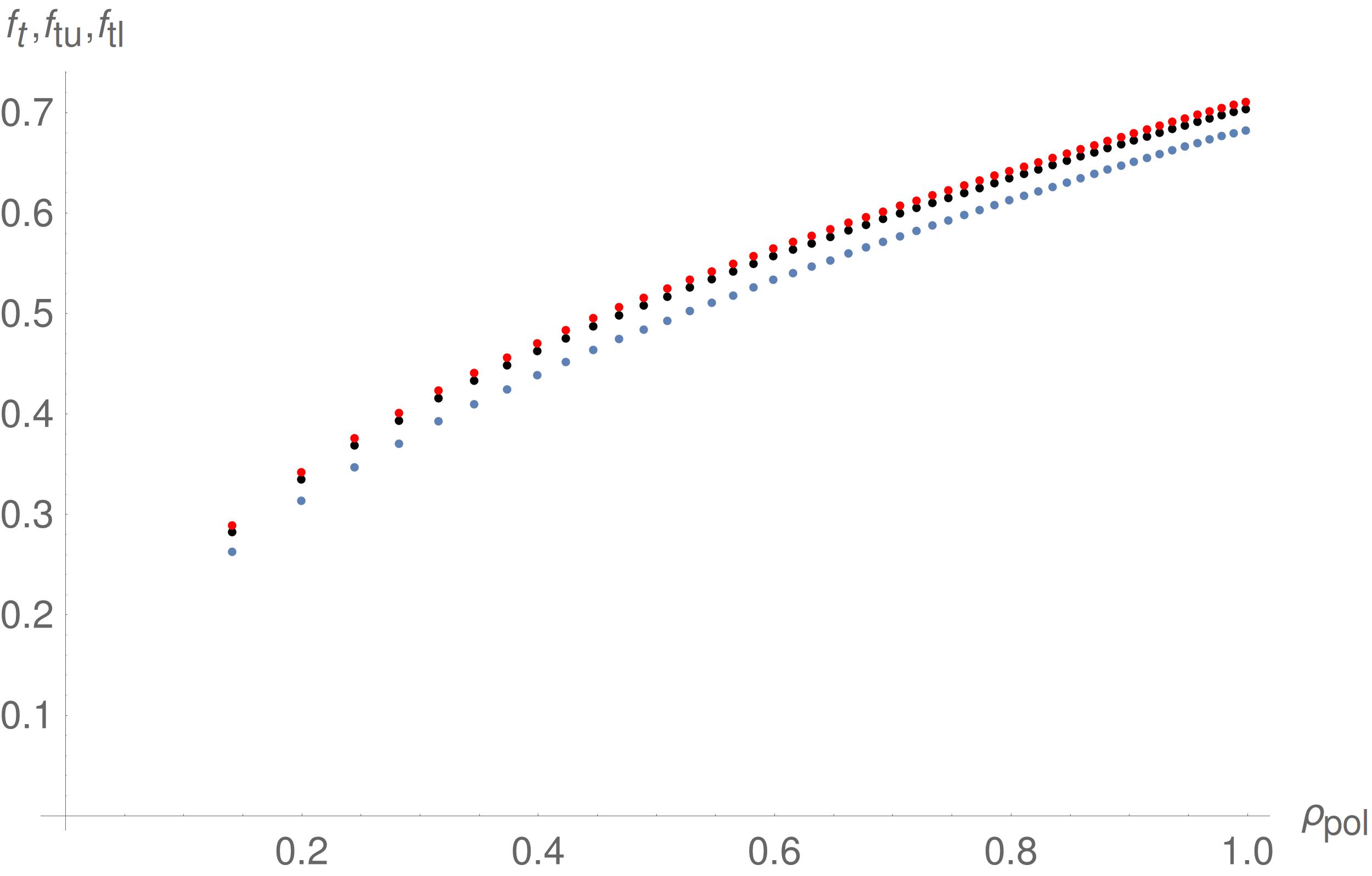}
\caption{Plot of the effective trapped particle fraction $f_{t}$ (in black) and its lower and and upper bounds respectively in red and blue for the DTT reference scenario.}
\label{fig:below}
\end{figure}

We introduce the effective trapped particle fraction:
\begin{equation}
\label{eq:76}
f_{t} = 1 - \frac{3}{4}\langle h^{2}\rangle \int_{0}^{1} \mathit{d} \lambda\frac{\lambda}{\langle( 1- \lambda h)^{1/2}\rangle_\psi}
\; ,
\end{equation}
where $h=B/B_{max}$ and $B_{max}$ is the maximum value of the magnetic field on a given flux surface. Thus $f_{t}$ is defined through a double integration which must be computed numerically for realistic geometries. This can be time consuming and non accurate and, therefore, several approximate formulas for $f_{t}$ have been proposed in literature. In particular, we will use the expression derived in \cite{lin1995upper}. In this article the Schwartz inequality is used in order to obtain an expression for an upper and a lower bound for $f_{t}$, i.e. respectively $f_{tu},f_{tl}$:
\begin{eqnarray}
\label{eq:77}
f_{tu}& = & 1 - \frac{\langle h^{2}\rangle_\psi}{\langle h\rangle_\psi^{2}}\left[ 1- (1- \langle h\rangle_\psi)^{1/2}\left(  1+ \frac{1}{2} \langle h \rangle_\psi\right) \right]\\
f_{tl}& = & 1 - \langle h^{2}\rangle_\psi \left \langle h^{-2} \left[ 1- (1-h)^{1/2} \left( 1 + \frac{1}{2}h \right) \right]\right \rangle_\psi.
\end{eqnarray}
It can be shown that a linear combination of these two functions is a good approximation for $f_{t}$. Using an analytic model with elliptical flux surfaces, it has been shown that the best approximation is achieved with this particular linear combination:
\begin{equation}
\label{eq:78}
f_{t}\approx \omega f_{tu} +(1-\omega)f_{tl}, \quad \omega \approx 0.75.
\end{equation}
In this work, we will use this expression in order to evaluate neoclassical banana fluxes. Using the data extracted from the \texttt{eqdsk} file describing the DTT reference scenario, i.e. a standard input for many numerical codes and also a normal way of storing experimental equilibrium data, we can calculate the values of these functions on each flux surface. In Fig. \ref{fig:below} we show a plot with the approximate value for $f_{t}$ and its lower and upper bounds for the DTT reference scenario.
The flux coordinate $\rho_{pol}$ will be defined in the next page. All the expressions for the transport coefficients obtained in \cite{angioni2000neoclassical} are function of $f_{t}$ and $f^{d}_{t}$. Therefore, in order to describe the low-collisionalility neoclassical transport, we need to calculate the latter which is defined as:
\begin{equation}
\label{eq:79}
f_{t}^{d}= 1 - \frac{3}{4}\langle B^{-2}\rangle^{-1}I_{\lambda}
\end{equation}
with:
\begin{equation}
\label{eq:80}
I_{\lambda}= (1- f_{t})\frac{4}{3}\frac{1}{\langle B^{2} \rangle}.
\end{equation}
This quantity can be computed numerically in terms of $f_{t}$ and of the magnetic field data extracted from the \texttt{eqdsk} file. All the collisional fluxes and, in particular the electronic particle flux, can be expressed in terms of $f^{d}_{t}$. In the following we will focus on the diffusion coefficients rather than the collisional fluxes and, for this reason, we will not show the expression for  $\langle \Gamma_{e} \cdot \bm{\nabla}r\rangle$ which, anyway, will be computed in order to derive the expressions for the diffusion coefficients,i.e. Eq. (\ref{eq:81}).
\section{Diffusion coefficients}
\label{sec-5-3}
\subsection{Normalizations}
\label{sec-5-3-1}
In order to compare our results with other works, we calculate the diffusion coefficients associated with each collisional flux. As an example of the calculations required, we derive here the diffusion coefficient $D_{c}$ associated with classical particles flux.

Using the properties of the flux surface average and the expressions already derived we can write the averaged continuity equation:
\begin{equation}
\label{eq:58}
\frac{\partial \langle n \rangle_{\psi}}{\partial t}  = - \frac{c}{e} \frac{1}{V'_{\psi}}\frac{\partial}{\partial \psi}\left( V'_{\psi} \langle {\bm F}_{\perp}\cdot {\bm \nabla} \phi R^{2}\rangle_{\psi} \right)
\end{equation}
with $V'_{\psi}= dV/d\psi$. From Eq. (\ref{eq:53}) we can see the diffusive nature of this equation, but we need some algebra in order to find a diffusion coefficient with the right physical dimensions, i.e. $area/time$. This is physically relevant because it allows to characterize the diffusion process in terms of a random walk with step-size $\Delta x$ and time-step $\Delta t$, describing the random motion of particles as they diffuse, i.e. $D \approx (\Delta x)^{2}/\Delta t$.  For this reason, we will express the gradients with respect to the variable $r$ instead of $\psi$. Thus, we define the normalized poloidal flux $\rho_{pol}$:
\begin{equation}
\label{eq:59}
\rho_{pol}= \sqrt{\frac{\psi}{\psi_{0}}}.
\end{equation}
where $\psi_{0}$ is the value of the poloidal flux $\psi$ calculated on the separatrix. We can express $V_{\psi}'$ in terms of $V'_{\rho_{pol}}= d V /d \rho_{pol}$ by applying the chain rule:
\begin{equation}
\label{eq:60}
V'_{\psi}= \frac{V'_{\rho_{pol}}}{2 \rho_{pol}\psi_{0}}.
\end{equation}
We re-write Eq. (\ref{eq:58}) in terms of $\rho_{pol}$:
\begin{equation}
\label{eq:61}
\frac{\partial \langle n \rangle_{\psi}}{\partial t}  = - \frac{c}{e} \frac{1}{V'_{\rho_{pol}}}\frac{\partial}{\partial \rho_{pol}}\left(\frac{1}{2 \rho_{pol}\psi_{0}} V'_{\rho_{pol}} \langle {\bm F}_{\perp}\cdot {\bm \nabla} \phi R^{2}\rangle_{\psi} \right).
\end{equation}
Using Eq. (\ref{eq:57}) and integrating over the volume enclosed by the flux surface $\psi$, we obtain the following expression:
\begin{equation}
\label{eq:62}
\int_{0}^{V} \mathit{d} \tilde{V} \frac{\partial}{\partial t} \langle n \rangle_{\psi} = - \int_{0}^{V} \mathit{d} \tilde{V}\frac{1}{V'_{\rho_{pol}}}\frac{\partial}{\partial \rho_{pol}}\left[\dots  \right].
\end{equation}
Doing the integration, we obtain the following formula for the derivative of the number of particles enclosed by the magnetic surface $\psi$:
\begin{equation}
\label{eq:63}
\frac{\partial N_{\rho_{pol}}}{\partial t} = - V'_{\rho_{pol}}\left \langle \Gamma_{c} \cdot {\bm \nabla}\rho_{pol} \right \rangle_{\psi}
\; ,
\end{equation}
where $\Gamma_{c}$ is the classical particle flux:
\begin{equation}
\label{eq:64}
\Gamma_{c}=-D_{c\rho_{pol}}\frac{1}{T_{e}}\left(  \frac{\partial P}{\partial \rho_{pol}} - \frac{3}{2}n_{e}\frac{\partial T_{e}}{\partial \rho_{pol}}\right)
\end{equation}
and:
\begin{equation}
\label{eq:65}
D_{c\rho_{pol}}= \frac{1}{2 \pi}\frac{T_{e}}{m_{e}\tau_{e} \left| \Omega_{e} \right|}\frac{1}{4 \rho_{pol}^{2}\psi_{0}^{2}}\int \mathit{d}\theta \frac{\left| {\bm \nabla \psi} \right|^{2}}{B \left| \Omega_{e} \right|}.
\end{equation}
The diffusion coefficient $D_{c\rho_{pol}}$ is still not expressed with the usual physical dimensions. For this reason we introduce the normalized toroidal flux:
\begin{equation}
\label{eq:66}
\rho_{tor}^{2}= \frac{r^2}{a^{2}}
\end{equation}
where $a$ is the value of $r$ on the separatrix and $r$ is the value of the radial polar coordinate at $\theta=0$. We obtain the following relation between $\rho_{tor}$ and $\rho_{pol}$:
\begin{equation}
\label{eq:67}
\rho_{tor}= \sqrt{\frac{\int_{0}^{\rho_{pol}} \mathit{d} \hat{\rho}_{pol}q}{\int_{0}^{1} \mathit{d} \hat{\rho}_{pol} \hat{\rho}_{pol}q}}= \sqrt{\frac{1}{\alpha} \int_{0}^{\rho_{pol}} \mathit{d} \hat{\rho}_{pol}q}.
\end{equation}
Using Eq. (\ref{eq:66}) and Eq. (\ref{eq:67}), we can calculate the classical diffusion coefficient obtaining an expression identical to Eq. (\ref{eq:65}) except for the factor:
\begin{equation}
\label{eq:68}
\frac{V'_{tor}}{(2 \rho_{pol}\psi_{0})^{2}}\left(  \frac{\alpha}{2 \rho_{tor}}q \rho_{pol} a\right)^{2}
\; ,
\end{equation}
instead of $(4 \rho_{pol}^{2}\psi_{0}^{2})^{-1}$, which multiplies the angular average.

The calculations for $D_{PS}$ are identical except for the geometrical factors and, therefore, will not be shown. The resulting diffusion coefficients for the DTT reference scenario will be plotted in the next section. The same procedure can be applied, using the expressions derived in \cite{angioni2000neoclassical}, in order to obtain $D_{ban}$. Strictly speaking, there is a diffusion coefficient for each thermodynamic force. These coefficients are identical except for a function of the effective trapped particle fraction $f^{d}_{t}$. In this work, without loss of generality, we will focus on the diffusion coefficient coupled with the temperature gradient. Using the same notation of \cite{angioni2000neoclassical} we write:
\begin{eqnarray}
\label{eq:81}
D_{ban}&=& \frac{1}{\tau_{e}}\frac{m_{e}T_{e}}{e^{2}}c^{2}F^{2}(\psi)\langle B^{-2}\rangle \left( \frac{d r}{d \psi} \right)^{2} \mathcal{L}^{e}_{12}
\; ,
\end{eqnarray}
where:
\begin{equation}
\label{eq:82}
\mathcal{L}^{e}_{12}= \mathcal{L}_{d}\left[B_{0}^{2}\langle B^{-2}\rangle \right]\mathcal{K}^{e}_{12}(f^{d}_{t})
\; ,
\end{equation}
and the function $\mathcal{L}_{d}$ is defined as:
\begin{equation}
\label{eq:83}
\mathcal{L}_{d}= \frac{n_{e} \rho^{2}_{ep}}{\tau_{e}}\left(\frac{d \psi}{d\rho}\right)^{2
}
\; ,
\end{equation}
with $\rho_{ep}$ the electronic poloidal Larmor radius:
\begin{equation}
\label{eq:84}
\rho_{ep}= \frac{v_{Te}}{\Omega_{ep}}= \frac{\sqrt{2 m_{e}T_{e}}}{|q_{e}|B_{po}(\rho
)}, \quad B_{po}= \frac{d\psi}{d \rho}\frac{B_{0}(\psi)}{I(\psi)}
\end{equation}
and $B_{0}(\psi)$ is an arbitrary function chosen to normalize the magnetic field on a given flux surface. The function $\mathcal{K}^{e}_{12}$ is given in \cite{angioni2000neoclassical} and reads:
\begin{eqnarray}
\label{eq:85}
\mathcal{K}^{e}_{12}& = &0.75 \left( 1+ \frac{0.9}{Z+0.5} \right)f^{d}_{t}- \frac{0.95}{Z + 0.5}(f^{d}_{t})^{2}+ \frac{0.3}{Z+0.5}(f^{d}_{t})^{3}-\frac{0.6}{Z+0.5}(f^{d}_{t})^{4}.
\end{eqnarray}
In the next section, we will evaluate these expressions using the data from the DTT reference scenario.
\subsection{Gyro-Bohm units}
\label{sec-5-3-2}
In this section we will introduce the gyro-Bohm diffusion coefficient $D_{gB}$ in order to normalize our results and facilitate the comparison with the existing literature. 

The maximum theoretical diffusion due to fluctuations that can be achieved in a magnetized plasma, which is due to the fluctuations, has been estimated by Bohm \cite{krommes2002fundamental} and reads:
\begin{equation}
\label{eq:73}
D_B^* = \frac{1}{16} D_B = \frac{1}{16} \frac{cT}{eB}.
\end{equation}
It is obtained modeling a particle with  a random walk with Larmor radius as step size and inverse cyclotron frequency as time step. Bohm diffusion is an upper bound for transport which would make controlled thermonuclear fusion practically impossible. Typical fusion experiments, due to the low frequency of the fluctuations with respect to $\Omega_{i}$ and due to the nonuniformity of the particle distribution functions,  are instead affected by gyro-Bohm diffusion, which is reduced by a factor  $\rho_s/a$ with respect to Bohm diffusion, where $\rho_s = c \,m_i^{1/2}T_e^{1/2}/eB$ is the ion Larmor radius computed at the sound speed and $a$ is the minor radius of the Tokamak. It is common practice to use the gyro-Bohm diffusion coefficient $D_{gB}$ as reference to express all other transport processes. In particular we can write:
\begin{equation}
\label{eq:74}
D_{gB} = \frac{1}{k_\perp \rho_s}\frac{\rho_s}{a} D_B
\; ,
\end{equation}
where $k_{\perp}$ is the typical perpendicular wave vector of the fluctuation spectrum. 
In the next section, we will normalize the diffusion coefficients with respect to $D_{e, gB}$.
\section{The DTT reference scenario}
In this section we will evaluate the expressions derived in the previous pages for the DTT. We choose the DTT mainly for two reasons: this is the most important Italian project proposal regarding magnetic confinement fusion and it might be one of the most important machines for the study of power exhaust problems \cite{kallenbach2013impurity}.
\begin{figure}[htb]
\centering
\includegraphics[width=.3\linewidth]{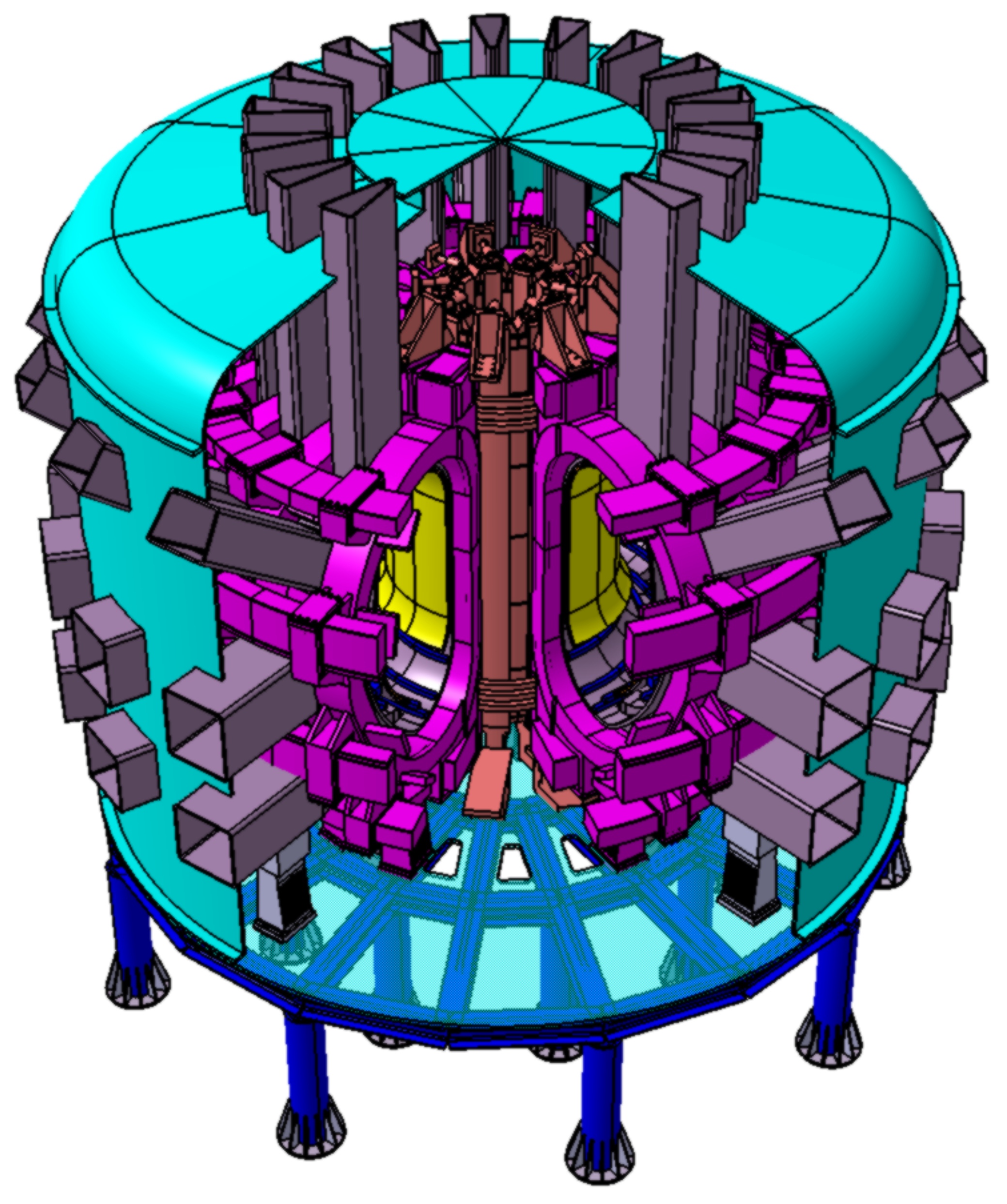}
\caption{View of the DTT (Courtesy of ENEA).}
\end{figure}
\subsection{The Divertor Tokamak Test facility}
\label{sec-5-4-2}
The DTT is a project sponsored by the EUROfusion consortium \cite{pizzuto2016dtt} with the goal of designing a new machine capable of eventually integrating all relevant physics and technologic issues concerning alternative power exhaust solutions for DEMO \cite{romanelli2013roadmap,ihli2007recent}. In the following table, we report the DTT parameters:
\begin{center}
  \begin{tabular}{|l|l|}
    \hline
  \multicolumn{2}{|c|}{DTT parameters} \\
\hline
major radius & $2.15 m$\\
\hline
aspect ratio ($R_0/a$) & $3.1$\\
\hline
toroidal field & $6 T$\\
\hline
plasma current & $6 MA$\\
\hline
additional power & $45 MW$\\
\hline
\end{tabular}
\end{center}
According to the European Road Map \cite{romanelli2013roadmap}, this machine should start the operations in 2022.
\subsection{Magnetic geometry}
\begin{figure}[htb]
\centering
\includegraphics[width=.4\linewidth]{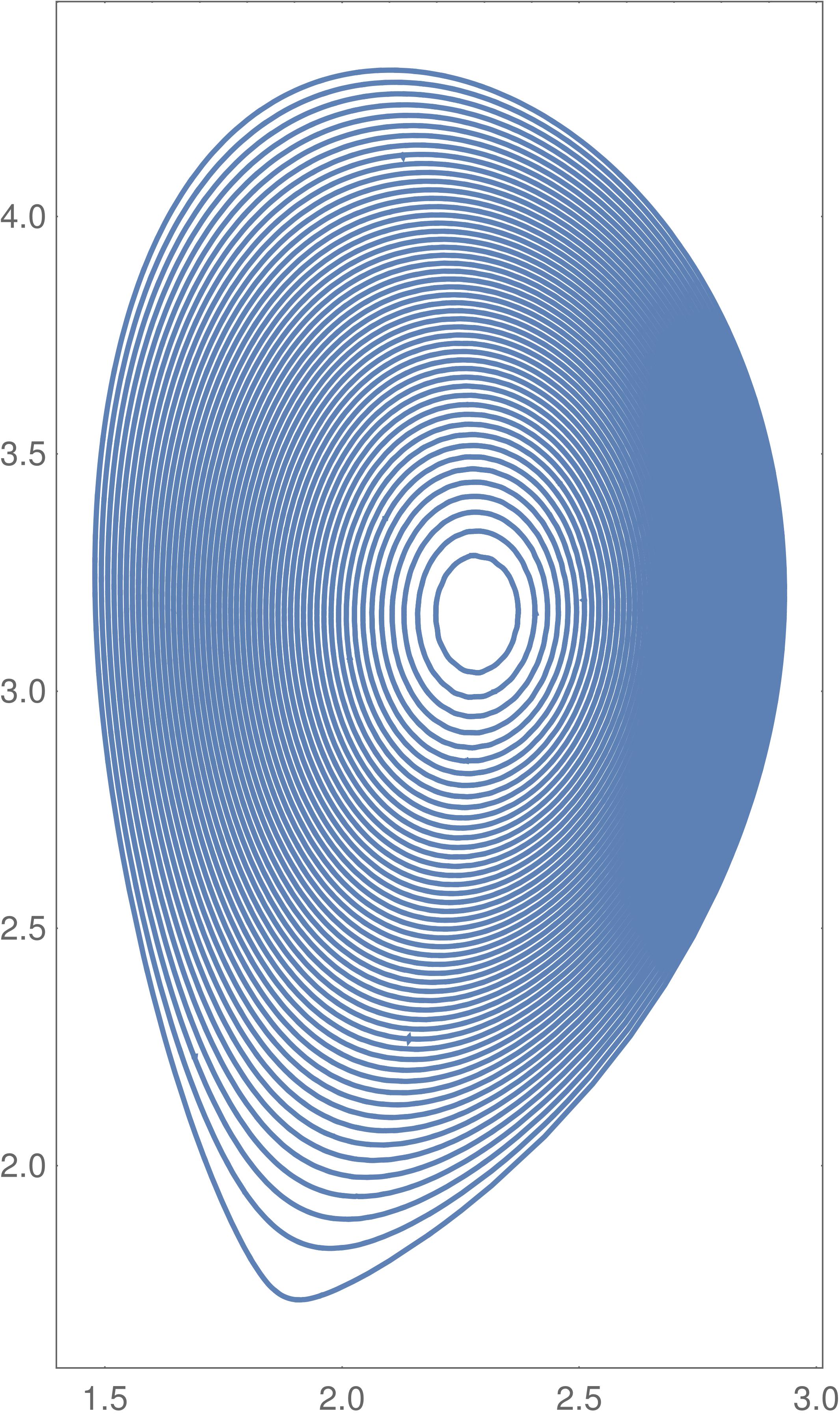}
\caption{Plot of the magnetic surfaces for the DTT reference scenario. The units on the axis are meters.}
\label{fig:nested}
\end{figure}

\begin{figure}[htb]
\centering
\includegraphics[width=.4\linewidth]{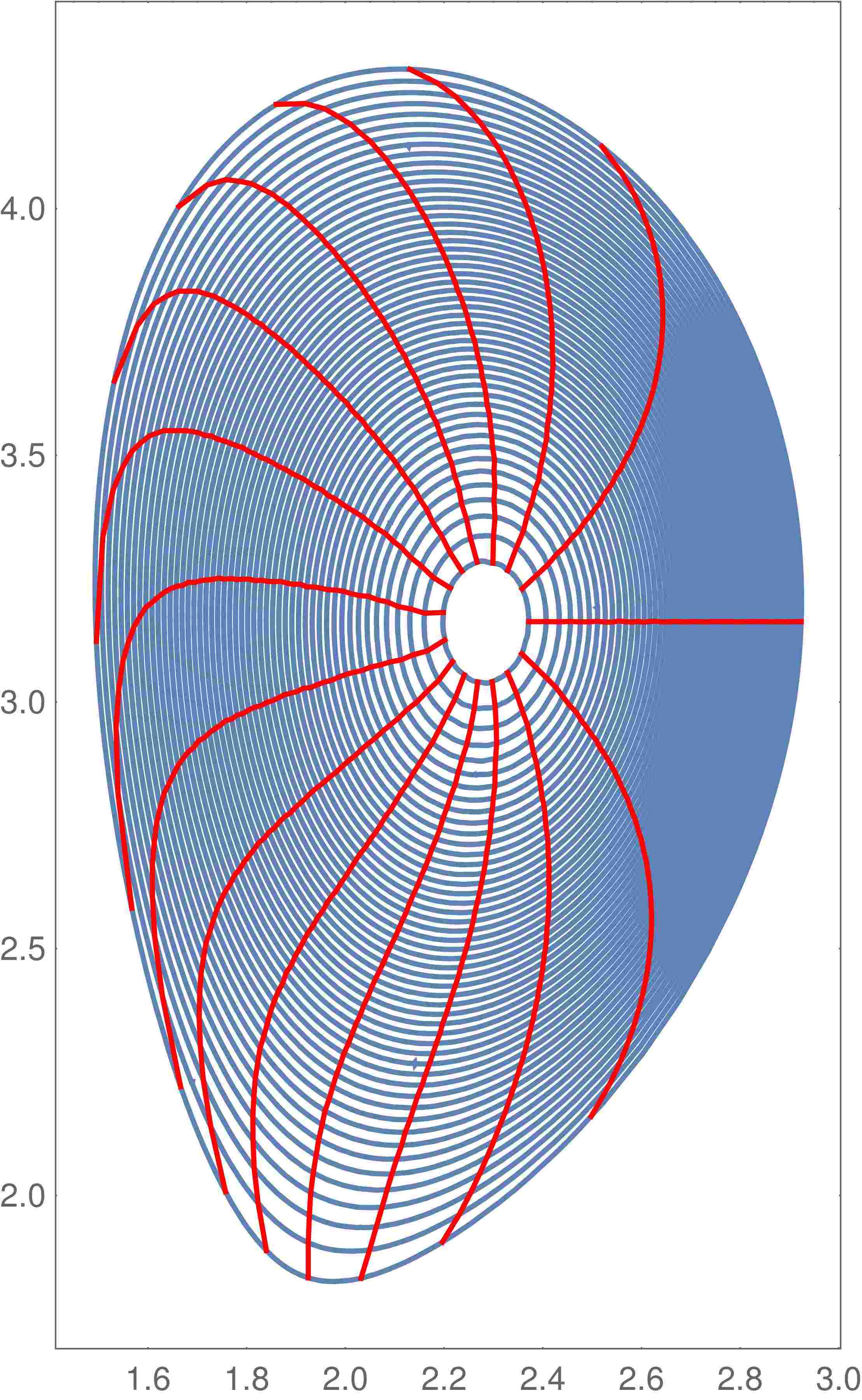}
\caption{Plot of the magnetic surfaces and of the the $\theta = const$ curves for the DTT reference scenario in Boozer coordinates.}
\label{fig:boozer}
\end{figure}

\begin{figure}[htb]
\centering
\includegraphics[width=.4\linewidth]{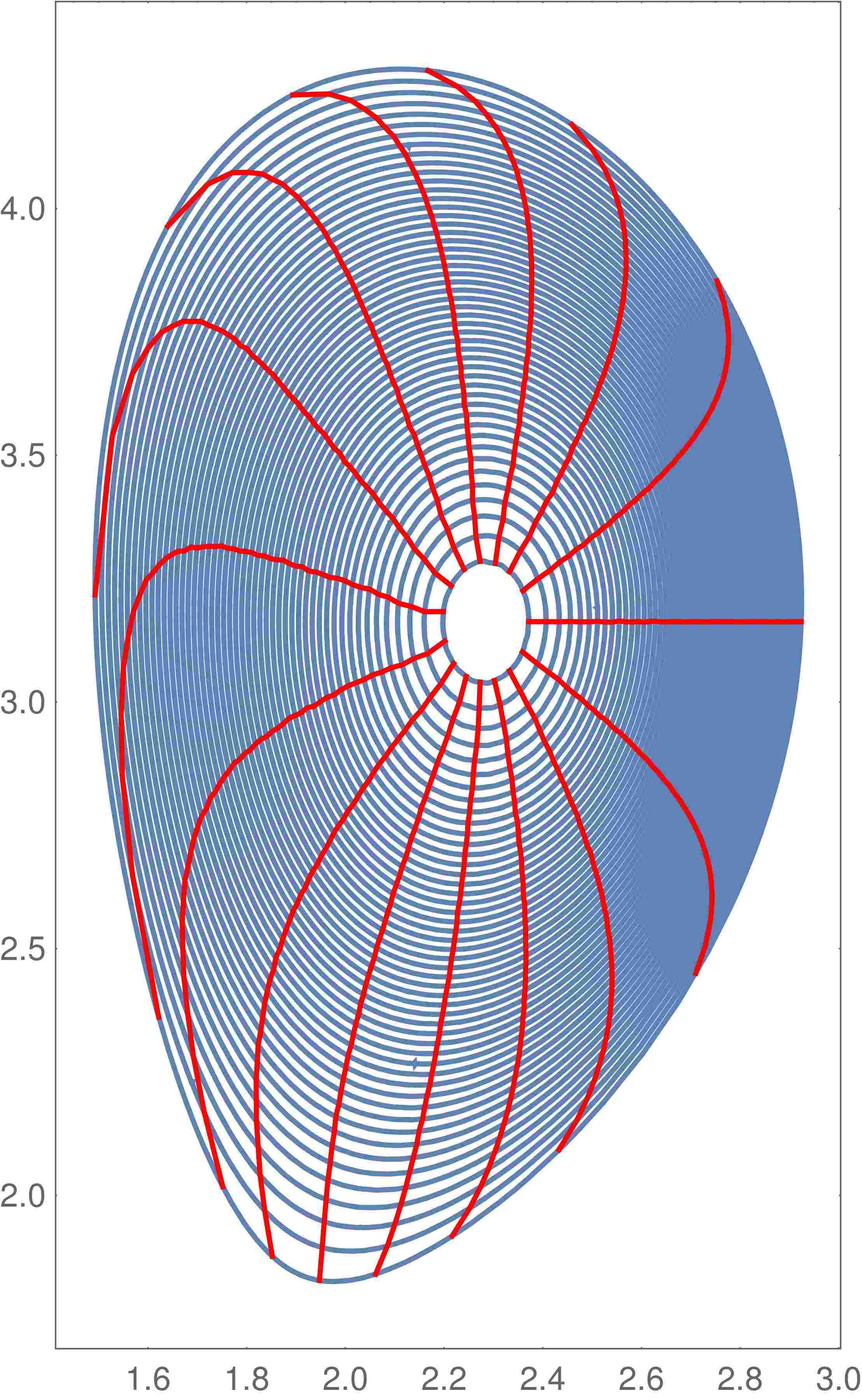}
\caption{Plot of the magnetic surfaces and of the the $\theta = const$ curves for the DTT reference scenario in Hamada coordinates.}
\label{fig:hamada}
\end{figure}

\begin{figure}[htb]
\centering
\includegraphics[width=.7\linewidth]{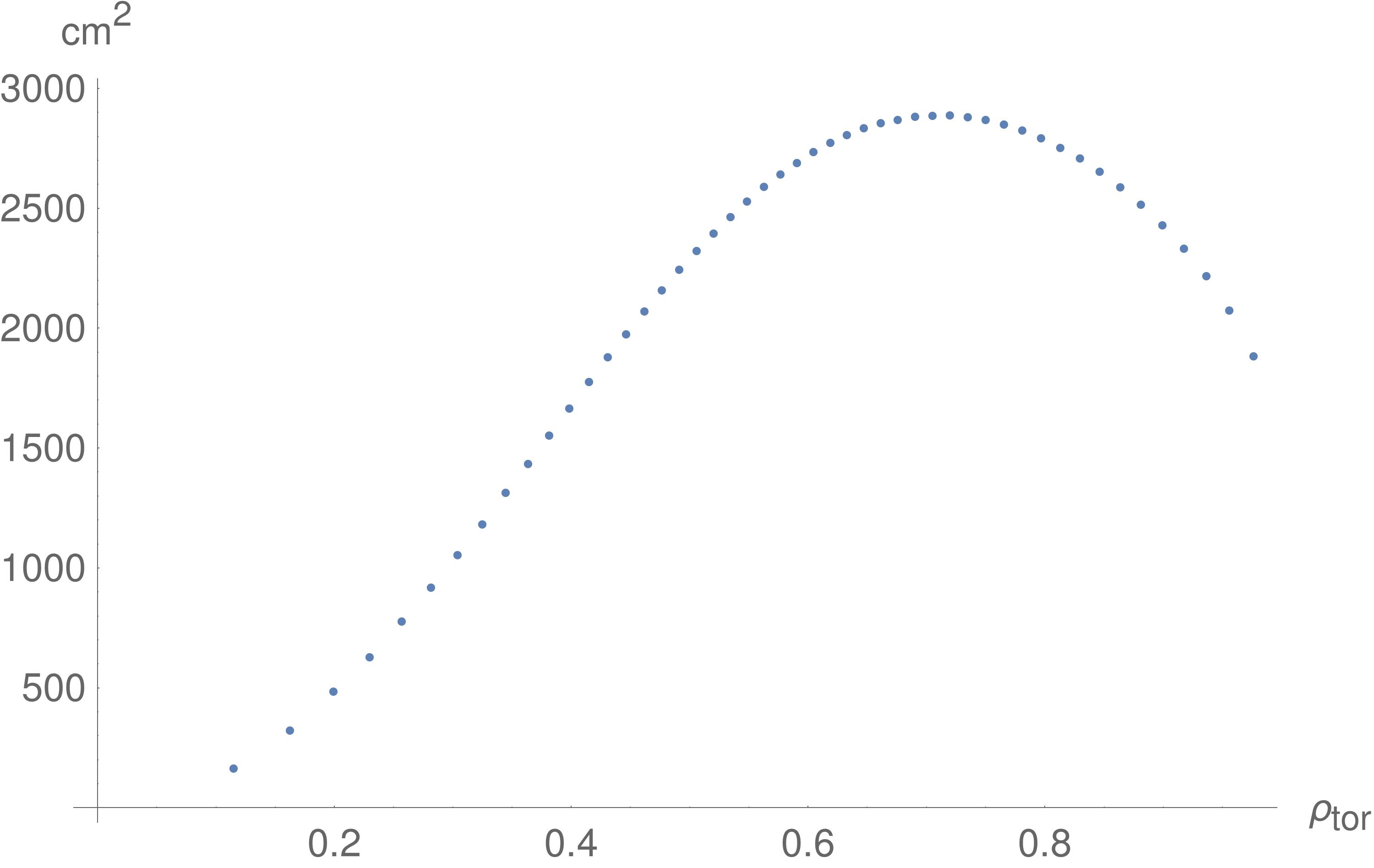}
\caption{Plot of the geometrical factor describing the classical particle flux in the DTT reference scenario, i.e. $\left \langle \left| {\bm \nabla}\psi \right|^2/B^{2}\right \rangle_\psi$.}
\label{fig:class}
\end{figure}

\begin{figure}[htb]
\centering
\includegraphics[width=.7\linewidth]{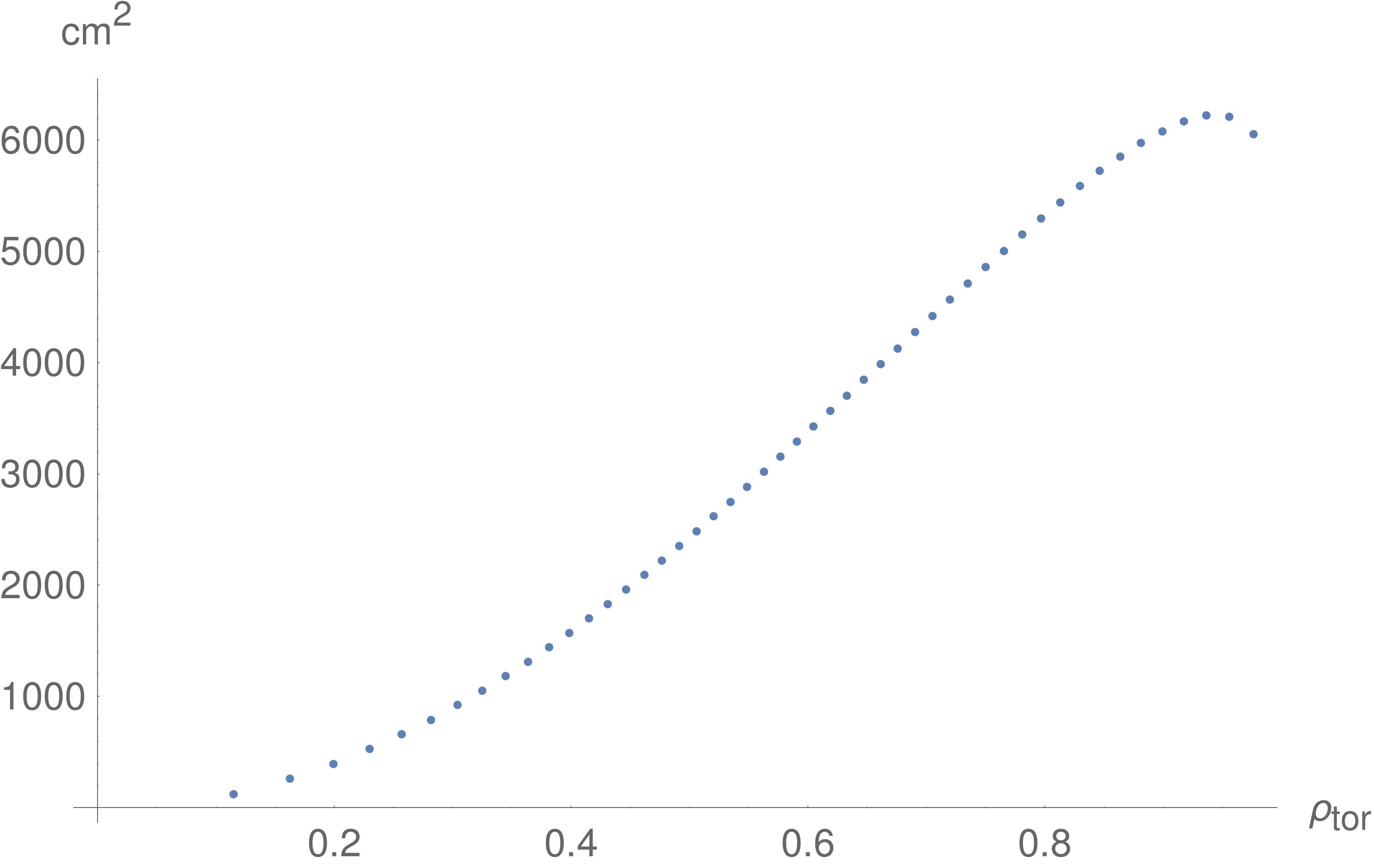}
\caption{Plot of the geometrical factor describing the Pfirsch Schl\"uter particle flux in the DTT reference scenario, i.e. $F^{2}\left(\left \langle 1/ B^{2} \right \rangle_{\psi}- 1/ \left \langle B^{2}\right \rangle_{\psi}\right)$.}
\label{fig:neoclass}
\end{figure}

The information about the equilibrium magnetic field for the DTT single null reference case (DTT reference scenario) are stored in a custom \texttt{.eqdsk} file. In Fig. \ref{fig:nested}, we show a plot with a section of the nested magnetic surfaces obtained from an elaboration of these data.
We now evaluate the expressions derived in the previous sections about Boozer coordinates. In particular, we estimate Eq. (\ref{eq:50}) with the data extracted from the \texttt{.eqdsk} file obtaining a plot of the Boozer coordinate grid, which is shown in Fig. \ref{fig:boozer}.
In the previous sections we have shown that the calculation required to obtain the diffusion coefficients are significantly simplified if we choose Hamada coordinates and, therefore, we will adopt this set of coordinates here. In Fig. (\ref{fig:hamada}), we show the grid associated with Hamada coordinates, which can be calculated by evaluating Eq. (\ref{eq:50}).
From the magnetic field data, we can also evaluate the geometrical factors, which enter into the calculation of the classical and the Pfirsch Schl\"uter particle flux. In the case of circular magnetic surfaces and large aspect ratio, these are analytically described by Eq. (\ref{eq:71}) and Eq. (\ref{eq:72}). In Fig. \ref{fig:class} and Fig. \ref{fig:neoclass},  we show the geometrical factor for the classical and Pfirsch Schl\"uter particle flux.
\subsection{Equilibrium parameters}
\begin{figure}[htb]
\centering
\includegraphics[width=.9\linewidth]{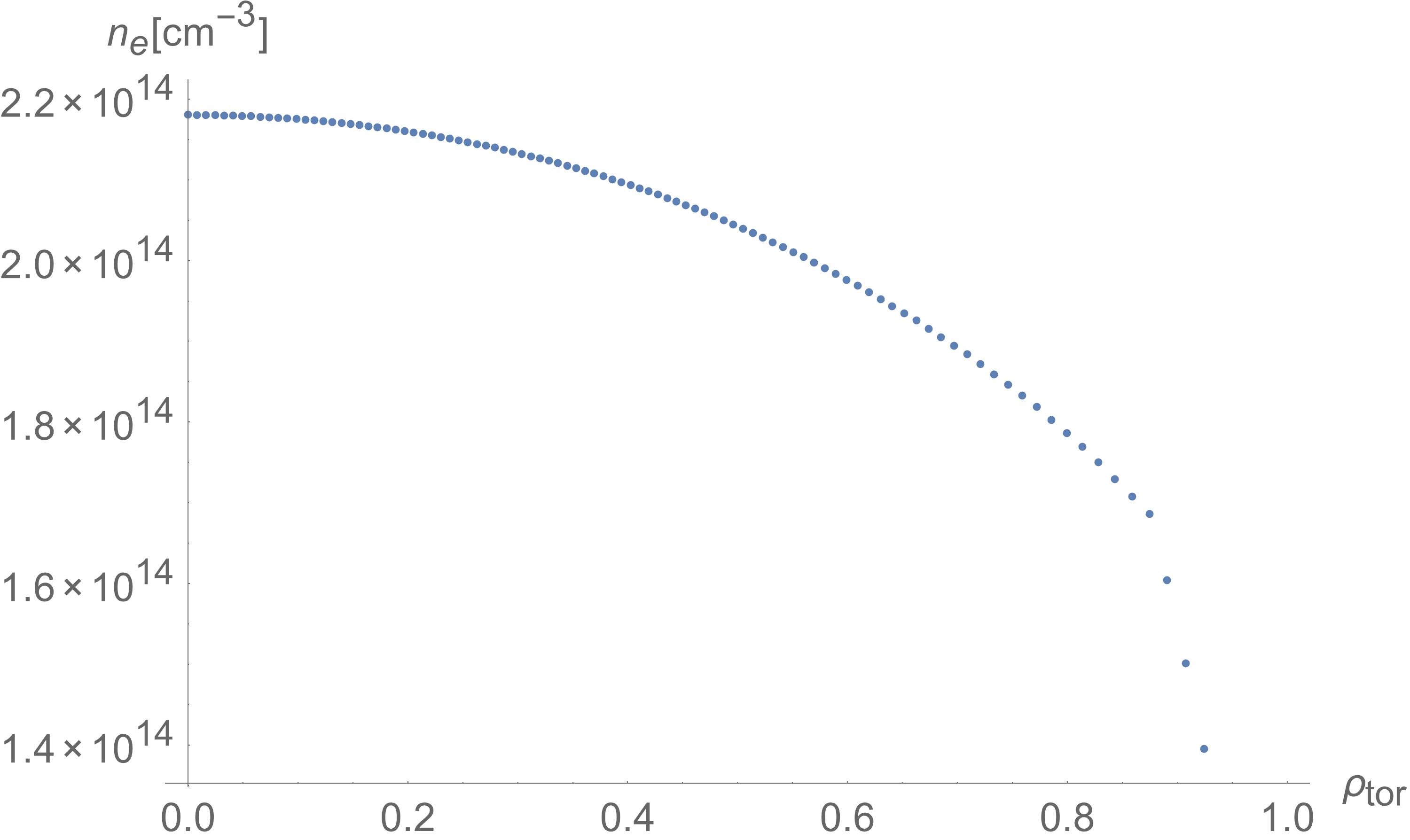}
\caption{Plot of the electronic density $n_{e}$.}
\label{fig:eldensity}
\end{figure}

\begin{figure}[htb]
\centering
\includegraphics[width=.9\linewidth]{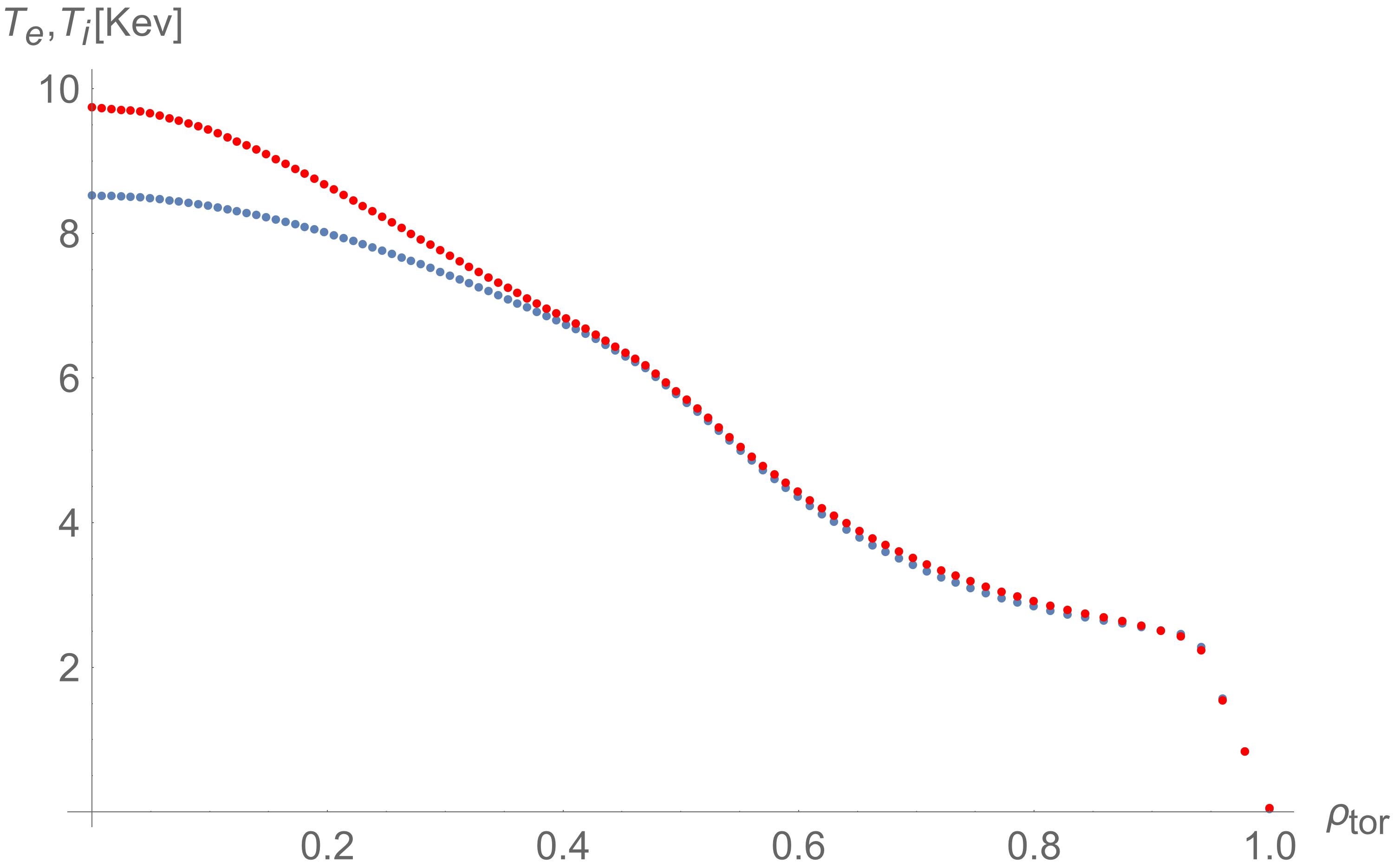}
\caption{Plot of $T_{i}$ and $T_{e}$ respectively in red and blue.}
\label{fig:eliotemperature}
\end{figure}

\begin{figure}[htb]
\centering
\includegraphics[width=.9\linewidth]{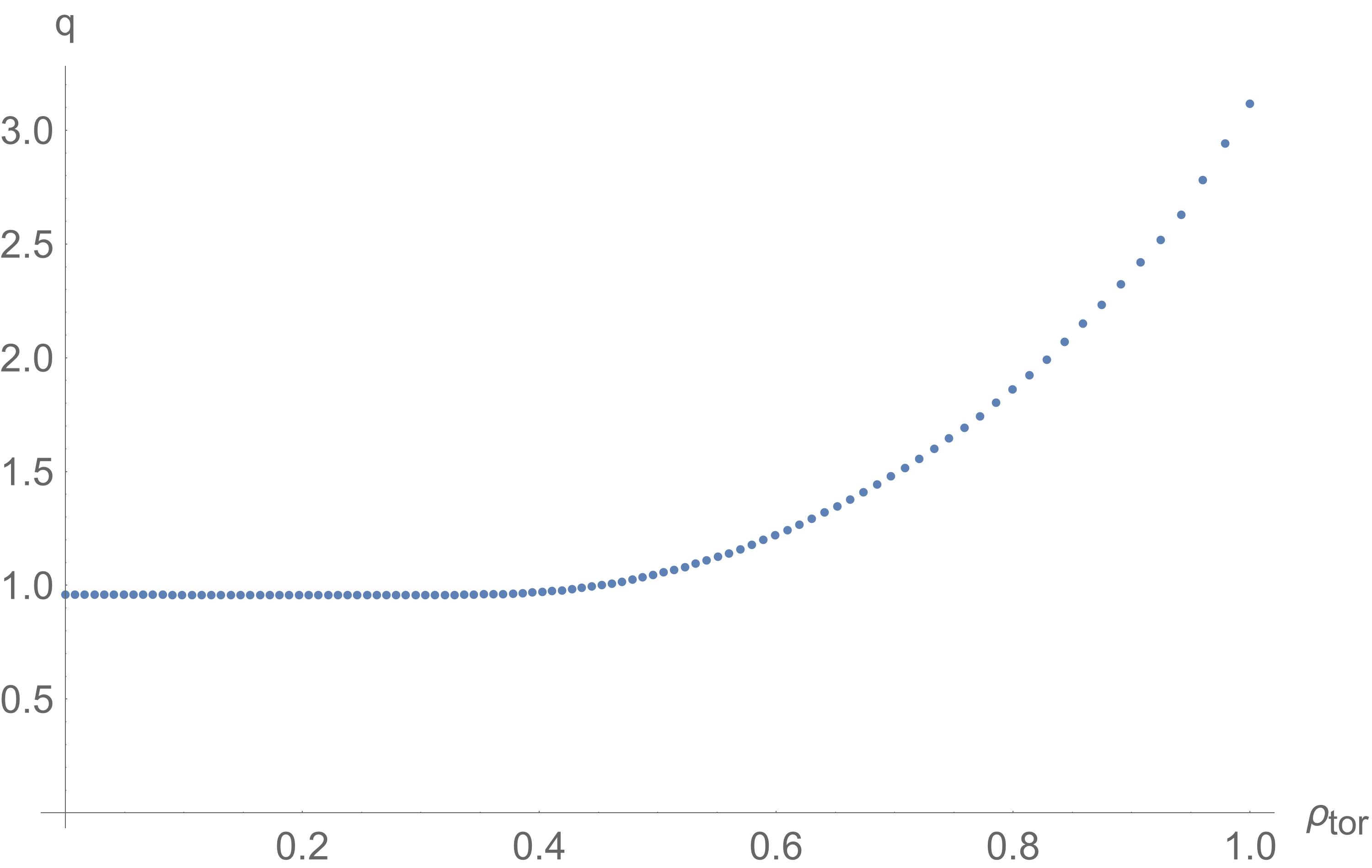}
\caption{Plot of the $q$ profile.}
\label{fig:qprofile}
\end{figure}

In the previous sections, we have shown that the expressions for the diffusion coefficients are the product of two factors: a geometrical one, which is related to the plots already shown, and a "physical" one with all the physical constants inside it. In the next pages we will show the plot of the factors that make up the latter, i.e. the plot of the electronic density in Fig. \ref{fig:eldensity}, the electronic and ionic temperature in Fig. \ref{fig:eliotemperature} and the $q$ profile for the DTT reference scenario in Fig. \ref{fig:qprofile}.
\clearpage
\subsection{Diffusion coefficients}
Taking the product of the factors discussed already we obtain the following plot of $D_{c}$ and $D_{PS}$:
\begin{figure}[htb]
\centering
\includegraphics[width=.8\linewidth]{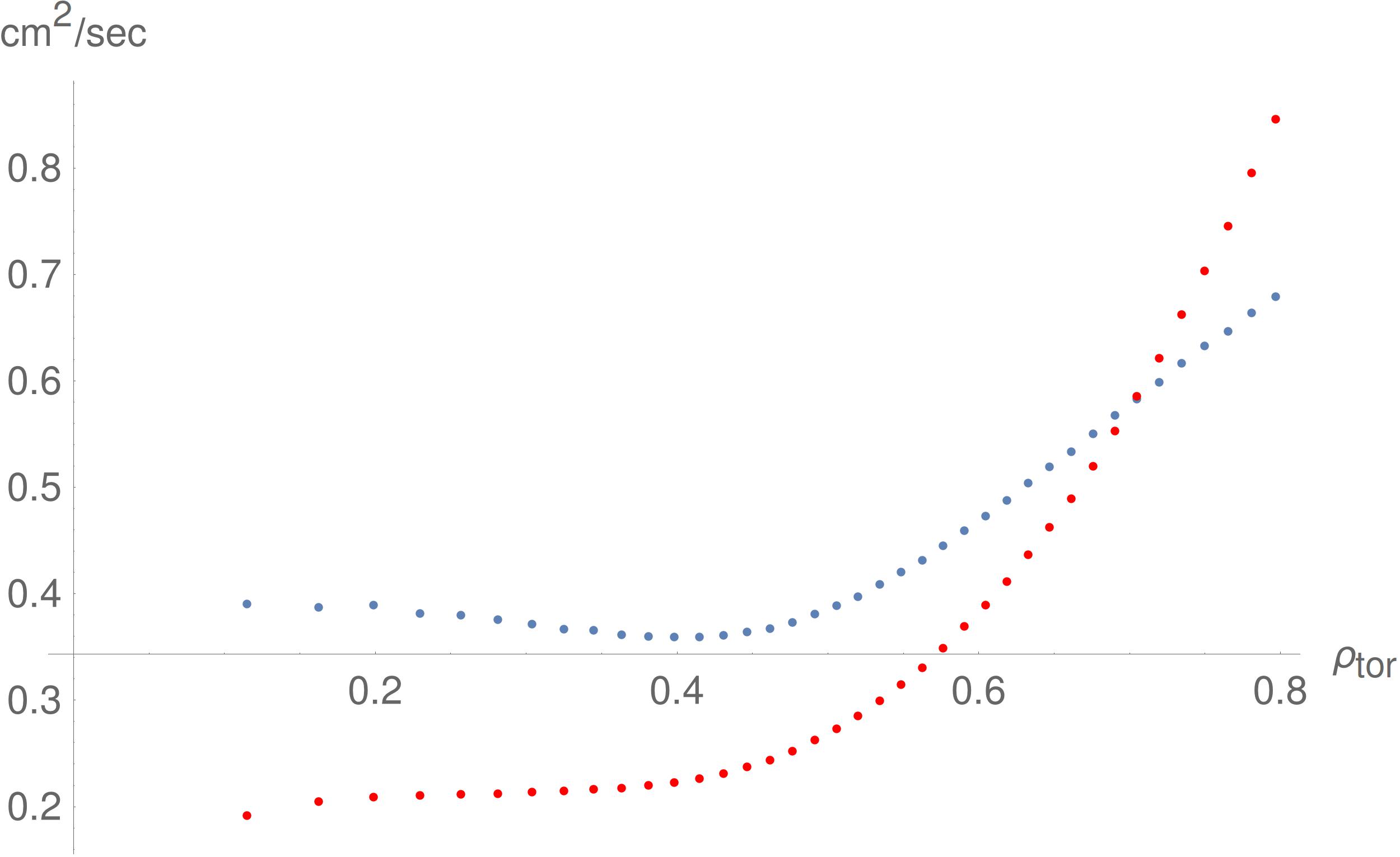}
\caption{Plot of the classical and Pfirsch Schl\"uter diffusion coefficients for the DTT reference scenario respectively in blue and red.}
\end{figure}

In the next figure we show $D_{ban}$ compared with the other diffusion coefficients:
\begin{figure}[htb]
\centering
\includegraphics[width=.8\linewidth]{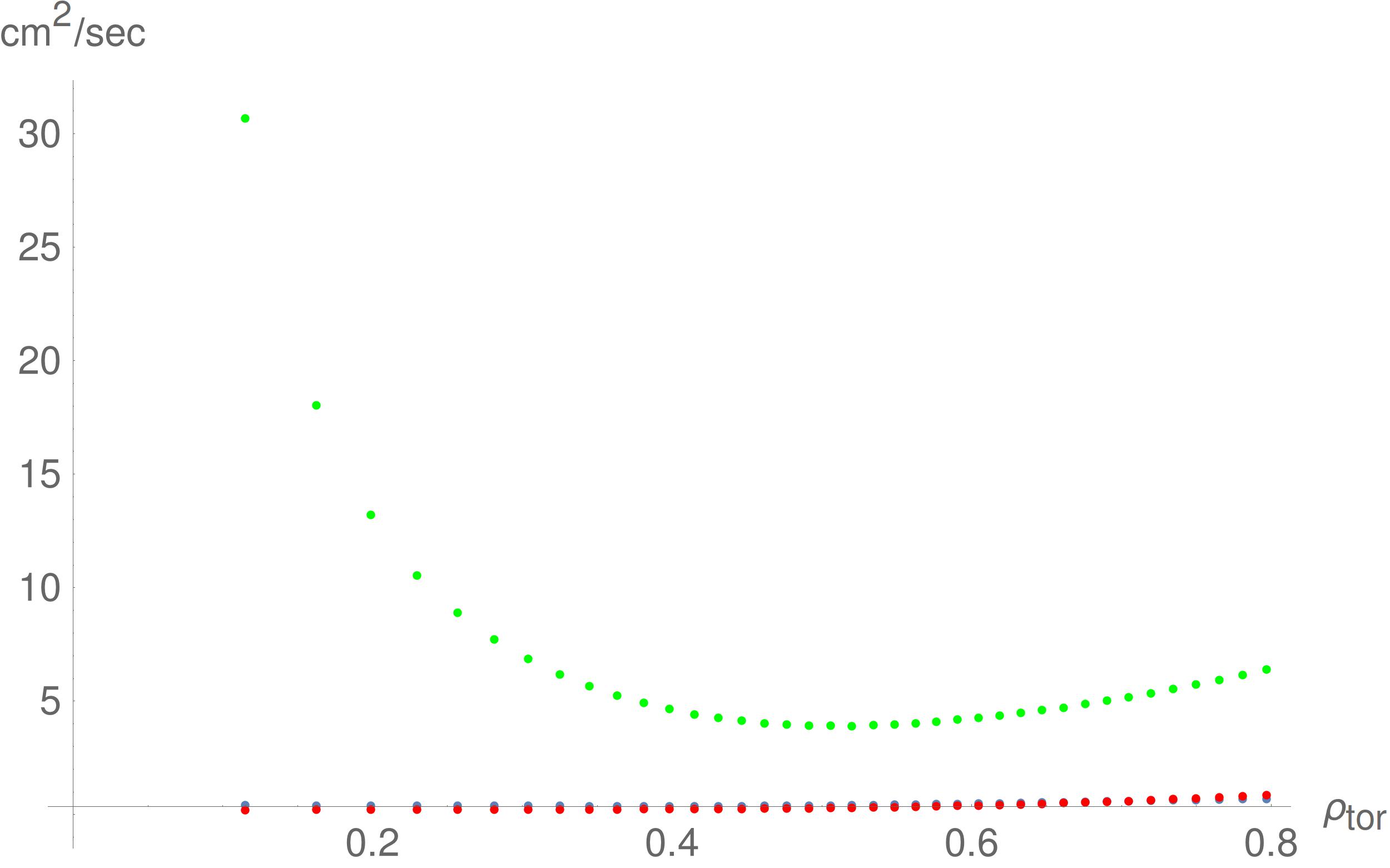}
\caption{Plot of the classical, Pfirsch Schl\"uter and banana diffusion coefficients for the DTT reference scenario respectively in blue, red and green.}
\end{figure}
\clearpage
From these plots, we can see that the classical diffusion coefficient $D_{c}$ near the magnetic axis is bigger with respect to $D_{PS}$. This result is not in agreement with the calculations for the circular magnetic surfaces and it is due to the elongation of the plasma. On the contrary, an analytic model with elliptical magnetic surfaces correctly describes this behavior. Normalizing the diffusion coefficients to $D_{e, GB}$ we obtain the following plots for $D_{c}$ and $D_{PS}$:
\begin{figure}[htb]
\centering
\includegraphics[width=.9\linewidth]{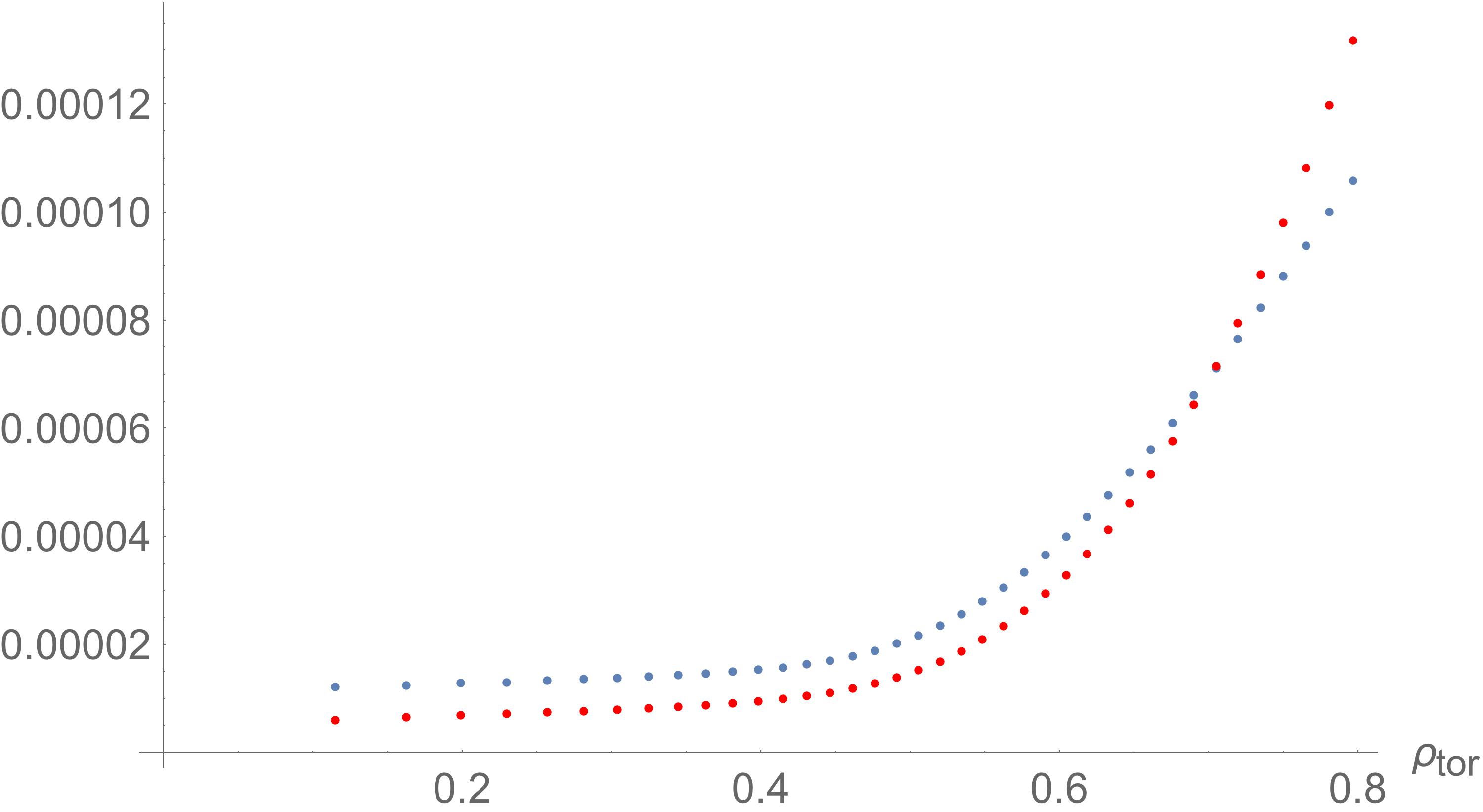}
\caption{Plot of the classical and Pfirsch Schl\"uter diffusion coefficients rescaled with respect to the electronic gyro-Bohm diffusion coefficient respectively in blue and red.}
\end{figure}

Plotting also $D_{ban}$ we obtain:
\begin{figure}[htb]
\centering
\includegraphics[width=.8\linewidth]{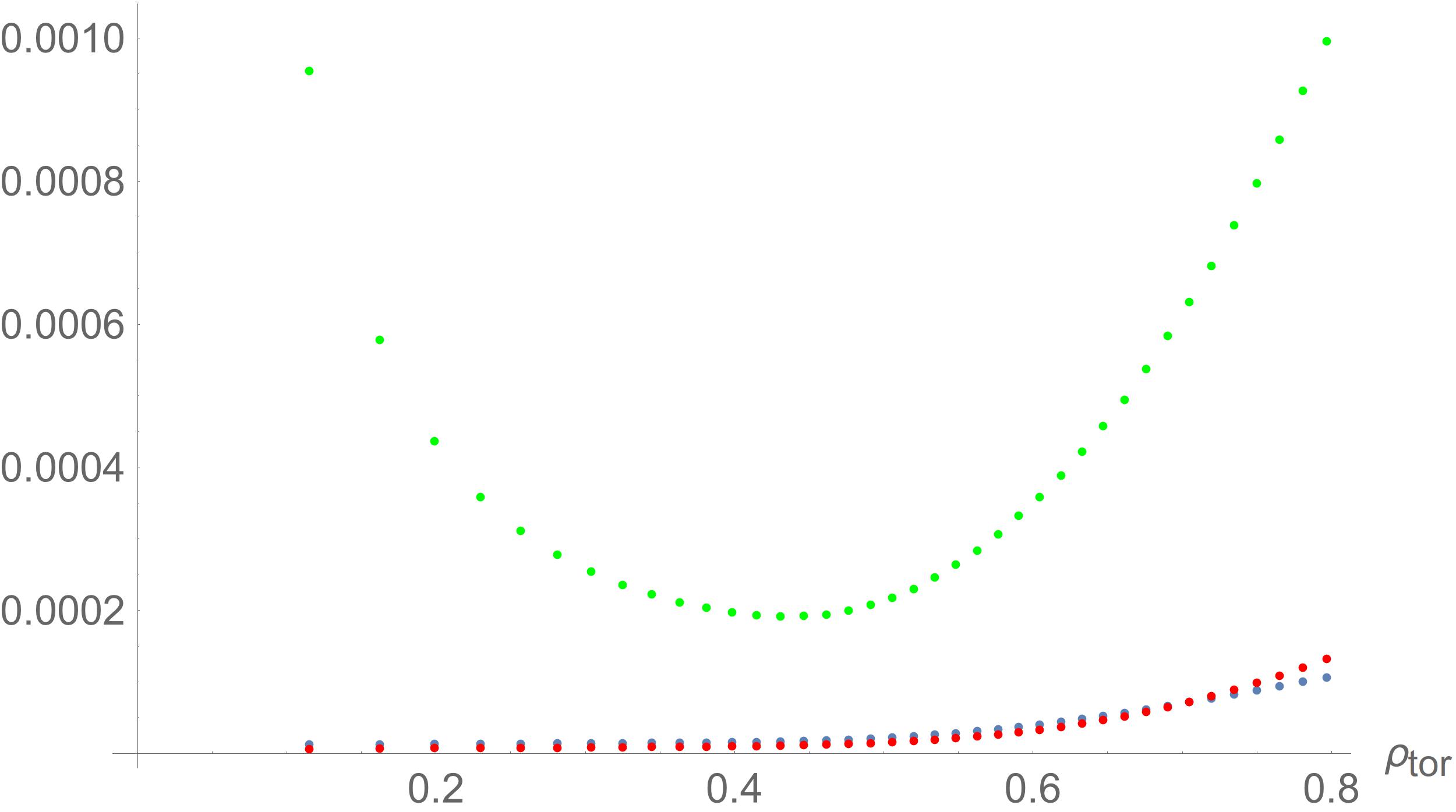}
\caption{Plot of the classical, Pfirsch Schl\"uter and banana diffusion coefficients rescaled with respect to the electronic gyro-Bohm diffusion coefficient respectively in blue red and green.}
\end{figure}

We can see that the collisional transport is dominated by the banana-transport. This is in agreement with the results for circular magnetic surfaces. In order to make this statement quantitative, we can calculate the ratio between $D_{ban}$ and $D_{PS}$ and compare it with the analytic result in the case of circular surfaces. In particular it can be shown that:
\begin{equation}
\label{eq:87}
\frac{D_{ban}}{D_{PS}}\approx \left( \frac{R_{0}}{r} \right)^{3/2}= \left(  \frac{R_0}{a}\right)^{3/2} \rho_{tor}^{-3/2}.
\end{equation}
In the next plot, we show that the ratio $D_{ban}/D_{PS}$ obtained for the DTT reference scenario is well fitted by the function $C \rho_{tor}^{-3/2}$ near the magnetic axis with a value for the constant $C$ in agreement with Eq. (\ref{eq:87}):
\begin{figure}[htb]
\centering
\includegraphics[width=.9\linewidth]{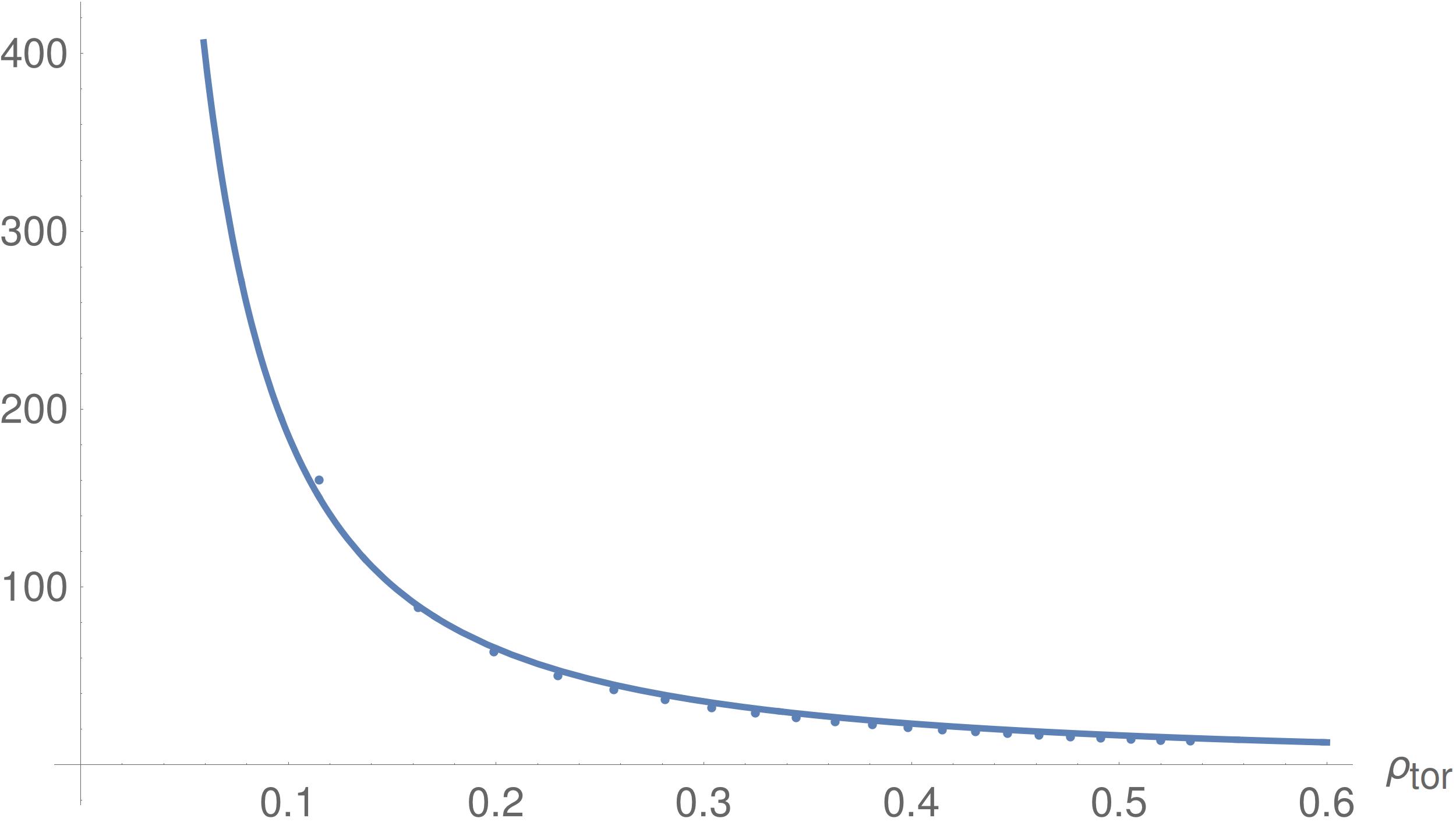}
\caption{Plot of the ratio between $D_{ban}$ and $D_{PS}$. The line is obtained by fitting the data with the function $C \, \rho_{tor}^{-3/2}$.}
\end{figure}
\chapter{Longer timescales}
\label{cap:longer}
In this thesis, using gyrokinetic theory, we have calculated neoclassical and anomalous particle transport in an axisymmetric tokamak plasma. A number of other works dealing with the same problem are: \cite{sugama1996transport,balescu1990anomalous,shaing1988neoclassical} and the more recent \cite{abel2013multiscale}. In all these works, and also in this thesis, collisional and turbulent fluxes are calculated up to $\mathcal{O}(\delta^{2})$ in the asymptotic expansion. Using the characteristic length and time-scales of a modern magnetic fusion device, we can estimate the corresponding time-scale of validity of the transport equations, for the fluxes which is of the order of the seconds. This is relatively short when compared with the expected duration of a pulse in the next generation Tokamaks, i.e. ITER, which is $>3000 s$ \cite{janeschitz2001plasma}. Therefore, in order to have predictive simulations, we need to describe collisional and fluctuation induced fluxes up to $\mathcal{O}(\delta^{3})$ and the  distribution function with an accuracy of $\mathcal{O}(\delta^{2})$. Furthermore, the characteristic length and time-scales considered so far typically apply to the core region of thermonuclear plasmas. Generally addressing the problem of the 
plasma transport as the edge plasma region is approached, where equilibrium magnetic field is modified from
closed to open field lines, poses even more severe issues. In fact, the relative ordering of spatiotemporal scale of turbulent fluctuation spectra and  transport phenomena is also modified and not so well separated as in the plasma core.  In particular the radial gradient scale length can be of the same order of the banana width of thermal ions \cite{chang2004numerical} in the so-called pedestal region, where plasma profile are characterized by sharp variations. Therefore the conventional neoclassical transport theory cannot be applied. For these reasons the study of higher order terms of the asymptotic expansion may be of crucial importance.

In this chapter, we will show the difficulties encountered in the treatment of the higher order collisional and turbulent fluxes. We remark that, even if a set of equations for the fluxes with a precision of $\mathcal{O}(\delta^{2})$ is not enough in order to predict the behavior of the plasma during a whole ITER plasma discharge, it could be used in order to build actuators \cite{gravesinternal,graves2012control} based on reduced models for the real time control of plasma dynamic evolution.

\section{Fluctuations induced transport}
\label{sec-1-2}
The aim of this thesis but, more generally, of gyrokinetic transport theory, is to study in a self-consistent way collisional and turbulent transport. Therefore, studying formal expressions of particle, momentum and heat fluxes that are valid on the time scale of an ITER discharge requires the parallel development of a gyrokinetic theory correct, at least, up to $\mathcal{O}(\delta^{2})$ and of a corresponding more accurate form of collisional fluxes. As we have shown in the previous chapters, gyrokinetics is based on an asymptotic theory, where the expansion parameter is defined as the ratio between the gyroradius and the characteristic length scale of variation of the equilibrium magnetic field. This is achieved in two steps: first, the fluctuating electromagnetic fields are ignored and only the background (equilibrium non-uniform) magnetic field  is considered; then, the turbulent fields are introduced and the corresponding plasma responses are calculated. Each step is based on an asymptotic expansion done with different perturbation parameters, which are respectively denoted \cite{brizard2007foundations} by $\epsilon_{B}$ and $\epsilon_{\delta}$. The gyrokinetic ordering typically assumes $\epsilon_{B}\sim\epsilon_{\delta}\sim \delta$. The asymptotic expansion in $\epsilon_{\delta}$ needs to be carried out at least at  up to second order to obtain an energy like invariant. Terms of order $\epsilon_{B}^{2}$ are usually neglected in practical applications because of their complexity  and, therefore, the gyrokinetic ordering is not carried out
on an equal footing with respect to fluctuation intensity and equilibrium magnetic field non-uniformity \cite{brizard2007foundations}. Generally, this is justified as $\epsilon_{B}$ is typically smaller than $\epsilon_{\delta}$ in cases of practical interest. Nonetheless, this issue is known in the fusion research community and efforts are being carried out to derive more accurate formulations of gyro-kinetics which may be applied on longer time scales, i.e. \cite{tronko2016second}, or in plasma conditions where expansion parameters underlying the asymptotic theory may be not as small as in typical burning plasma core region. This is, e.g., the situation of fusion plasmas in the edge region, as anticipated above, where the presence of material walls surrounding the core plasma volume and of sharp spatial gradients may challenge the standard approach to gyrokinetic theory \cite{xu2007edge}. In general, the perturbative expansions have been consistently carried out up to the second order in $\delta$ only in the electrostatic case, i.e. where the turbulent fluctuation spectrum does not significantly affect the magnetic field\cite{parra2011phase}. The more general case of a fully electromagnetic fluctuation spectrum in non-uniform toroidal plasmas has not been addressed to date. Therefore, the analogous form of the pullback of the distribution function, i.e. Eq.\ref{eq:102}, up to the second order in $\delta$ has not been given. We stress the fact that, in principle, the non-canonical perturbation theory \cite{cary1983noncanonical}, which is described in \autoref{cap:appA} of this thesis, allows to formally derive the desired  pullback operator at any order of the asymptotic expansion \cite{brizard2007foundations}. However the calculation becomes very convoluted already for the second order electrostatic case.

\section{Collisional transport}
\label{sec-1-1}
It is well known \cite{hinton1976theory,balescu1988transport} that neoclassical transport theory as well as classical transport theory deal with a linear collision operator which approximates the Landau collision integral. These theories show that the approximated collision operator is consistent with a positive production of entropy and the Onsager symmetry \cite{onsager1931reciprocal} in the linear relations connecting the thermodynamic forces and the fluxes. These are linear closure relations and, therefore, they have a clear interpretation in terms of non-equilibrium thermodynamics. In the study of higher orders of the asymptotic expansion, discussed earlier in this chapter we need to deal, in general, with nonlinear closure relations. In transport theory the nonlinear closure relations and the (nonlinear) Landau collision operator have been studied with different approaches. In \cite{chang2004numerical}, this problem has been addressed by means of numerical simulations, while an analytic approach has been carried out by G.Sonnino in a series of works: \cite{sonnino2007geometrical,sonnino2007minimum,sonnino2008nonlinear,sonnino2009nonlinear,sonnino2014thermodynamic} and the more recent \cite{PhysRevE.94.042103}. This author introduces and describes the Thermodynamic Field Theory as a useful tool to derive corrections to the linear closure relations with applications to plasma physics.

In the previous chapters, we have shown how nonlinear closure relations spontaneously arise when considering the joint effect of collisions and fluctuations even at $\mathcal{O}(\delta^{2})$. Therefore the development of a theory capable of describing these effects coherently with a thermodynamic description of the plasma is of crucial importance. For this reason we have collaborated with G.Sonnino in the development of the TFT as part of the research activity carried out in this thesis work. In this context, we have described the TFT in terms of group theory: \cite{PhysRevE.94.042103,chileproceeding}.

In the next section, following \cite{sonnino2009nonlinear}, we will introduce the Thermodynamic Field Theory and we will calculate the Noether currents associated with the symmetry of the TFT action under linear transformations of the thermodynamic forces, which form a subgroup of the thermodynamic coordinate transformation (TCT).

\subsection{Thermodynamic field theory}
\label{sec-1-1-1}
In the previous chapters, we have recalled that, starting from the kinetic equation, it is possible to write down a set of equations for the moments of the single particle distribution function $f$. The system of equations obtained is infinite and, in order to know the time evolution of a certain moment, the knowledge of a moment of the next order is required. Therefore, some simplifying assumption needs to be introduced in order to truncate this hierarchy at a certain point. The mathematical theory dealing with this problem is called transport theory \cite{balescu1988transport}. The resulting equations will have a number of undetermined quantities, which need to be computed through the closure relations. A class of closure relations is constituted by the transport equations relating the thermodynamic forces with the dissipative fluxes of the system. Close to the equilibrium, the transport equations of a thermodynamic system are provided by the Onsager theory. The Onsager relations read:
\begin{equation}
\label{eq:TFT1}
J_{\mu}= \tau_{0 \mu \nu}X^{\nu}
\; ,
\end{equation}
with $\tau_{0 \mu \nu }$ being the components of the transport coefficient matrix. The matrix of the transport coefficients can be decomposed in a symmetric $L_{\mu \nu}$ and an skew symmetric part $f_{0 \mu \nu}$. The second principle of thermodynamics imposes that $L_{ \mu \nu}$ is positive definite. Near the thermodynamic equilibrium of the system, in the so called Onsager region, $\tau_{\mu \nu}$ are independent of $X$ and thus:

\begin{equation}
\label{eq:TFT2}
\frac{\partial \tau_{ 0 \mu \nu}}{\partial X^{\lambda}}=0.
\end{equation}
In \cite{sonnino2007geometrical} and in \cite{sonnino2007minimum}, a mathematical framework to generalize the Onsager relations, i.e. Eq. (\ref{eq:TFT1}), has been introduced which, near the thermodynamic equilibrium of the system, recovers the relations mentioned already. This theory is called Thermodynamic Field Theory. It is purely macroscopic and postulates the second principle of thermodynamics and the transport equations; i.e., it does not deal with their derivation from the microscopic dynamics. The evolution of the system takes place in the thermodynamic space, which is covered by the $n$ independent thermodynamic forces $X^{\mu}$. The evolution equations are not derived from the microscopic dynamics but are obtained by postulating three geometrical principles:

\begin{itemize}
\item The shortest path principle;
\item the Thermodynamic Covariance Principle (TCP);
\item the principle of least action.
\end{itemize}
The TCP states that thermodynamic systems obtained by a transformations of thermodynamic forces and fluxes in such a way that the entropy production and the Glansdorff-Prigogine dissipative quantity \cite{glansdorff1973thermodynamic}:
\begin{equation}
\label{eq:TFT6}
P \equiv J_{\mu}\frac{d X^{\mu}}{dt}
\; ;
\end{equation}
remains unaltered, are thermodynamically equivalent. The transformations connecting equivalent systems, are called Thermodynamic Coordinate Transformations (TCT). By definition of equivalent systems these transformations must leave the equations of motion on the Thermodynamic space unaltered, leading to their covariance.

The analysis starts from the following observation: let's consider a system which is relaxing to the steady state inside the Onsager region. The universal criterion of evolution, which must be satisfied during the dynamics, reads $P\leq 0$ and the Onsager relations, i.e. Eq. (\ref{eq:TFT1}), hold. We can write:

\begin{equation}
\label{eq:TFT8}
P \equiv J_{\mu}\dot{X}^{\mu} \equiv J_{\mu} \frac{d}{d\zeta}X^{\mu}= \tau_{0 \mu \nu}X^{\nu}\dot{X}^{\mu}= (L_{\mu \nu} + f_{0\mu \nu})X^{\nu}\dot{X}^{\mu}
\; , 
\end{equation}
where with the dot we have indicated the operation of derivation with respect to the parameter $\zeta$ which is the curvilinear coordinate parametrizing the motion of the system in the thermodynamic space that satisfies the following metric relation:
\begin{equation}
\label{eq:TFT9}
d \zeta^{2}= L_{\mu \nu} dX^{\mu}dX^{\nu}.
\end{equation}
Therefore, in the Onsager region, the metric is assumed to be Euclidean. Thus the equation for the shortest path, which we postulate to describe the evolution of the system in the thermodynamic space, is the following:
\begin{equation}
\label{eq:TFT10}
\ddot
{X}^{\mu}=0
\; ,
\end{equation}
which can be solved in order to obtain the trajectory of the system in the thermodynamic space:
\begin{equation}
\label{eq:TFT11}
X^{\mu}= a^{\mu}\zeta + b^{\mu}.
\end{equation}

This can be substituted into (\ref{eq:TFT8}) obtaining:

\begin{equation}
\label{eq:TFT12}
J_{\mu}\dot{X}^{\mu}= \tau_{0 \mu \nu}X^{\nu}\dot{X}^{\mu} = (L_{\mu \nu} + f_{0 \mu\nu})(a^{\mu}a^{\nu}\zeta + a^{\mu}b^{\nu})= \zeta + \tau_{0 \mu \nu}a^{\mu} b^{\nu}
\; ,
\end{equation}
where we have used the antisymmetry of $f_{0\nu}$ and the relation $L_{\mu\nu}a^{\mu}a^{\nu}=1$. If we now choose $\zeta$ in such a way that when the system reach the stationary state, $\zeta = 0$ and at the beginning of the evolution $\zeta = l$, we have that:

\begin{equation}
\label{eq:TFT13}
J_{\mu}\dot{X}^{\mu}  |_{steady}=0 \ \Rightarrow \ l= -\tau_{0 \mu \nu}a^{\mu}b^{\nu}
\; ,
\end{equation}

and therefore:

\begin{equation}
\label{eq:TFT14}
P = -(l - \zeta) \leq 0.
\end{equation}

In summary, having postulated that:

\begin{itemize}
\item the system evolves over the shortest path;
\item the metric of the thermodynamic space is flat inside the Onsager region
\item the validity of the Onsager relations
\end{itemize}
we obtain that the Universal criterion of evolution is satisfied. Also $\dot{\sigma}= 2P \leq 0$.

In the following, we will extend this result outside the Onsager region using an appropriate metric to describe the thermodynamic space and, thus the evolution of the system along the shortest path. The metric is required to be Euclidean, i.e. flat, inside the Onsager region, while it can be curved outside.

In particular, a curved space is introduced, whose geometry is constructed in such a way that:
\begin{itemize}
\item the theorems which are valid when a generic thermodynamic system relaxes to equilibrium are satisfied;
\item the nonlinear closure relations, i.e. the generalization of the Onsager relations, are covariant under TCT.
\end{itemize}

We assume that a system, which is driven away from the equilibrium by a set of $n$ independent thermodynamics forces $X^{\mu}$ with $\mu = \{1,2...,n \}$ satisfies the following relations between the thermodynamic forces and the thermodynamic fluxes $J^{\mu}$:

\begin{equation}
\label{eq:TFT15}
J_{\mu}= \tau_{\mu \nu}(X)X^{\nu
}
\; ,
\end{equation}

which are analogous to the Onsager relations except for the dependence of the transport coefficient matrix on the thermodynamic forces. We can decompose $\tau_{\mu \nu}$ (from now on we remove from the notation the dependence from the thermodynamic forces, which is assumed) in its symmetric and skew symmetric part:

\begin{equation}
\label{eq:TFT16}
\tau_{\mu \nu}= \frac{1}{2}\left(\tau_{\mu \nu}+\tau_{\nu \mu}\right )+\frac{1}{2}\left(\tau_{\mu \nu}-\tau
_{\nu \mu}\right) \equiv g_{\mu \nu}+ f_{\mu \nu}
\; , 
\end{equation}
with the following symmetry properties:
\begin{equation}
\label{eq:TFT17}
g_{\mu \nu}=g_{\nu \mu}, \quad f_{\mu \nu}= - f_{\nu \mu}.
\end{equation}

We define two classes of objects starting from the elements of the transport matrix $\tau_{\mu \nu}$:

\begin{itemize}
\item operators such as $\sigma = \sigma(X)$ and $\tilde{P}=\tilde{P}(X)$ ;
\item tensorial objects such as $f$ and $g$ which transform according to the prescription specified below.
\end{itemize}
The entropy production operator $\sigma (X)$ and the dissipative quantity operator $\tilde{P}$ are scalar operators acting on the thermodynamic forces as follows:
\begin{eqnarray}
  \sigma(X) & \equiv & XgX^{T} \label{eq:TFT18a}
  \; ,\\
\tilde{P}(X) & \equiv & X \tau \left[\frac{d X}{d \rho}\right ]^{T}
\; ,
\end{eqnarray}
where the superscript indicates the transpose operation. Eq. (\ref{eq:TFT18a}) ensures the validity of the second principle of thermodynamics, i.e. that the entropy production $\sigma$ is positive, if the matrix $g$ is positive definite. Thermodynamics states such that:
\begin{equation}
\label{eq:TFT19}
\left [\tilde{P}\frac{d \varrho}{dt}\right]_{X=X_{steady}}=0
\; ,
\end{equation}
are referred to as steady states \cite{sonnino2009nonlinear}. They must be invariant under thermodynamic coordinate transformations.

According to the De Donder-Prigogine statement, we will consider equivalent two systems that can be mapped one into the other, i.e. $X^{\mu} \rightarrow X ^{'\mu}$ and $J_{\mu}\rightarrow J^{'}_{\mu}$ through thermodynamic transformation such that $\sigma$ and $\tilde{P}$ are the same for the two systems. This is verified if $X^{'\mu}$ and $J^{'}_{\mu}$ are obtained with a transformation such that:

\begin{eqnarray}
\label{eq:TFT21}
  X^{'\mu} & = & \frac{\partial X^{' \mu}}{\partial X^{\nu}} X^{\nu}
  \; ,\\
J^{'}_{\mu} & = & \frac{\partial X^{\nu}}{\partial X^{'\mu}} J_{\nu} .
\end{eqnarray}

Assuming that Eq. (\ref{eq:TFT15}) holds, these transformations imply that:

\begin{equation}
\label{eq:TFT22}
\sigma= J_{\mu}X^{\mu}= \tau_{\mu \nu} X^{\mu}X^{\nu}= g_{\mu \nu} X^{\mu}X^{\nu}= g^{'}_{\mu \nu}X^{' \mu}X^{'\nu}= \sigma^{'}
\; ,
\end{equation}

and, thus, that $g$ transform as a second rank tensor. The same statement can be derived for the skew symmetric part of $\tau$, i.e. $f$. The covariant and contravariant vectors under TCT are:

\begin{eqnarray}
\label{eq:TFT23}
  dX^{' \mu} &  = & \frac{\partial X^{' \mu}}{\partial X^{\nu}} d X^{\nu}
  \; ,\\
\frac{\partial}{\partial X^{'\mu}} &  = & \frac{\partial X^{\nu}}{\partial X^{'\mu}}\frac{\partial}{\partial X^{\nu}}.
\end{eqnarray}

Every quantity that is obtained by taking a contraction of a covariant and a contravariant vector (tensor in general) is a scalar under TCT. For example, the parameter $\zeta$ defined as:

\begin{equation}
\label{eq:TFT24}
d \zeta^{2}= g_{\mu \nu} d X^{\mu}d X^{\nu}
\end{equation}

is a scalar under TCT. The operator $\mathcal{O}$, i.e. Lie derivative, defined as

\begin{equation}
\label{eq:TFT25}
\mathcal{O}\equiv X^{\mu}\frac{\partial}{\partial X^{\mu}}
\; ,
\end{equation}

is invariant under TCT because is obtained with a contraction between a covariant and a contravariant tensor. 


The general solution of the TCT transformation, introduced in Eq. (\ref{eq:TFT21}), reads:

\begin{equation}
\label{eq:TFT27}
X^{' \mu}= X^{1}F^{\mu} \left ( \frac{X^{2}}{X^{1}}, \frac{X^{3}}{X^{2}}, \dots \frac{X^n}{X^{n-1}
  }\right)
\; ,
\end{equation}
and, thus, the general expression of a TCT transformation can be quite convoluted. Following \cite{PhysRevE.94.042103,chileproceeding}, it is possible to give a characterization in terms of group theory of the TCT transformations. We note that the function $F^{\mu}$ is invariant under homogeneous scaling of the coordinates, and that the ratio of the coordinates $X^{\mu}/X^{\mu-1}$ for all the different values of $\mu$ are the coordinates of a different space, i.e. the coordinates of the projective space $\mathbb{RP}^{n-1}$. TCT transformations are mapping lines passing through the origin, e.g. in a two dimensional space $X^{2}= \beta X^{1}$, into lines passing through the origin, e.g. $X^{'2}= \beta'X^{'1}$. Therefore, the TCT yields a map from $\mathbb{RP}^{n-1}$ into itself:
\begin{equation}
\label{eq:111}
\mathbb{RP}^{n-1} \rightarrow \mathbb{RP}^{n-1}: \frac{X^{\mu}}{X^{\mu-1}} \rightarrow \frac{X^{'\mu}}{X^{'\mu-1}}= \frac{F^{\mu} \left ( \frac{X^{2}}{X^{1}}, \frac{X^{3}}{X^{2}}, \dots \frac{X^n}{X^{n-1}
}\right)}{F^{\mu-1} \left ( \frac{X^{2}}{X^{1}}, \frac{X^{3}}{X^{2}}, \dots \frac{X^n}{X^{n-1}
}\right)}.
\end{equation}
Identifying all the points belonging to a line with a point on the unit circle defined by the intersection of the two objects, TCT are mapping the unit circle into itself. We are now tempted to say that the group of transformations satisfied by the TCT is the group of diffeomorphisms from $\mathbb{RP}^{n-1}$ into itself, but the information about the ratios $X^{'\mu}/X^{'\mu-1}$ are not enough to reconstruct all the new coordinates $X^{'\mu}$. In fact we need $X^{'1}$ or equivalently, $F^{1}$. For this reason, the transformation is the product between $dif f(\mathbb{RP}^{n-1})$ and the scalar functions $F: \mathbb{RP}^{n-1} \rightarrow \mathbb{R},^{\times}$ i.e. the non vanishing scalar functions defined over the projective space.  

In the Onsager region, closure relations have to reduce to:

\begin{equation}
\label{eq:TFT28}
X^{' \mu}= c^{\mu}_{\nu}X^{\nu}
\; ,
\end{equation}

where $c^{\mu}_{\nu}$ is independent of the thermodynamic forces. We describe the space of the thermodynamic forces through the affine connection $\Gamma^{\mu}_{\alpha\beta}$. Following \cite{sonnino2009nonlinear}, we introduce the transformation law for the components of $\Gamma$:

\begin{equation}
\label{eq:TFT73}
\Gamma^{'\mu}_{\alpha\beta}= \Gamma^{\nu}_{\lambda k}\frac{\partial X^{' \mu}}{\partial X^{\nu}}\frac{\partial X^{\lambda}}{\partial X^{'\alpha}}\frac{\partial X^{k}}{\partial X^{'\beta}} + \frac{\partial X^{'\mu}}{\partial X^{\nu}}\frac{\partial^{2}X^{\nu}}{\partial X^{' \alpha}\partial X^{' \beta}}.
\end{equation}

Using $\Gamma^{\mu}_{\alpha\beta}$, it is possible to define the absolute derivative of an arbitrary thermodynamic contravariant vector $T^{\mu}$

\begin{equation}
\label{eq:TFT74}
\frac{\delta T^{\mu}}{\delta \zeta} = \frac{d T^{\mu}}{d \zeta} + \Gamma^{\mu}_{\alpha\beta}T^{\alpha}\frac{d X^{\beta}}{d \zeta
}.
\end{equation}
Having introduced the affine connection, it is possible to define the notion of shortest path between two points, i.e. a curve connecting the two points such that a vector initially tangent to the curve remains tangent at every point. The vector tangent to a curve in the thermodynamic space parametrized by $X^{\alpha}(\zeta)$ reads:

\begin{equation}
\label{eq:TFT77}
V^{\alpha}= \frac{d X^{\alpha}}{d \zeta}
\end{equation}
and, using Eq. (\ref{eq:TFT74}), we obtain the equation that defines a geodetic, i.e. the shortest path connecting two points of the thermodynamic space:

\begin{equation}
\label{eq:TFT78}
\frac{\delta V^{\alpha}}{\delta \zeta}= \frac{d^{2}X^{\alpha}}{d \zeta^{2}}+ \Gamma^{\alpha}_{\gamma \eta} \frac{d X^{\gamma}}{d \zeta}\frac{d X^{\eta}}{d \zeta}=0.
\end{equation}


In \cite{sonnino2009nonlinear}, the least action principle, which describes the evolution of the thermodynamic system along the shortest path, and the relative TCT invariant action are introduced:
\begin{equation}
\label{eq:TFT30}
I = \int d^{n}X\, \sqrt{g}\left[ R_{\mu\nu}g^{\mu\nu} - \left( \Gamma^{\lambda}_{\mu\nu}- {\tilde \Gamma^{\lambda}_{\mu\nu}}
\right)S_{\lambda}^{\mu\nu} \right]
\; ,
\end{equation}
where every tensor is defined in terms of $g_{\mu\nu}$, $f_{\mu\nu}$ and $\Gamma^{\kappa}_{\mu\nu}$ and their first derivatives. The action can be quite complex \cite{sonnino2009nonlinear} but, in the case of confined plasmas the following simpler relations hold:
\begin{align*}
\label{eq:TFT3}
  &R = R_{\mu\nu}g^{\mu\nu}
  \; , \\
  &R_{\mu\nu}=\Gamma^{\kappa}_{\nu\kappa,\mu}-\Gamma^{\kappa}_{\nu\mu,\kappa}+\Gamma^{\kappa}_{\nu\lambda}\Gamma^{\lambda}_{\kappa\mu}-\Gamma^{\lambda}_{\kappa\lambda}
  \; , \\
  &S^{\mu\nu}_{\lambda}=\Psi^{\nu_{\lambda\alpha}}g^{\nu\alpha}+\Psi^{\mu}_{\lambda\alpha}g^{\mu\alpha}-\frac{1}{2}\Psi^{\mu}_{\alpha\beta}g^{\alpha\beta}\delta^{\nu}_{\lambda}-\frac{1}{2}\Psi^{\nu}_{\alpha\beta}g^{\alpha\beta}\delta^{\mu}_{\lambda}
  \; , \\
& \tilde{\Gamma}^{\mu}_{\alpha\beta}=
\begin{Bmatrix}
\mu\\
\alpha \beta
\end{Bmatrix}
  + \frac{1}{2\sigma}X^\mu X^\kappa g_{\alpha\beta,\kappa}-\frac{X^\kappa X^\lambda}{2(n+1)\sigma}(\delta^m_\alpha g_{\beta\kappa,\lambda}+\delta^\mu_\beta g_{\alpha\kappa,\lambda})
  \; , \\
& \begin{Bmatrix}
\mu\\
\alpha \beta
\end{Bmatrix}= \frac{1}{2}g^{\mu \lambda}(g_{\lambda\alpha,\beta}+g_{\lambda\beta,\alpha}- g_{\alpha\beta,\lambda})
  \; , \\
& \Delta\Gamma^{\mu}_{\alpha\beta}\equiv \Gamma^{\mu}_{\alpha\beta}-\begin{Bmatrix}
\mu\\
\alpha \beta
\end{Bmatrix}
  \;  ,
\end{align*}
where the comma stand for the partial derivative along the direction identified by the following greek letter. The nonlinear closure relations are obtained imposing that $I$ is stationary with respect to arbitrary variations in these fields and reads:
\begin{eqnarray*}
\label{eq:TFT31}
  R_{\mu\nu}- \frac{1}{2}g_{\mu\nu}R &=& - S^{\alpha\beta}_{\lambda}\frac{\delta \tilde{\Gamma}^{\lambda}_{\alpha\beta}}{\delta g^{\mu\nu}}
  \; ,\\
S^{\alpha\beta}_{\lambda}\frac{\delta \tilde{\Gamma}^{\lambda}_{\alpha\beta}}{\delta f^{\mu\nu}}&=&0\\
g_{\mu\nu|\lambda}&=&-\Psi^{\alpha}_{\mu \lambda}g_{\alpha\nu}-\Psi^{\alpha}_{\nu\lambda}g_{\alpha\mu}
\; , 
\end{eqnarray*}
where the absolute derivative with respect to one index is denoted with the vertical bar. In \cite{sonnino2008nonlinear}, these equations have been solved in the weak field limit, i.e. close to the Onsager region, for the particular case of magnetically confined plasmas, obtaining an amplification of the neoclassical transport coefficients.
\subsection{Linear transformations and Noether currents}
\label{sec-1-1-2}
The TFT action is invariant under TCT transformations, i.e. Eq. (\ref{eq:TFT27}) and, in particular, under linear transformations, i.e. Eq. (\ref{eq:TFT28}). We can apply the Noether's theorem in order to obtain the Noether currents associated to the symmetry under linear transformations:

The TFT action can be written in the following compact form:

\begin{equation}
\label{eq:TFTcurr1}
I[\Gamma^{\mu}_{\alpha\beta}, g^{\alpha\beta}, f^{\alpha\beta}]= \int d^{n}X\,\sqrt{g} F[ \Gamma^{\mu}_{\alpha\beta}, \partial_{\eta} \Gamma^{\mu}_{\alpha\beta}; g^{\alpha\beta}, \partial_{\eta}g^{\alpha\beta}; f^{\alpha\beta}, \partial_{\eta} f^{\alpha\beta}]
\end{equation}

where $F=R_{\mu\nu}g^{\mu\nu}-\left( \Gamma^{\lambda}_{\mu\nu}- \tilde{\Gamma}^{\lambda}_{\mu\nu} \right)S^{\mu\nu}_{\lambda}$. In the following we will use a slightly different notation with respect to this section  where the covariant derivative of a tensor is expressed with $\nabla_{\gamma}$ and the partial derivative is expressed with $\partial_{\gamma}$. The Lie derivative with respect to a vector field with components $\xi^{\gamma}$ will be indicated with $\delta_{\xi}$. We can derive the following equations:
\begin{align}
  &\delta_{\xi}F  =   \xi^{\gamma}\partial_{\gamma} F = \xi^{\gamma}\nabla_{\gamma} F \label{eq:TFTcurr2}
  \; ,\\
  &\delta_{\xi}g^{\alpha\beta} = \xi^{\gamma}\partial_{\gamma}g^{\alpha\beta} - g^{\gamma \beta}\partial_{\gamma}\xi^{\alpha} - g^{\alpha\gamma}\partial_{\gamma} \xi^{\beta} = \xi^{\gamma}\nabla_{\gamma}g^{\alpha\beta} - g^{\gamma \beta}\nabla_{\gamma}\xi^{\alpha} - g^{\alpha\gamma}\nabla_{\gamma} \xi^{\beta} \label{eq:TFTcurradded1}
  \; ,\\
&\delta_{\xi} \Gamma^{\mu}_{\alpha \beta}  =  \xi^{\gamma}\partial_{\gamma} \Gamma^{\mu}_{\alpha\beta} - \Gamma^{\gamma}_{\alpha\beta} \partial_{\gamma} \xi^{\mu} + \Gamma^{\mu}_{\gamma\beta} \partial_{\alpha} \xi^{\gamma} +    \Gamma^{\mu}_{\alpha\gamma} \partial_{\beta} \xi^{\gamma}= \xi ^{\gamma}\nabla_{\gamma}\Gamma^{\mu}_{\alpha\beta} \dots \label{eq:TFTcurradded2}
\; ,
\end{align}

Equation (\ref{eq:TFTcurradded2}), in general, is not rigorous (or strictly correct) because $\Gamma^{\mu}_{\alpha\beta}$ is not a tensor. For the moment, we will deal only with linear transformations and, in this particular case, we can verify from Eq. (\ref{eq:TFT73}) that it behaves like a tensor. Thus, we can compute its Lie derivative with the usual expression.

In order to obtain the equation of motion of the fields, we need to show that the Lie derivative commutes with the partial derivative for every field. For the special case of the linear transformation, $\partial_{\eta} A_{\mu}$ but, more in general, the derivative of a tensor, is a tensor and the formulas introduced for the Lie derivative of tensors can be used to show the commutation property. Taking the variation of the action and integrating by parts, we can write:

\begin{eqnarray*}
\label{eq:TFTcurr3}
\delta_{\xi} I & =  & \int d^{n}X\,\sqrt{g} \left\{ \frac{\delta_{\xi} \sqrt{g}}{\sqrt{g}}F + \frac{\delta F}{ \delta g^{\alpha \beta}} \delta_{\xi} g^{\alpha\beta} + \frac{\delta F}{ \delta f^{\alpha \beta}} \delta_{\xi} f^{\alpha\beta} +  \frac{\delta F}{ \delta \Gamma^{\mu}_{\alpha\beta}} \delta_{\xi} \Gamma^{\mu}_{\alpha\beta}  + \right. \\
&+ &\left. \partial_{\eta} \left[ \frac{\partial F}{\partial\partial_{\eta}g^{\alpha\beta}}\delta_{\xi}g^{\alpha\beta} + \frac{\partial F}{\partial\partial_{\eta}f^{\alpha\beta}}\delta_{\xi}f^{\alpha\beta} + \frac{\partial F}{\partial\partial_{\eta}\Gamma^{\mu}_{\alpha\beta}}\delta_{\xi}\Gamma^{\mu}_{\alpha\beta}\right] \right \}.
\end{eqnarray*}

Using the equation of motion of the fields, i.e. on shell, this expression becomes:

\begin{equation}
\label{eq:TFTcurr4}
\delta_{\xi} I  =   \int d^{n}X\,\sqrt{g} \left \{ \frac{\delta_{\xi} \sqrt{g}}{\sqrt{g}}F  + \partial_{\eta} \left[ \frac{\partial F}{\partial\partial_{\eta}g^{\alpha\beta}}\delta_{\xi}g^{\alpha\beta} + \frac{\partial F}{\partial\partial_{\eta}f^{\alpha\beta}}\delta_{\xi}f^{\alpha\beta} + \frac{\partial F}{\partial\partial_{\eta}\Gamma^{\mu}_{\alpha\beta}}\delta_{\xi}\Gamma^{\mu}_{\alpha\beta}\right] \right\}.
\end{equation}

We can also write the Lie derivative of the action as:

\begin{equation}
\label{eq:TFTcurr5}
\delta_{\xi}I = \int d^{n}X\,\sqrt{g} \left[ \frac{\delta_{\xi} \sqrt{g}F}{\sqrt{g}} + \delta_{\xi} F\right].
\end{equation}

Combining this expression with (\ref{eq:TFTcurr4}) we obtain:

\begin{equation}
\label{eq:TFTcurr6}
\int d^{n}X\, \sqrt{g} \delta_{\xi} F = \int d^{n}X\, \sqrt{g} \partial_{\eta} \left[ \frac{\partial F}{\partial\partial_{\eta}g^{\alpha\beta}}\delta_{\xi}g^{\alpha\beta} + \dots \right]
\; .
\end{equation}
Expanding $\delta_{\xi}F$ using Eq. (\ref{eq:TFTcurr2}) and integrating by parts, we obtain:
\begin{equation}
\label{eq:TFTcurr7}
\int d^{n}X\, \sqrt{g} \partial_{\eta} \left( \xi^{\eta} F\right) = \int d^{n}X\, \sqrt{g} \left \{ F \partial_{\eta} \xi^{\eta} + \partial_{\eta} \left[ \frac{\partial F}{\partial\partial_{\eta}g^{\alpha\beta}}\delta_{\xi}g^{\alpha\beta} + \frac{\partial F}{\partial\partial_{\eta}f^{\alpha\beta}}\delta_{\xi}f^{\alpha\beta} + \frac{\partial F}{\partial\partial_{\eta}\Gamma^{\mu}_{\alpha\beta}}\delta_{\xi}\Gamma^{\mu}_{\alpha\beta}\right]\right\}
\; .
\end{equation}
Expanding the Lie derivatives of the tensors using Eq. (\ref{eq:TFTcurradded1}) and Eq. (\ref{eq:TFTcurradded2})we obtain:
\begin{equation}
\label{eq:TFTcurr8}
\int d^{n}X\,\sqrt{g} \left( \xi^{\rho} \Lambda^{\eta}_{\rho}\right)= \int d^{n}X\, \sqrt{g} \left[ F \partial_{\eta} \xi^{\eta} - \partial_{\eta}\left[\left(\partial_{\rho}\xi^{\beta}\right)\Pi^{\eta\rho}_{\beta}\right]\right]
\; .
\end{equation}
with:
\begin{eqnarray}
\label{eq:TFTcurr9}
  \Lambda^{\eta}_{\rho} & = & \delta ^{\eta}_{\rho}F - \frac{\partial F}{\partial\partial_{\eta}g^{\alpha\beta}}\partial_{\rho} g^{\alpha\beta} - \frac{\partial F}{\partial\partial_{\eta}f^{\alpha\beta}}\partial_{\rho} f^{\alpha\beta} - \frac{\partial F}{\partial\partial_{\eta}\Gamma^{\mu}_{\alpha\beta}}\partial_{\rho} \Gamma^{\mu}_{\alpha\beta}
  \; .\\
\Pi^{\eta\rho}_{\beta} &=  & 2 \frac{\partial F}{\partial \partial_{\eta} f^{\alpha\beta}}f^{\alpha\rho}  + 2 \frac{\partial F}{\partial \partial_{\eta}g^{\alpha\beta}} g^{\alpha\rho} + \frac{\partial F}{\partial \partial_{\eta} \Gamma^{\beta}_{\sigma\tau}} \Gamma^{\rho}_{\sigma \tau} - 2 \frac{\partial F}{\partial \partial_{\eta}\Gamma^{\mu}_{\rho\tau}}\Gamma^{\mu}_{\beta\tau}
\; .
\end{eqnarray}
Remembering that we are studying linear transformations. Therefore, $\partial_{\eta}\partial_{\rho}\xi^{\beta} =0$ we can rewrite (\ref{eq:TFTcurr8}):
\begin{equation}
\label{eq:TFTcurr10}
\int d^{n}X\,\sqrt{g} \left \{ \partial_{\eta} \xi^{\rho}\left[\Lambda^{\eta}_{\rho} + \partial_{\beta}\Pi^{\beta\eta}_{\rho} - \delta^{\eta}_{\rho}F \right]  + \xi^{\rho}\left[ \partial_{\eta}\Lambda^{\eta}_{\rho} \right] \right \}=0.
\end{equation}
This result is independent, from the domain of integration and thus:
\begin{equation}
\label{eq:TFTcurr11}
\partial_{\eta} \xi^{\rho}\left[\Lambda^{\eta}_{\rho} + \partial_{\beta}\Pi^{\beta\eta}_{\rho} - \delta^{\eta}_{\rho}F \right]  + \xi^{\rho}\left[ \partial_{\eta}\Lambda^{\eta}_{\rho} \right]=0.
\end{equation}
This relation is valid for arbitrary $\xi^{\rho}$ and thus, the coefficients of $\xi^{\rho}$ and $\partial_{\eta}\xi^{\rho}$ which are functionally independent need to vanish separately, yielding:
\begin{equation}
\label{eq:TFTcurr12}
\partial_{\eta} \Lambda^{\eta}_{\rho}=0, \quad \Lambda^{\eta}_{\rho} + \partial_{\beta} \Pi^{\beta \eta}_{\rho} - \delta^{\eta}_{\rho}F=0.
\end{equation}
The first equation defines $n$ Noether currents and is one contribution to the development of nonlinear collisional transport theory obtained in this thesis work.

\clearpage

\chapter{Conclusions \& future developments}
\label{cap:concl}
\section{Summary}
\label{sec-1-2}
In this thesis we have conducted a self-consistent study of particle and energy transport on the energy confinement time scale in a magnetized plasma taking into account both the contributions of Coulomb collisions and turbulence on the same footing. We have applied gyrokinetic field theory \cite{brizard2007foundations,frieman1982nonlinear} in order to describe fluctuation induced fluxes and compare the result with the theory of phase space zonal structures \cite{chen2007theory,zonca2006physics,zonca2015nonlinear}. As an application, we have calculated collisional fluxes for the DTT reference scenario \cite{pizzuto2016dtt} and compared it with the fluctuation induced ones. In the last chapter of the thesis we have addressed the problem of extending the self-consistent study of transport processes on longer time scales, which is necessary in order to describe the full duration of a pulse in a modern magnetic confinement experiment. Finally, we have addressed the problem of non-linear closure relations \cite{sonnino2009nonlinear} and its fundamental importance in the derivation of a self-consistent model, which takes into account both collisional and fluctuation induced fluxes. In particular, we have presented the Thermodynamic Field Theory \cite{sonnino2009nonlinear}, which until now has been applied in order to study collisional fluxes in quiescent plasmas, as an effective tool to derive the set of non-linear closure relations; and we have discussed its classification in terms of group theory.
\section{Main results}
\label{sec-1-3}
We have derived a set of evolution equations describing the transport of particles and energy in strongly magnetized plasmas (\autoref{cap:gyrotransp}). These equations hold at every point in space and they do not involve any radial averaging operation. This is the main difference with the previous works on this topic \cite{abel2013multiscale,plunk2009theory,barnes2009trinity} based on the systematic scale separation between fluctuating and equilibrium quantities. Another element of novelty is the derivation technique, which uses the moment approach \cite{hazeltine2003plasma} and the gyrokinetic push-forward representation of the fluid moments \cite{brizard2007foundations}. This approach is illuminating, since it shows the natural spatiotemporal scales brought about by the nonlinear evolution of plasma profiles; i.e.
the corrugation of plasma equilibrium on mesoscales \cite{chen2007theory,zonca2015nonlinear}. This approach also allows the comparison with the the theory of phase space zonal structures. These equations show that, analyzing separately collisional and fluctuation induced transport, thus neglecting mutual interactions between the two, we commit an error, which in accurate and, in principle, yields non negligible effects on the transport time scale. The resulting equations also show that fluctuation induced fluxes are produced only by toroidal symmetry breaking perturbations. We have shown that the long wave length scale limit of the evolutive equation for phase space zonal structures produce fluctuation induced fluxes identical to the terms of the transport equations calculated previously via the moment equation. Therefore, fluctuation induced cross-field transport across flux surfaces is solely determined by the dynamics of phase space zonal structures. Furthermore the resulting fluxes have the same formal expression for both particle and energy transport, suggesting thus, and illuminating further, that the fundamental objects governing turbulent transport processes are phase space zonal structures. Taking the proper moment of their evolutive equation give information about the related turbulent flux. The evolutive equation for phase space zonal structures shows that multiple time and length scales can be generated by turbulent mode-mode couplings eventually invalidating the hypothesis of nonlinear evolution of plasma profiles and the corresponding phase space structures. Thus, separation of scales between equilibrium and turbulence assumed in a number of works \cite{abel2013multiscale,plunk2009theory,barnes2009trinity} may break down. As illustrative application, collisional fluxes have been calculated for a modern Tokamak machine (\autoref{cap:coll}), i.e. the DTT, which is the flagship Italian proposal for experimental studies of the power exhaust issues in next step burning plasma experiments. Various plasma scenarios have been explored, corresponding to various regimes of collisionality showing, as expected \cite{hinton1976theory}, that the transport is dominated by neoclassical low collisional regime. In particular we have calculated the resulting diffusion coefficients in terms of Gyro-Bohm units in order to facilitate a comparison with the possible experimental observations. In \autoref{cap:longer} we report our contribution to the development of Thermodynamic Field Theory \cite{sonnino2009nonlinear}, calculating the Noether currents associated to a particular subgroup of the thermodynamic transformations, i.e. the linear transformations.
\section{Future developments}
\label{sec-1-4}
In order to describe plasma evolution on time scales that are relevant for the operation of a fusion reactor, such as ITER, an important element is the derivation of a set of transport equations, analogous to what has been derived in this thesis, which should hold on a time scale $\sim \delta^{-1}\tau_{transp}$. This requires second order gyrokinetic field theory in the asymptotic expansion parameter, which, at the present moment, is formally derived but not fully carried out for generic electromagnetic fluctuations in 
non-uniform toroidal plasmas \cite{brizard2007foundations}. The difficulty is mostly technical and not conceptual; furthermore, technical complications of formal derivations \cite{brizard2007foundations} are often reflected in the necessary approximations that are routinely used in numerical implementations of nonlinear gyrokinetic theory. In particular, some progress have been made with the derivation of a second order electrostatic model \cite{catto2009extension,parra2011phase,krommes2009report}. A very challenging issue is also posed by the necessity of dealing with global description of plasma transport, where underlying instabilities may occur on micro- (Larmor radius) and macro-scales (equilibrium)
and nonlinear evolution of plasmas profiles mediates the interplay of these phenomena on the meso- spatiotemporal scales. In this thesis work, the relative ordering of temporal and spatial scales, as well as fluctuation amplitudes, has been assumed consistent with gyrokinetic field theory of the core region of magnetized thermonuclear plasmas. As the edge plasma region is approached, where equilibrium magnetic field is modified from
closed to open field lines, the relative ordering of spatiotemporal scale of turbulent fluctuation spectra and transport phenomena is also modified and not so well separated as in the plasma core. The development of a gyrokinetic field theory that encompasses these different ordering within a unified framework is the current effort of leading scientists worldwide. Any progress in this research field will be of crucial importance in order to describe transport processes on long time scales. Also collisional fluxes must be studied to an higher level of accuracy with respect to standard neoclassical transport theory and, therefore, the usual linearization of the collision integral is not sufficient to study transport with the required accuracy. Thermodynamic Field Theory \cite{sonnino2009nonlinear} could be used to calculate corrections to neoclassical closure relations, which are consistent with the theorems of non equilibrium thermodynamics.

Another important field of research concern the derivation of transport equations as long wave length scale limit of the corresponding kinetic equations. In general, we should be able to obtain both collisional and fluctuation induced fluxes by studying the evolutive equations for phase space zonal structures including an appropriate collisional term. This requires the introduction of a gyrokinetic collision integral \cite{brizard2009orbit,brizard2004guiding,burby2015energetically}. Gyrokinetics codes, based on Lagrangian particle-in-cell approaches, such as \cite{lin1998turbulent,idomura2003global,grandgirard2007global,dimits1996scalings,wang2007nonlocal,parker1993fully,sydora1996fluctuation,bottino2011global} as well as Eulerian descriptions \cite{kotschenreuther1995comparison,jenko2000electron,candy2003eulerian,gorler2011global,merz2008gyrokinetic,sugama2006collisionless,peeters2004effect,idomura2008conservative,garbet2010gyrokinetic} could be used in order to calculate fluctuation induced fluxes and the correction to neoclassical fluxes due to the presence of fluctuations, thus obtaining, a complete description of the transport processes in magnetized plasmas.
\appendix
\chapter{Noncanonical Hamiltonian methods for particle motion in a strong magnetic field}
\label{cap:appA}
\section{Non Canonical perturbation theory}
\label{sec-1-1}
In this appendix, following the instructive article by Cary and Littlejohn: \cite{cary1983noncanonical} we introduce noncanonical perturbation theory. Differently from the article which is devoted to the study of magnetic field line flow we apply this theory to the motion of charged particles inside a strong magnetic field and we calculate the pullback of the gyrocenter distribution up to $\mathcal{O}(\delta)$. This result will be used in the main body of the thesis for the evaluation of fluctuation-induced fluxes. In this chapter we have used the original notation introduced in \cite{cary1983noncanonical} which, normally, is not used in gyrokinetic field theory. This notation, which explicitly distinguish between functions and values, is more precise but at the same time less compact. For this reason in the main body of the thesis we choose to adopt standard gyrokinetics notation \cite{brizard2007foundations}. The conversion between the two notation is straightforward.
\section{Lie transform}
We define a generic change of coordinate in the extended phase space \cite{arnold2007mathematical}:
\begin{equation}
\label{eq:appendix1}
\bar{z}^{\nu}= \bar{z}^{\nu}_{f}(z)
\end{equation}
and the following backward transformation:
\begin{equation}
\label{eq:appendix2}
z^{\nu}= \bar{z}^{\nu}_{b}(\bar{z}),
\end{equation}
where we have chosen a notation which is explicit in terms of the differences between functions, i.e. $\bar{z}_{f}^{\nu}$, and values, i.e. $\bar{z}^{\nu}$. The identity function $I$ acts on the coordinates in the following way:
\begin{equation}
\label{eq:appendix3}
I^{\nu}(z)= z^{\nu}, \quad I^{\nu}(\bar{z})= \bar{z}^{\nu}.
\end{equation}
A near identity transformation is a transformation which can be written in the following form:
\begin{equation}
\label{eq:appendix4}
\bar{z}^{\mu}= \bar{z}^{\mu}_{f}(z)=I^{\mu}(z)
+\epsilon \bar{z}^{\mu}_{1f}(z)+\epsilon^{2}\bar{z}^{\mu}_{2f}(z) + \dots
\end{equation}
A Lie transformation is a near identity transformation defined by a parameter $\epsilon$ satisfying the following dynamical system:
\begin{equation}
\label{eq:appendix6}
\left\{\begin{matrix}
\partial_\epsilon \bar{z}^{\mu}_f(z,\epsilon)=G^{\mu}(\bar{z}_f(z,\epsilon))\\
\bar{z}^{\mu}_f(z,0)=I^{\mu}(z)
\end{matrix}\right.
\end{equation}
where $G^{\mu}$ are the generators of the Lie transformation. Applying $\partial_{\epsilon}$ to:
\begin{equation}
\label{eq:appendix7}
z^{\mu}= \bar{z}^{\mu}_{b}(\bar{z}_{f}(z,\epsilon),\epsilon)
\end{equation}
we get the useful relation:
\begin{equation}
\label{eq:appendix5}
\partial_{\epsilon}\bar{z}_{b}^{\mu}(\bar{z}, \epsilon)= - G^{\nu}(\bar{z})\frac{\partial \bar{z}^{\mu}_{b}}{\partial \bar{z}^{\nu}}(\bar{z},\epsilon).
\end{equation}
Under a coordinate transformation a scalar function $S$ transform in the following way:
\begin{equation}
\label{eq:appendix8}
S(z)= \bar{S}(\bar{z})
\end{equation}
and, in particular, under a Lie transformation:
\begin{equation}
\label{eq:appendix9}
\bar{S}(\bar{z},\epsilon)=S(\bar{z}_{b}(\bar{z},\epsilon)).
\end{equation}
Comparing Eq. (\ref{eq:appendix8}) and Eq. (\ref{eq:appendix9}) we can see that only in the second expression both sides are calculated on the same value, i.e. $\bar{z}$. Expression with this feature are relations between functions, i.e. $\bar{S}= S \circ \bar{z}_{b} $. Applying $\partial_{\epsilon}$ to both sides of Eq. (\ref{eq:appendix9}) and using Eq. (\ref{eq:appendix5}) we get:
\begin{equation}
\label{eq:appendix10}
\partial_{\epsilon}\bar{S}(\bar{z},\epsilon)= -G^{\mu}(\bar{z})\frac{\partial S}{\partial \bar{z}^{\mu}}(\bar{z},\epsilon)
\end{equation}
which holds for every value of $\bar{z}$ and therefore is a relation between functions. Using functional notation, i.e. removing all the values dependencies, we can write:
\begin{equation}
\label{eq:appendix11}
\partial_{\epsilon}\bar{S}=-G^{\mu}\partial_{\mu}\bar{S} \equiv - L_{G}\bar{S}.
\end{equation}
From Eq. (\ref{eq:appendix11}) we get by recursion:
\begin{equation}
\label{eq:appendix12}
\partial_{\epsilon}^{n}\bar{S}=\left(-L_{G}\right)^{n}\bar{S}.
\end{equation}
Taylor expanding $S(\bar{z},\epsilon)$ around $\epsilon =0$ and using Eq. (\ref{eq:appendix12}), we get:
\begin{equation}
\label{eq:appendix13}
\bar{S}(\bar{z},\epsilon)= \sum_{n=0}^{\infty}\frac{\epsilon^{n}}{n !} \left(-L_{G}\right)^{n} \bar{S}\big |_{\epsilon=0}(\bar{z})= e^{- \epsilon L_{G}}\bar{S}\big |_{\epsilon=0} (\bar{z})= e^{- \epsilon L_{G}} S (\bar{z})
\end{equation}
which is again a relation between functions and which defines the pull-back operator acting on the function $S$:
\begin{equation}
\label{eq:appendix14}
\bar{S}= e^{-\epsilon L_{G}}S.
\end{equation}
Using the identity function $I$ we get a relation connecting the coordinates before and after the Lie transformation in terms of the pull-back operator:
\begin{equation}
\label{eq:appendix15}
z^{\nu}=\bar{z}^{\nu}_{b}(\bar{z},\epsilon)= I^{\nu}(z) \Rightarrow z^{\nu}= (e^{- \epsilon L_{G}}I^{\nu} )(\bar{z}).
\end{equation}
An analogue procedure can be carried out for the covariant vectors. Assuming the transformation rule:
\begin{equation}
\label{eq:appendix16}
\bar{\gamma}_{\mu}(\bar{z},\epsilon)= \frac{\partial \bar{z}_{b}^{\nu}}{\partial \bar{z}^{\mu}}(\bar{z},\epsilon)\gamma_{\nu}(\bar{z}_{b}(\bar{z},\epsilon))
\end{equation}
and deriving with respect to $\epsilon$ we get:
\begin{equation}
\label{eq:appendix17}
\partial_{\epsilon}\bar{\gamma}_{\mu}(\bar{z},\epsilon)=-G^{\lambda}(\bar{z})\left[\frac{\partial \bar{\gamma}_{\mu}}{\partial \bar{z}^{\lambda}}(\bar{z},\epsilon)- \frac{\partial \bar{\gamma}_{\lambda}}{\partial \bar{z}^{\mu}}(\bar{z},\epsilon)  \right]- \frac{\partial}{\partial \bar{z}^{\mu}}\left[G^{\lambda}(\bar{z})\gamma_{\lambda}(\bar{z},\epsilon)\right].
\end{equation}
Using the following definitions:
\begin{equation}
\label{eq:appendix18}
\left(L_{G} \xi \right)_{\mu} \equiv G^{\lambda}\left[\partial_{\nu}\xi_{\mu}-\partial_{\mu}\xi_{\nu}\right], \quad \left(d f\right) _{\mu} \equiv \partial_{\mu}f
\end{equation}
Eq. (\ref{eq:appendix17}) can be written in functional form:
\begin{equation}
\label{eq:appendix19}
\partial_{\epsilon}\bar{\gamma}_{\mu} = - \left(L_{G} \bar{\gamma} \right)_{\mu} - \left( d \left(G^{\lambda}\gamma_{\lambda}\right)\right)_{\mu}.
\end{equation}
Recursively we get:
\begin{equation}
\label{eq:appendix20}
\partial_{\epsilon}^{n} \bar{\gamma}_{\mu}= \left[\left(- L_{G}\right)^{n} \bar{\gamma}\right]_{\mu}- \left[d \left(G^{\lambda} \partial_{\epsilon}^{n-1} \gamma_{\lambda} \right)\right]_{\mu}.
\end{equation}
Applying the same procedure used for the scalar function $S$ and we get the following relation:
\begin{equation}
\label{eq:appendix21}
\bar{\gamma}_{\mu}= \left(e^{-\epsilon L_{G}} \gamma \right)_{\mu} + \partial_{\mu} S
\end{equation}
where the function $\partial_{\mu} S$ is defined in terms of the derivatives of $\bar{\gamma}$ and of $G^{\mu}$.
\section{Noncanonical perturbation theory}
We assume that the motion that we want to study is governed by a Lagrangian in the extended phase:
\begin{equation}
\label{eq:appendix22}
L(z)= \left ( \gamma_{0\mu}(z) + \epsilon \gamma_{1\mu}(z) + \dots\right )V^{\mu}(z) - K(z)
\end{equation}
where we assume that the motion described by $\gamma_{\nu}=\gamma_{0\nu}$ is well understood in terms of integral of motion and symmetries and, eventually, integrable. Under a transformation of coordinates $\bar{z}_{f}$ the Lagrangian becomes:
\begin{equation}
\label{eq:appendix23}
\bar{L}(\bar{z})= \left ( \bar{\gamma}_{0\mu}(\bar{z}) + \epsilon \bar{\gamma}_{1\mu}(\bar{z}) + \dots\right )\bar{V}^{\mu}(\bar{z}) - \bar{K}(\bar{z})
\end{equation}
where $\gamma_{0\mu}=\bar{\gamma}_{0\mu}$. All the changes of coordinates such that $\bar{\gamma}_{n\nu}=0 \, \, \forall n >0$ leave unchanged the simplectic part of the Lagrangian and therefore the structure of the equation of motion. Thus if $\bar{K}$ has the same number of cyclic coordinates of $K$, the number of integral of motion remain unchanged for the perturbed system.
The change of coordinates such that $\gamma_{n\nu}=0$ must be defined for each value of $n$. We introduce the following operator:
\begin{eqnarray}
\label{eq:appendix24}
T & = & \dots T_{3}T_{2}T_{1} \\
T_{n} & = & e^{-\epsilon^{n}L_{G_{n}}}\equiv e^{-\epsilon^{n}L_{n}}
\end{eqnarray}
where we are concatenating Lie transformations and $T$ is the pull-back operator of the resulting transformation. In the previous section we showed that the following relation hold:
\begin{equation}
\label{eq:appendix27}
\bar{\gamma}_{\nu}(\bar{z})= \left(T \gamma\right)_{\nu}(\bar{z})+\frac{\partial S}{\partial \bar{z}^{\nu}}(\bar{z}).
\end{equation}
Expanding both members of Eq. (\ref{eq:appendix27}) we get:
\begin{eqnarray}
\label{eq:appendix28}
\bar{\gamma}_{0\nu} (\bar{z})& = & \gamma_{0\nu}(\bar{z}) \\
\bar{\gamma}_{1\nu} (\bar{z})& = & \gamma_{1\nu}(\bar{z})+(-L_{1}\gamma_{0})_{\mu}(\bar{z})+ \frac{\partial S_{1}}{\partial \bar{z}^{\mu}}(\bar{z})\\
&\vdots&\\\label{eq:667}
\bar{\gamma}_{n\nu} (\bar{z})& = & \gamma_{n\nu}(\bar{z})+(-L_{n}\gamma_{0})_{\mu}(\bar{z})+ \frac{\partial S_{n}}{\partial \bar{z}^{\mu}}(\bar{z}) + C_{n\nu}(\bar{z}).
\end{eqnarray}
We obtain the following relation:
\begin{equation}
\label{eq:appendix29}
(-L_{n}\gamma_{0})_{\nu}(\bar{z})= G^{\mu}_{n}(\bar{z})\left(\frac{\partial \gamma_{0\mu}}{\partial \bar{z}^{\mu}}(\bar{z})- \frac{\partial \gamma_{0\nu}}{\partial \bar{z}^{\nu}}(\bar{z}) \right)=-G^{\mu}_{n}(\bar{z})\omega_{0\mu\nu}(\bar{z})
\end{equation}
where we have defined the Lagrange matrix $\omega_{\mu\nu}$:
\begin{equation}
\label{eq:appendix26}
\omega_{\mu\nu} = \left( \frac{\partial \gamma_{0\nu}}{\partial \bar{z}^\mu} - \frac{\partial \gamma_{0\mu}}{\partial \bar{z}^{\nu}}  \right).
\end{equation}
From Eq. (\ref{eq:appendix29}), we can obtain and expression in terms of the inverse of $\omega$, i.e. the Poisson Matrix $\Pi_{0}^{\mu\nu}$:
\begin{equation}
\label{eq:appendix30}
G^{\alpha}_{n}(\bar{z})=\Pi_{0}^{\nu\alpha}(\bar{z})\left[\frac{\partial S_{n}}{\partial \bar{z}^{\nu}}(\bar{z})-\bar{\gamma}_{n\nu}(\bar{z})+C_{n\nu}(\bar{z})\right].
\end{equation}
This relation holds for every value and thus is a relation between functions:
\begin{equation}
\label{eq:appendix31}
G^{\alpha}_{n}=\Pi_{0}^{\nu\alpha}\left[\partial_{\nu}S_{n} +C_{n\nu} - \bar{\gamma}_{n\mu} \right].
\end{equation}
We already showed that, for a scalar function $K$:
\begin{equation}
\label{eq:appendix32}
\bar{K}(\bar{z})= \left( T K\right)(\bar{z}).
\end{equation}
By expanding both sides we get:
\begin{eqnarray}
\label{eq:appendix33}
\bar{K}_{0}(\bar{z}) & = & K_{0}(\bar{z}) \\
\bar{K}_{1}(\bar{z}) & = & K_{1}(\bar{z})+(-L_{1}K_{0})(\bar{z}) \\
&\vdots&\\
\label{eq:666}\bar{K}_{n}(\bar{z}) & = & D_{n}(\bar{z})+(-L_{n}K_{0})(\bar{z})=D_{n}(\bar{z}) -G^{\alpha}_{n}(\bar{z})\frac{\partial K_0}{\partial \bar{z}^{\alpha}}(\bar{z}).
\end{eqnarray}
Eq. (\ref{eq:666}) can be substituted into Eq. (\ref{eq:667}) obtaining:
\begin{equation}
\label{eq:appendix34}
\bar{K}_{n}(\bar{z})= D_{n}(\bar{z})- \left\{\Pi_{0}^{\nu\alpha}(\bar{z})\left[\frac{\partial S_{n}}{\partial \bar{z}^{\nu}}(\bar{z})+ C_{n\nu}(\bar{z}) -\bar{\gamma}_{n\nu}(\bar{z})\right]\right\}\frac{\partial K_{0}}{\partial \bar{z}^{\alpha}}(\bar{z})
\end{equation}
which can be written in functional form:
\begin{equation}
\label{eq:appendix35}
\bar{K}_{n}= D_{n}- \Pi_{0}^{\nu\alpha}\left( \partial_{\nu}S_{n}+ C_{n\nu} - \bar{\gamma}_{n\nu}\right) \partial_{\alpha}K_{0}.
\end{equation}
We now show that Eq. (\ref{eq:appendix34}) can be interpreted as an evolutive equation (along the unperturbed motion):
\begin{equation}
\label{eq:appendix37}
\Pi_{0}^{\nu\alpha}(\bar{z})\frac{\partial S_{n}}{\partial \bar{z}^{\nu}}(\bar{z})\frac{\partial K_{0}}{\partial \bar{z}^{\alpha}}(\bar{z})= V^{\alpha}_{u}(\bar{z})\frac{\partial S_{n}}{\partial \bar{z}^{\alpha}}(\bar{z})= \left[\left(\frac{d}{d\tau}\right)_{u}S_{n}\right](\bar{z})
\end{equation}
where $V_{u}^{\alpha}$ are the components of the contravariant vector field satisfying  the least action principle with an extended phase space Lagrangian $L = \gamma_{0\nu}V^{\nu}-K_{0}$, i.e. $V^{\nu}=V_{u}^{\nu}=\Pi_{0}^{\nu\alpha}\partial_{\alpha}K_{0}$ and $\tau$ is the curvilinear coordinate parametrizing the motion in the extended phase space.
Therefore the evolutive equation for $S_{n}$ reads:
\begin{equation}
\label{eq:appendix38}
V^{\alpha}_{u}(\bar{z})\frac{\partial S_{n}}{\partial \bar{z}^{\alpha}}(\bar{z})=V_{u}^{\nu}(\bar{z})\left[\bar{\gamma}_{n}(\bar{z})-C_{n\nu}(\bar{z})\right] + D_{n}(\bar{z})-\bar{K}_{n}(\bar{z})
\end{equation}
which can be rewritten as:
\begin{equation}
\label{eq:appendix39}
\left[\left(\frac{d}{d\tau}\right)_{u}S_{n}\right](\bar{z})= \left[\left(\frac{d}{d\tau}\right)_{u}I^{\nu}\right](\bar{z})\left(\bar{\gamma}_{n}(\bar{z})-C_{n\nu}(\bar{z})\right)+D_{n}(\bar{z})- \bar{K}_{n}(\bar{z}).
\end{equation}
We stress that, at fixed $\tau$, the action of the operator $d/d\tau$ on a function $S(\bar{z})$ (or $I^{\mu}$) is:
\begin{equation}
\label{eq:appendix40}
\left(\frac{d}{d\tau} S\right)(\bar{z})= V^{\nu}(\bar{z})\frac{\partial S}{\partial \bar{z}^{\nu}}(\bar{z}).
\end{equation}
If the unperturbed motion: $\bar{z}_{u}^{\nu}(\tau)$, solution of $d\bar{z}^{\nu}_{u}(\tau)/d\tau=V_{u}(\bar{z}(\tau))$, has some periodicities $\tau_{i}$ (in particular when the motion is integrable), we need to require $\left \langle \left(\frac{d}{d\tau} \right)_{u}S_{n} \right \rangle_{\tau_{i}} =0 \, \, \forall \bar{z}_{u}(0)$. Otherwise, after each period $\tau_{i}$, $S_{n}$ will increase invalidating our ordering after a sufficient amount of time.

Until now we have developed a perturbation theory which, exploiting the arbitrariness of the gauge function $S_{n}$, allows to find a particular set of coordinates such that the simplectic part of the Lagrangian of the perturbed system is identical to the simplectic part of the unperturbed Lagrangian in terms of the initial coordinates. This allows to preserve the number of integrals of motion of the unperturbed system if the following conditions are satisfied for each value of $\nu$ and $a$:
\begin{eqnarray}
\label{eq:appendix41}
 \left \langle \left(\frac{d}{d\tau} \right)_{u}S_{n} \right \rangle_{\tau_{i}} & = & 0 \\
\frac{\partial K_{0}}{\partial \bar{z}^{a}}(\bar{z})=0 & \Rightarrow &  \frac{\partial \bar{K}_{n}}{\partial \bar{z}^{a}}(\bar{z})=0 \\
\frac{\partial K_{0}}{\partial \bar{z}^{a}}(\bar{z})=0 & \Rightarrow &  \frac{\partial \bar{\gamma}_{\nu}}{\partial \bar{z}^{a}}(\bar{z})=0.
\end{eqnarray}

\section{Motion of a particle in an electromagnetic field}
The motion of a charged particle in an electromagnetic field can be described using  guiding-center coordinates \cite{littlejohn1983variational},\cite{cary2009hamiltonian}: $(x^{i},U,\mu,\theta,-w,t)$ which are chosen in order to have a Lagrangian describing the motion of a charged particle in the equilibrium fields such that:
\begin{equation}
\label{eq:appendix25}
\gamma_{0\nu}= \left(\frac{e}{c}A^{*}_{0\nu} \big |_{\nu=1},\frac{e}{c}A^{*}_{0\nu} \big |_{\nu=2}, \frac{e}{c}A^{*}_{0\nu} \big |_{\nu=3},0,0,0, \frac{m c}{e}I^{5},I^{7}\right)
\end{equation}
where $A_{0\nu}^{*}= A_{0 \nu}+ (c/e)b_{0\nu}I^{4}$ for $\nu = 1,2,3$ and $0$ for $\nu > 3$, $K_{0}= (1/2m)(I^{4})^{2}+I^{3}B_{0}+I^{7}$.
In this section we address the question of how $\gamma$ changes if we want to consider fluctuating fields. We can write:
\begin{eqnarray}
\label{eq:appendix42}
\hat{A}_{0\nu} & \rightarrow & \hat{A}_{\nu} + \delta \hat{A}_{\nu} \\
\hat{K} & \rightarrow & \hat{K}_{\nu} + e\, \delta \hat{\phi}
\end{eqnarray}
where we have introduced the guiding center change of coordinates (from the guiding center set of coordinates to the particle ones): $\hat{z}^{\nu}=\hat{z}_{f}(z)$ and the related pull-back operator $e^{\mathbf{\rho} \cdot \nabla}$. It follows that:
\begin{equation}
\label{eq:appendix43}
\delta \hat{K}(\hat{z})=\left(e^{\mathbf{\rho} \cdot \nabla}\delta K\right)(\hat{z}), \;\; \;\delta K (z)= \delta \hat{K}(\hat{z})=\delta \hat{K}\left(\left(e^{\mathbf{\rho} \cdot \nabla} I \right)(z)\right)
\end{equation}
where $\hat{z}^{\nu}=\left(e^{\mathbf{\rho} \cdot \nabla} I \right)(z) = z^{\nu} + \rho^{\nu}$ for $\nu=1,2,3$ and $\hat{z}^{\nu}=\left(e^{\mathbf{\rho} \cdot \nabla} I \right)(z) = z^{\nu}$ for $\nu >3$. We can calculate the effect of the fluctuations on $\gamma_{\nu}$:
\begin{align}
\label{eq:appendix44}
&\hat{\gamma}_{1\nu}(\hat{z})\frac{d \hat{z}^{\nu}}{d \tau}= \hat{\gamma}_{1\nu}(\hat{z})\hat{V}^{\nu}(\hat{z})= \hat{\gamma}_{1\nu}\frac{\partial \hat{z}^{\nu}_{f}}{\partial z^{\eta}}(\hat{z}_{b}(\hat{z}))V^{\eta}(\hat{z}_{b}(\hat{z}))=\\
&=\nonumber \frac{e}{c}\delta \hat{A}_{i}(\hat{z})\frac{\partial \hat{z}_{f}}{\partial z^{\eta}}(\hat{z}_{b}(\hat{z}))V^{\eta}(\hat{z}_{b}(\hat{z}))=  \\
&\nonumber =\frac{e}{c}\delta \hat{A}_{i}\left(\left(e^{\mathbf{\rho} \cdot \nabla} I \right)(z)\right) \left(\delta^{i}_{j}V^{j}(z)+ \frac{\partial \rho^{i}}{\partial z^{\eta}}(z) V^{\eta}(z)\right)= \\
&\nonumber  =  \gamma_{1\nu}(z)V^{\nu}(z)  = \gamma_{1\nu}(z)\frac{d z^{\nu}}{d \tau}
\end{align}
It can be showed that $\rho^{i}(\bar{z})= \rho^{i}(\bar{\mu},\bar{\theta})$ and, therefore, we get the expression for $\gamma_{1\nu}$:
\begin{equation}
\label{eq:appendix45}
\nonumber \left(\frac{e}{c} \delta \hat{A}_{\nu} \left( \left(e^{\mathbf{\rho} \cdot \nabla} I \right)\right)\Big |_{\nu=1},\dots, 0, \frac{e}{c}\delta \hat{A}_{i}\left( \left(e^{\mathbf{\rho} \cdot \nabla} I \right)\right)\partial_{\mu}\rho^{i} , \frac{e}{c}\delta \hat{A}_{i}\left( \left(e^{\mathbf{\rho} \cdot \nabla} I \right)\right)\partial_{\theta}\rho^{i}, 0\right).
\end{equation}
For the unperturbed motion described by $\gamma_{0\nu}$ in a toroidal and axisymmetric configuration there are actually three cyclic coordinates and therefore the motion is integrable. The most general perturbation destroys all these conserved quantities and, even using perturbation theory, it is not possible to recover these integrals without putting strong limitations on the functional shape of the perturbations in order to remove the secularities in Eq. (\ref{eq:appendix41}). In particular we will focus on the fact that while $\partial_{\theta} \gamma_{0\nu}=0$ it easy to see that $\partial_{\theta} \gamma_{\nu} \neq 0$ and thus $I^{5}$ is not a conserved quantity. The gyrocenter transformation of coordinates is used to find a set of coordinates such that, for perturbations with a characteristic frequency $\omega \ll \Omega$, $I^{5}$ is a conserved quantity. Physically this is possible because the limitation on the perturbation characteristic frequency $\omega$ eliminates any resonances with the gyromotion of the particles. Anyway resonances between the perturbation and the slow motion of the particles, i.e. guiding center motion,  are still possible and this is why the other integrals of motions will not be recovered without making further assumptions. Eq. (\ref{eq:appendix38}) with $n=1$ reads:
\begin{equation}
\label{eq:appendix46}
V^{\alpha}_{u}(\bar{z})\frac{\partial S_{1}}{\partial \bar{z}^{\alpha}}(\bar{z})=V_{u}^{\nu}(\bar{z})\left[\bar{\gamma}_{1\nu}(\bar{z})-\gamma_{1\nu}(\bar{z})\right] + K_{1}(\bar{z})-\bar{K}_{1}(\bar{z}).
\end{equation}
Using Eq. (\ref{eq:appendix44}) the term $V^{\nu}_{u}(\bar{z})\gamma_{1\nu}(\bar{z})$ is readily evaluated:
\begin{equation}
\label{eq:appendix49}
V^{\nu}_{u}(\bar{z})\gamma_{1\nu}(\bar{z})= \frac{e}{c}\delta \hat{A}_{i}\left( \left(e^{\mathbf{\rho} \cdot \nabla} I \right)(\bar{z})\right)\hat{V}_{u}^{i}\left( \left(e^{\mathbf{\rho} \cdot \nabla} I \right)(\bar{z})\right).
\end{equation}
Eq. (\ref{eq:appendix46}) can be written in compact form using the Poisson brackets:
\begin{equation}
\label{eq:appendix50}
\bar{K}_{1}(\bar{z})= -\{S_{1},K_{0}\}(\bar{z})+ e \delta \psi(\bar{z})
\end{equation}

where, without loss of generality, we have set $\bar{\gamma}_{1}=0$ and :
\begin{equation}
\label{eq:appendix51}
\delta \psi(\bar{z}) = \left(e^{\mathbf{\rho}\cdot \nabla}\delta\hat{\phi}\right)(\bar{z})- \frac{1}{c}\delta \hat{A}_{i}\left( \left(e^{\mathbf{\rho} \cdot \nabla} I \right)(\bar{z})\right)\hat{V}_{u}^{i}\left( \left(e^{\mathbf{\rho} \cdot \nabla} I \right)(\bar{z})\right).
\end{equation}
As already stated the notation adopted in this appendix is not consistent with the main body of the thesis. In particular the physical fields are indicated with the hat, e.g. $\delta \hat{A}_{i}$ while the fields without apex are obtained from these ones by means of the guiding center pull-back operator. In the main body of the thesis the physical fields have no apex while the fields obtained by applying the pull-back operator have the subscript $_{gc}$. As an example of comparison between the two notations we note that Eq. (\ref{eq:appendix51}) is equivalent to Eq. (\ref{eq:103}). In order to satisfy  Eq. (\ref{eq:appendix41}), Eq. (\ref{eq:appendix42}) and  Eq. (\ref{eq:appendix43}) (only with respect to the $\bar{\theta}$ variable):
\begin{equation}
\label{eq:appendix52}
\{S_{1},K_{0}\}_{u}(\bar{z})=e \left[\delta \psi(\bar{z})- \left \langle \delta \psi (\bar{z})\right \rangle_{\bar{\theta}}\right ].
\end{equation}
We now use the assumption that $\Omega \gg \omega$ in order to expand this relation and, up to the leading order, we obtain:
\begin{equation}
\label{eq:appendix53}
\{S_{1},K_{0}\}_{u}(\bar{z})\sim \Omega \frac{\partial S_{1}}{\partial \bar{\theta}}(\bar{z}).
\end{equation}
Thus we obtain an explicit expression for $S_{1}$:
\begin{equation}
\label{eq:appendix54}
S_{1}(\bar{z})= e\Omega^{-1}\int^{\bar{\theta}}d \bar{\theta}'\left(\delta \psi(\bar{z})- \left \langle \delta \psi(\bar{z}) \right \rangle \right)
\end{equation}
and for $\bar{K}_{1}$:
\begin{equation}
\label{eq:appendix55}
\bar{K}_{1}(\bar{z})=\left \langle \delta \psi (\bar{z})\right \rangle_{\bar{\theta}}.
\end{equation}
We can finally calculate:
\begin{align}
\label{eq:appendix56}
&\bar{F}_{1}(\bar{z}) =  F_{1}(\bar{z}) - G_{1}^{\alpha}(\bar{z})\frac{\partial F_{0}}{\partial \bar{z}^{\alpha}}= \\
\nonumber&=F(\bar{z}) + \{F_{0},S_{1}\}_{u}(\bar{z})-\gamma_{1\nu}(\bar{z})\Pi^{\nu\alpha}_{u}(\bar{z})\frac{\partial F_0}{\partial \bar{z}^{\alpha}}(\bar{z}).
\end{align}
After some calculations and using the relation:
\begin{equation}
\label{eq:appendix59}
\frac{\partial \rho^{i}}{\partial \bar{\theta}}(\bar{z})=\Omega^{-1}\epsilon^{ijk}b_{j}\hat{V}_{uk}\left( \left(e^{\mathbf{\rho} \cdot \nabla} I \right)(\bar{z})\right)
\end{equation}
we obtain the following functional relation:
\begin{align}
\nonumber&\bar{F}_{1}=F_{1}- \frac{1}{mB}\partial_{\mu}F_{0}\Big[e \left(e^{\mathbf{\rho} \cdot \nabla} \delta \hat{\phi}\right) - \frac{e}{c}\hat{V}_{u \parallel}^{i} \left(e^{\mathbf{\rho} \cdot \nabla} I \right)\delta\hat{A}_{\parallel i}  \left(e^{\mathbf{\rho} \cdot \nabla} I \right) + \label{eq:appendix57a}\\
\nonumber&- e \left \langle \left(e^{\mathbf{\rho} \cdot \nabla} \delta \hat{\phi}\right) - \frac{1}{c}\hat{V}_{u}^{i} \left(e^{\mathbf{\rho} \cdot \nabla} I \right)\delta\hat{A}_{i}  \left(e^{\mathbf{\rho} \cdot \nabla} I \right)\right \rangle_{\theta}\Big]- \frac{1}{B^{*}_{\parallel}}\epsilon^{kji}b_{k}\delta\hat{A}_{i}  \left(e^{\mathbf{\rho} \cdot \nabla} I \right)\partial_{j}F_{0}+\\
\nonumber&-\frac{e}{mc}\delta \hat{A}_{\parallel}\left(e^{\mathbf{\rho} \cdot \nabla} I \right)\partial_{U}F_{0}.
\end{align}
In this work we will use the set of coordinates $(x^{i},\mathcal{E},\mu,\theta,-w,t)$ with $\mathcal{E}= \frac{1}{2}v^{2}$ and thus we need another change of coordinates. Applying the chain rule we obtain the following relations:
\begin{equation}
\label{eq:appendix36}
\partial_{\mu}|_{U}= \partial_{\mu}|_{\mathcal{E}} + B \partial_{\mathcal{E}}|_{\mathcal{\mu}}, \quad  \partial_{U}|_{\mu}= m U \partial_{\mathcal{E}}|_{\mathcal{\mu}}
\end{equation}
and, by substitution into the previous expression, it follows the formula for $\bar{F}_{1}$ used in this work for the calculation of the transport induced by fluctuations, i.e. Eq. (\ref{eq:102}).
\chapter{Push-forward representation of the moments of the distribution function}
\label{cap:appB}
In this appendix we calculate the push-forward representation of the first two moments of the distribution function by using the expressions for the pull-back/push-forward operators derived in the previous chapters. The resulting expressions are necessary for the calculation of the fluctuation-induced particles flux in terms of the gyro-center distribution function. The calculations required in order to obtain the analogue expressions for the energy flux are identical and therefore not treated in this work.
\section{Density}
In the previous appendix we have shown that, for a generic scalar function S, the following relation hold:
\begin{equation}
\label{eq:apppush1}
e^{{\bm \rho }\cdot {\bm \nabla}} S(z)= S\left(e^{{\bm \rho }\cdot {\bm \nabla}} I(z)\right)=S(z+\rho(z)).
\end{equation}
Analogously we can show that:
\begin{equation}
\label{eq:apppush2}
S(z-\rho)= e^{-{\bm \rho }\cdot {\bm \nabla}}  S(z).
\end{equation}
Taking the gyro-average we obtain:
\begin{align*}
\left \langle e^{-{\bm \rho }\cdot {\bm \nabla}}  S(z) \right \rangle_{\theta} &= \left \langle S(z-\rho)\right \rangle_{\theta}= \\
&= \frac{1}{2 \pi}\int \mathit{d} \theta \,  \int d {\bm k}\, e^{i k_{l} (z^{l}- \rho^{l})}S_{{\bm k}}
\end{align*}
which can be written in a compact form choosing properly the gyro-angle $\theta$ such that ${\bm \rho}\cdot {\bm k}_{\perp}= \lambda \cos \theta$ obtaining:
\begin{align*}
\left \langle e^{-{\bm \rho }\cdot {\bm \nabla}}  S(z) \right \rangle_{\theta} & = \frac{1}{2 \pi} \int d {\bm k}\, e^{i k_{l}z^{l}} S_{{\bm k}} \int d \theta\, e^{i\lambda_{k}\cos \theta }= \\
&= \sum_{{\bm k}} e^{i k_{l}z^{l}}S_{{\bm k}}J_{0}(\lambda_{k}) = \hat{I}_{0} S(z)
\end{align*}
where $J_{0}$ is the zeroth order Bessel function and $\hat{I}_{0}$ is an integral operator defined in the thesis.

We can use this relation, combined with the pull-back representation of the distribution function in order to calculate the push-forward representation of the density:
\begin{align*}
n(\hat{z})= \left  \langle \hat{F}_{0}(\hat{z})+ \hat{F}_{1}(\hat{z}) \right \rangle_{\hat{v}}.
\end{align*}
Using the pull-back representation of the distribution function, i.e. Eq. (\ref{eq:102}) we can write:
\begin{align*}
\left \langle \hat{F}_{1} (\hat{z})\right \rangle_{\hat{v}}&= \left \langle \left( e^{-{\bm \rho }\cdot {\bm \nabla}}\bar{F}_{1} \right )(\hat{z})  + \dots \right \rangle_{\hat{v}} = \\
&= \int dU d\mu\, J(z) 2 \pi \left \langle  e^{-{\bm \rho }\cdot {\bm \nabla}} \bar{F}_{1} (\hat{z})  + \dots \right \rangle_{\theta} = \\
&= \int dU d \mu J(z) 2 \pi \left ( \hat{I}_{0} \bar{F}_{1} (\hat{z}) + \dots\right )=\\
&= - 2 \pi \int dU d\mu\, J(z) \hat{I}_{0} \left[ \bar{F}_{1}(\hat{z}_{f}(z)) - \frac{e}{m} \frac{\partial \bar{F}_{0}}{\partial \mathcal{E}} \gav{e^{{\bm \rho}\cdot {\bm \nabla}}\left(\delta \hat{\phi} - \frac{1}{c}\hat{V}^{i}\delta \hat{A}_{i}\right)} \right] + \\
&+  \frac{e}{m} \delta \phi \left \langle\frac{\partial \bar F_{0}}{\partial \cal E} \right \rangle_{\hat{v}}   + \epsilon^{kji}\frac{b_{k}}{B}\delta \hat{A}_{i}(\hat{z})\frac{\partial \bar{F}_{0}}{\partial \hat{z}^{j}}(\hat{z})
\end{align*}
where we are using the symbol $\bar{F}$ to indicate the gyrocenter distribution function in the phase space with $\mathcal{E}$ instead of $U$ as coordinate and we have made the assumption that $\frac{\partial \bar{F}_{0}}{\partial \mu}=0$. Summing this term with the lower order one we obtain the expression used in the thesis.
\section{Flux of particles}
In order to calculate the push-forward representation of the first moment of the distribution function we need to calculate:
\begin{equation}
\label{eq:apppush3}
\left \langle \hat{V}^{i}(z)\left(e^{-{\bm \rho }\cdot {\bm \nabla}}S \right)(z)\right \rangle_{\theta} = \left \langle \hat{V}^{i}(z)S(z-\rho)\right \rangle_{\theta}.
\end{equation}
We can apply the same procedure used for the density for the parallel component of the velocity because it is not depending on the gyrophase obtaining:
\begin{equation}
\label{eq:apppush4}
\left \langle \hat{V}^{i}_{\parallel}(z)\left(e^{-{\bm \rho }\cdot {\bm \nabla}}S \right)(z)\right \rangle_{\theta} = \hat{V}^{i}_{\parallel}(z)\hat{I}_{0}S(z).
\end{equation}
The calculations for the perpendicular velocity are similar and after some algebra we obtain the following expression for the two components of $\hat{V}_{\perp}$:
\begin{align*}
&\left \langle \hat{V}^{1}_{\perp}(z)S(z-\rho)\right \rangle_{\theta}= \frac{mc}{e}\mu \hat{I}_{1} \frac{\partial S}{\partial z^{2}}(z)\\
&\left \langle \hat{V}^{2}_{\perp}(z)S(z-\rho)\right \rangle_{\theta}= -\frac{mc}{e}\mu \hat{I}_{1} \frac{\partial S}{\partial z^{1}}(z).
\end{align*}
We recall that we are using a set of coordinates aligned to the magnetic field and, therefore, we can write:
\begin{equation}
\label{eq:apppush5a}
\left \langle \hat{V}^{i}_{\perp}(z)S(z-\rho)\right \rangle_{\theta}= - \frac{mc}{e} \mu \hat{I}_{1}\epsilon^{ijk}b_{j}\frac{\partial S}{\partial z^{k}}(z).
\end{equation}
We can use this result in order to calculate the perpendicular flux of particles. We obtain the following expression for the first order term:
\begin{equation*}
\left \langle \hat{V}^{i}_{\perp}(\hat{z}) \hat{F}_{1}(\hat{z})\right \rangle= \int \mathit{d} \hat{v} \hat{V}^{i}_{\perp}(\hat{z})\left\{e^{-{\bm \rho }\cdot {\bm \nabla}}\left[\bar{F}_{1}- \frac{e}{m} \frac{\partial \bar{F}_{0}}{\partial \mathcal{E}}(\bar{z}_{f}(\hat{z})) \gav{e^{{\bm \rho }\cdot {\bm \nabla}}\left(  \delta \hat{\phi} - \frac{1}{c} \hat{V}^{i}_{\parallel}\delta \hat{A}_{i}\right)(\hat{z})} \right]\right\}
\end{equation*}
where we have made the assumption that $\hat{F}_{0}$ is an even function of the velocity and that $\frac{\partial \bar{F}_{0}}{\partial \mu}=0$. Summing the zeroth-order term with the previous expression and substituting Eq. (\ref{eq:apppush5a}) we obtain:
\begin{align}
\label{eq:apppush8}
\left \langle \hat{V}^{i}_{\perp}(\hat{z}) \hat{F}_{1}(\hat{z})\right \rangle_{\theta} = - 2 \pi &\int \mathit{d} U \mathit{d} \mu J(z) \frac{mc}{e} \mu \hat{I}_{1} \epsilon^{ijk}b_{k} \times \\ \nonumber
\times& \frac{\partial}{\partial z^{k}}\left[ \bar{F}(\hat{z}_{f}(z)) - \frac{e}{mc} \frac{\partial \bar{F}}{\partial \mathcal{E}}(\bar{z}_{f}(\hat{z}_{f}(z))) \gav{e^{{\bm \rho }\cdot {\bm \nabla}}\left(\delta \hat{\phi}- \frac{1}{c}\hat{V}^{i}_{\parallel}\delta \hat{A}_{\parallel i} \right)(\hat{z}_{f}(z))}\right].
\end{align}
This formula combined with the expression for the parallel flux of particles which can be obtained obtained from Eq. (\ref{eq:apppush4}) gives the expression for the flux of particles used in the thesis.
\chapter{Derivation of the fluctuation induced particle flux}
\label{cap:appC}
In this appendix we calculate the flux of particles across magnetic surfaces induced by fluctuations. This term, together with the flux induced by collisions, determines the evolutive equation for the surface-averaged density.\vspace*{1\baselineskip}

We need to evaluate the following expression:
\begin{equation*}
\left\langle (e/c) \left\langle \bm v f \right\rangle_v \cdot R^2 \bm \nabla \phi \times  (\bm \nabla \times \delta \bm A)  \right\rangle_\psi.
\end{equation*}
We can write this expression in term of the gyrocenter distribution function using the expression for the perpendicular flux of particles derived in the previous appendix, i.e. Eq. (\ref{eq:apppush8}), obtaining:
\begin{align*}
  &\sav{\frac{e}{c} \vav{\hat{I}_{0}\left[  \bar{F} - \frac{e}{m} \gav{\delta \psi_{gc}}\frac{\partial \bar{F}}{\partial \mathcal{E}}\right]v_{\parallel} \wb{b}}\cdot R^{2}\wb{\nabla}\phi\times \left(\wb{\nabla}\times\delta \wb{A}  \right)} +\\
+  & \sav{\vav{ m \mu \hat{I}_{1} \wb{b}\times \wb{\nabla} \delta \bar{G}}\cdot R^{2} \wb{\nabla}\times
(\delta B_{\parallel}\wb{b} + \delta \wb{B}_{\perp})}.
\end{align*}
Using the following identity:
\begin{equation*}
\wb{b}\cdot \wb{\nabla}\times \delta\wb{B}= B_{0}^{-1}R^{-2}\wb{\nabla}\psi\cdot \delta \wb{B}
\end{equation*}
we can re-write the first term as:
\begin{equation*}
\sav{\vav{\frac{e v_{\parallel}}{c}\hat{I}_{0}\delta \bar{G}}\frac{\wb{\nabla}\psi\cdot \delta\wb{B}}{B_{0}}}.
\end{equation*}
From the drift ordering we know that at the leading order $\delta B_{\perp}=\wb{\nabla}\delta A_{\parallel} \times \wb{b}$ and, therefore, we can write:
\begin{equation*}
\sav{\vav{\frac{e v_{\parallel}}{c}\hat{I}_{0}\delta \bar{G}} B_{0}^{-1}\wb{\nabla}\psi\cdot(\wb{\nabla}\delta A_{\parallel}\times \wb{B}) }.
\end{equation*}
Using the identity $\wb{b}\times \wb{\nabla}\psi= F \wb{b} - R^{2}B \wb{\nabla}\phi$ we finally obtain:
\begin{equation*}
-e \sav{\vav{\wb{\nabla}\left( \frac{\delta A_{\parallel}v_{\parallel}}{c} \right)\hat{I}_{0} \delta \bar{G}}\cdot R^{2} \wb{\nabla}\phi}.
\end{equation*}
The second term to calculate is the following:
\begin{equation}
\label{eq:appdc3a}
\sav{\vav{ m \mu \hat{I}_{1} \wb{b}\times \wb{\nabla} \delta \bar{G}}\cdot R^{2} \wb{\nabla}\times
(\delta B_{\parallel}\wb{b} + \delta \wb{B}_{\perp})}
\end{equation}
which is the sum of two contributions. We can show that:
\begin{equation*}
  \wb{b}\times \wb{\nabla}\delta \bar{G}\cdot R^{2} \wb{\nabla}\phi \times \delta \wb{B}_{\perp}= \frac{F}{B_{0}}\delta \wb{B}_{\perp}- \wb{\nabla}_{\perp}\delta \bar{G}
\end{equation*}
and, therefore, we can re-write the second term of Eq. (\ref{eq:appdc3a}):
\begin{equation}
\label{eq:appdc3}
\sav{\vav{m \mu \hat{I}_{1} \frac{F}{B_{0}} \delta \wb{B}_{\perp}\cdot \wb{\nabla}_{\perp}\delta \bar{G}}}.
\end{equation}
Analogously we can show that $(\wb{b}\times \wb{\nabla} \delta\bar{G})\cdot (\wb{\nabla}\phi \times \wb{b}) = - \wb{\nabla}\phi \cdot \wb{\nabla}_{\perp}\delta \bar{G}$ and we can re-write the first term of Eq. (\ref{eq:appdc3a}) as:
\begin{equation}
\label{eq:appdc5}
-m \sav{\vav{\left(\delta B_{\parallel}R^{2}\wb{\nabla}\phi- \frac{F}{B_{0}}\delta\wb{B}_{\perp}\right)\cdot \mu \hat{I}_{1}\wb{\nabla}_{\perp}\delta \bar{G}}}.
\end{equation}
This is again the sum of two terms. The first one:
\begin{equation*}
-m \sav{\vav{\delta B_{\parallel}R^{2}\wb{\nabla}\phi \cdot \mu \hat{I}_{1}\wb{\nabla}_{\perp}\delta \bar{G}}}
\end{equation*}
can be written at the leading order as:
\begin{equation*}
-m \sav{\vav{\delta B_{\parallel}R  \mu \hat{I}_{1} \frac{\partial \delta \bar{G}}{\delta \phi}}}.
\end{equation*}
This can be written, remembering that the surface average involves an average over the angular coordinate $\phi$, as:
\begin{equation}
\label{eq:appdc5}
m \sav{ \vav{ R^{2} \wb{\nabla}\phi \cdot \wb{\nabla}(\delta B_{\parallel}\mu \hat{I}_{1}\delta \bar{G})}}.
\end{equation}
It can be demonstrated that the second term of Eq. (\ref{eq:appdc5}) is of higher order with respect to the others.

\newpage
\phantomsection
\bibliographystyle{my_style2}
 \addcontentsline{toc}{chapter}{Bibliography}
 \footnotesize
 \bibliography{/home/falematte/Dropbox/updatedbib.bib}
\end{document}

%% file: TeX_style.tex
\usepackage[default]{gfsbodoni}

\usepackage{amsmath}
\usepackage{amsfonts}
\usepackage{amssymb}
\usepackage{bbm}
\usepackage{bm}
\usepackage{nicefrac}
\usepackage{amssymb}
\usepackage{amsthm}
\usepackage{multirow}
\usepackage{float}
\usepackage{graphicx}
\usepackage{cite}
\usepackage[top=2.3cm,bottom=2.3cm,left=2.3cm,right=2.3cm]{geometry}
\DeclareGraphicsExtensions{.pdf,.png,.jpg,.mps}
 \usepackage{quotchap} 

\usepackage{multicol} 
\usepackage{ifthen}
\usepackage{bbold}
\usepackage[tight,italian]{minitoc}
\usepackage{amsfonts}
\usepackage{chemarrow}
\usepackage{listings}
\usepackage{slashed}
\usepackage{rotate}
\usepackage{color, colortbl}
\usepackage[table]{xcolor}
\usepackage{lscape}
\usepackage{mathrsfs}
\usepackage[vcentermath]{youngtab}
\usepackage{lettrine}
\usepackage{newlfont}
\usepackage[toc]{appendix}
\usepackage{sidecap}
\usepackage{booktabs} 
\usepackage[small,bf]{caption} 
\usepackage{pgfplots}
\pgfplotsset{compat=1.5}
\usepackage{tikz}
\usetikzlibrary{calc,matrix,backgrounds,fit}
\pgfdeclarelayer{myback}
\pgfsetlayers{myback,background,main}

\usepackage{hyperref}

\usepackage{fancyheadings}
\fancypagestyle{plain}{%
\fancyhf{} 
\fancyfoot[C]{--~\thepage~--} 

}

%% file: phdfalessiarxiv.bbl
\begin{thebibliography}{100}
\providecommand{\url}[1]{\texttt{#1}}
\providecommand{\urlprefix}{URL }
\expandafter\ifx\csname urlstyle\endcsname\relax
  \providecommand{\doi}[1]{doi:\discretionary{}{}{}#1}\else
  \providecommand{\doi}{doi:\discretionary{}{}{}\begingroup
  \urlstyle{rm}\Url}\fi
\providecommand{\eprint}[2][]{\url{#2}}

\bibitem{chen2016physics}
\textsc{L.~Chen} and \textsc{F.~Zonca}.
\newblock \emph{Physics of Alfv{\'e}n waves and energetic particles in burning
  plasmas}.
\newblock Reviews of Modern Physics \textbf{88}(1), 015008, (2016).

\bibitem{frieman1982nonlinear}
\textsc{E.~Frieman} and \textsc{L.~Chen}.
\newblock \emph{Nonlinear gyrokinetic equations for low-frequency
  electromagnetic waves in general plasma equilibria}.
\newblock Physics of Fluids (1958-1988) \textbf{25}(3), 502--508, (1982).

\bibitem{hinton1976theory}
\textsc{F.~Hinton} and \textsc{R.~Hazeltine}.
\newblock \emph{Theory of plasma transport in toroidal confinement systems}.
\newblock Reviews of Modern Physics \textbf{48}(2), 239, (1976).

\bibitem{abel2013multiscale}
\textsc{I.~Abel}, \textsc{G.~Plunk}, \textsc{E.~Wang}, \textsc{M.~Barnes},
  \textsc{S.~Cowley}, \textsc{W.~Dorland} and \textsc{A.~Schekochihin}.
\newblock \emph{Multiscale gyrokinetics for rotating tokamak plasmas:
  fluctuations, transport and energy flows}.
\newblock Reports on Progress in Physics \textbf{76}(11), 116201, (2013).

\bibitem{plunk2009theory}
\textsc{G.~G. Plunk}.
\newblock \emph{The theory of gyrokinetic turbulence: A multiple-scales
  approach},  (ProQuest2009).

\bibitem{sugama1996transport}
\textsc{H.~Sugama}, \textsc{M.~Okamoto}, \textsc{W.~Horton} and
  \textsc{M.~Wakatani}.
\newblock \emph{Transport processes and entropy production in toroidal plasmas
  with gyrokinetic electromagnetic turbulence}.
\newblock Physics of Plasmas (1994-present) \textbf{3}(6), 2379--2394, (1996).

\bibitem{balescu1990anomalous}
\textsc{R.~Balescu}.
\newblock \emph{Anomalous fluxes in the plateau regime for a weakly turbulent,
  magnetically confined plasma}.
\newblock Physics of Fluids B: Plasma Physics (1989-1993) \textbf{2}(9),
  2100--2112, (1990).

\bibitem{shaing1988neoclassical}
\textsc{K.-C. Shaing}.
\newblock \emph{Neoclassical quasilinear transport theory of fluctuations in
  toroidal plasmas}.
\newblock Physics of Fluids (1958-1988) \textbf{31}(8), 2249--2265, (1988).

\bibitem{balescu1988transport}
\textsc{R.~Balescu}.
\newblock \emph{Transport processes in plasmas}  (1988).

\bibitem{sonnino2008nonlinear}
\textsc{G.~Sonnino} and \textsc{P.~Peeters}.
\newblock \emph{Nonlinear transport processes in tokamak plasmas. I. The
  collisional regimes}.
\newblock Physics of Plasmas (1994-present) \textbf{15}(6), 062309, (2008).

\bibitem{brizard2007foundations}
\textsc{A.~Brizard} and \textsc{T.~Hahm}.
\newblock \emph{Foundations of nonlinear gyrokinetic theory}.
\newblock Reviews of modern physics \textbf{79}(2), 421, (2007).

\bibitem{chen2007theory}
\textsc{L.~Chen} and \textsc{F.~Zonca}.
\newblock \emph{Theory of Alfv{\'e}n waves and energetic particle physics in
  burning plasmas}.
\newblock Nuclear Fusion \textbf{47}(10), S727, (2007).

\bibitem{zonca2015nonlinear}
\textsc{F.~Zonca}, \textsc{L.~Chen}, \textsc{S.~Briguglio},
  \textsc{G.~Fogaccia}, \textsc{G.~Vlad} and \textsc{X.~Wang}.
\newblock \emph{Nonlinear dynamics of phase space zonal structures and
  energetic particle physics in fusion plasmas}.
\newblock New Journal of Physics \textbf{17}(1), 013052, (2015).

\bibitem{zonca2014energetic}
\textsc{F.~Zonca}, \textsc{L.~Chen}, \textsc{S.~Briguglio},
  \textsc{G.~Fogaccia}, \textsc{A.~V. Milovanov}, \textsc{Z.~Qiu},
  \textsc{G.~Vlad} and \textsc{X.~Wang}.
\newblock \emph{Energetic particles and multi-scale dynamics in fusion
  plasmas}.
\newblock Plasma Physics and Controlled Fusion \textbf{57}(1), 014024, (2014).

\bibitem{barnes2009trinity}
\textsc{M.~A. Barnes}.
\newblock \emph{Trinity: A unified treatment of turbulence, transport, and
  heating in magnetized plasmas},  (ProQuest2009).

\bibitem{sonnino2009nonlinear}
\textsc{G.~Sonnino}.
\newblock \emph{Nonlinear closure relations theory for transport processes in
  nonequilibrium systems}.
\newblock Physical Review E \textbf{79}(5), 051126, (2009).

\bibitem{pizzuto2016dtt}
\textsc{A.~Pizzuto}.
\newblock \emph{DTT Divertor Tokamak Test facility Project Proposal}  (ISBN:
  978-88-8286-318-0).

\bibitem{iter91}
\textsc{K.~Tomabechi}, \textsc{J.~Gilleland}, \textsc{Y.~A. Sokolov},
  \textsc{R.~Toschi}, \textsc{I.~Team} \emph{et~al.}
\newblock \emph{ITER conceptual designWork performed under the auspices of the
  IAEA.}
\newblock Nuclear Fusion \textbf{31}(6), 1135, (1991).

\bibitem{aymar97}
\textsc{R.~AYMAR}, \textsc{V.~CHUYANOV} and \textsc{M.~HUGUET}.
\newblock \emph{R. PARKER, Y. SHIMOMURA and the ITER JOINT CENTRAL TEAM and
  HOME TEAMS}.
\newblock In \emph{Fusion Energy 1996: Proceedings of the Sixteenth
  International Conference on Fusion Energy}, volume~1, p.~1,  (International
  Atomic Energy Agency1997).

\bibitem{chapman11}
\textsc{I.~Chapman}, \textsc{R.~Kemp} and \textsc{D.~Ward}.
\newblock \emph{Analysis of high $\beta$ regimes for DEMO}.
\newblock Fusion Engineering and Design \textbf{86}(2), 141--150, (2011).

\bibitem{nicholson1983introduction}
\textsc{D.~R. Nicholson} and \textsc{D.~R. Nicholson}.
\newblock \emph{Introduction to plasma theory},  (Cambridge Univ Press1983).

\bibitem{bogoliubov1962studies}
\textsc{N.~Bogoliubov} \emph{et~al.}
\newblock \emph{Studies in Statistical Mechanics}.
\newblock Studies in Statistical Mechanics I  (1962).

\bibitem{balescu1963statistical}
\textsc{R.~Balescu}.
\newblock \emph{Statistical mechanics of charged particles}, volume~4,
  (Interscience New York1963).

\bibitem{landau1965collected}
\textsc{L.~D. Landau}.
\newblock \emph{Collected papers of LD Landau},  (Intl Pub Distributor
  Inc1965).

\bibitem{alexandre2004landau}
\textsc{R.~Alexandre} and \textsc{C.~Villani}.
\newblock \emph{On the Landau approximation in plasma physics}.
\newblock In \emph{Annales de l'IHP Analyse non lin{\'e}aire}, volume~21, pp.
  61--95 (2004).

\bibitem{klimontovich2012statistical}
\textsc{Y.~L. Klimontovich}.
\newblock \emph{Statistical Theory of Open Systems: Volume 1: A Unified
  Approach to Kinetic Description of Processes in Active Systems}, volume~67,
  (Springer Science \& Business Media2012).

\bibitem{spohn2012large}
\textsc{H.~Spohn}.
\newblock \emph{Large scale dynamics of interacting particles},  (Springer
  Science \& Business Media2012).

\bibitem{atkinson1929frage}
\textsc{R.~d.~E. Atkinson} and \textsc{F.~G. Houtermans}.
\newblock \emph{Zur Frage der Aufbaum{\"o}glichkeit der Elemente in Sternen}.
\newblock Zeitschrift f{\"u}r Physik \textbf{54}(9-10), 656--665, (1929).

\bibitem{von1938uber}
\textsc{C.~F. Von~Weizs{\"a}cker}.
\newblock \emph{Uber Elementumwandlungen in Innern der Sterne. II}.
\newblock Physikalische Zeitschrift \textbf{39}, 633, (1938).

\bibitem{wesson2011tokamaks}
\textsc{J.~Wesson} and \textsc{D.~J. Campbell}.
\newblock \emph{Tokamaks}, volume 149,  (Oxford University Press2011).

\bibitem{lawson1957some}
\textsc{J.~D. Lawson}.
\newblock \emph{Some criteria for a power producing thermonuclear reactor}.
\newblock Proceedings of the Physical Society. Section B \textbf{70}(1), 6,
  (1957).

\bibitem{freidberg1982ideal}
\textsc{J.~P. Freidberg}.
\newblock \emph{Ideal magnetohydrodynamic theory of magnetic fusion systems}.
\newblock Reviews of Modern Physics \textbf{54}(3), 801, (1982).

\bibitem{boozer2005physics}
\textsc{A.~H. Boozer}.
\newblock \emph{Physics of magnetically confined plasmas}.
\newblock Reviews of modern physics \textbf{76}(4), 1071, (2005).

\bibitem{helander2005collisional}
\textsc{P.~Helander} and \textsc{D.~J. Sigmar}.
\newblock \emph{Collisional transport in magnetized plasmas}, volume~4,
  (Cambridge University Press2005).

\bibitem{yushmanov1990scalings}
\textsc{P.~Yushmanov}, \textsc{T.~Takizuka}, \textsc{K.~Riedel},
  \textsc{O.~Kardaun}, \textsc{J.~Cordey}, \textsc{S.~Kaye} and
  \textsc{D.~Post}.
\newblock \emph{Scalings for tokamak energy confinement}.
\newblock Nuclear Fusion \textbf{30}(10), 1999, (1990).

\bibitem{koide1994internal}
\textsc{Y.~Koide}, \textsc{M.~Kikuchi}, \textsc{M.~Mori}, \textsc{S.~Tsuji},
  \textsc{S.~Ishida}, \textsc{N.~Asakura}, \textsc{Y.~Kamada},
  \textsc{T.~Nishitani}, \textsc{Y.~Kawano}, \textsc{T.~Hatae} \emph{et~al.}
\newblock \emph{Internal transport barrier on q= 3 surface and poloidal plasma
  spin up in JT-60U high-$\beta$ p discharges}.
\newblock Physical review letters \textbf{72}(23), 3662, (1994).

\bibitem{borgogno2011barriers}
\textsc{D.~Borgogno}, \textsc{D.~Grasso}, \textsc{F.~Pegoraro} and
  \textsc{T.~Schep}.
\newblock \emph{Barriers in the transition to global chaos in collisionless
  magnetic reconnection. I. Ridges of the finite time Lyapunov exponent field}.
\newblock Physics of Plasmas (1994-present) \textbf{18}(10), 102307, (2011).

\bibitem{falessi2015lagrangian}
\textsc{M.~Falessi}, \textsc{F.~Pegoraro} and \textsc{T.~Schep}.
\newblock \emph{Lagrangian coherent structures and plasma transport processes}.
\newblock Journal of Plasma Physics \textbf{81}(05), 495810505, (2015).

\bibitem{d2012flux}
\textsc{W.~D. D'haeseleer}, \textsc{W.~N. Hitchon}, \textsc{J.~D. Callen} and
  \textsc{J.~L. Shohet}.
\newblock \emph{Flux coordinates and magnetic field structure: a guide to a
  fundamental tool of plasma theory},  (Springer Science \& Business
  Media2012).

\bibitem{boozer1981plasma}
\textsc{A.~H. Boozer}.
\newblock \emph{Plasma equilibrium with rational magnetic surfaces}.
\newblock Physics of Fluids (1958-1988) \textbf{24}(11), 1999--2003, (1981).

\bibitem{boozer1982establishment}
\textsc{A.~H. Boozer}.
\newblock \emph{Establishment of magnetic coordinates for a given magnetic
  field}.
\newblock Physics of Fluids \textbf{25}(3), 520, (1982).

\bibitem{hamada1962hydromagnetic}
\textsc{S.~Hamada}.
\newblock \emph{Hydromagnetic equilibria and their proper coordinates}.
\newblock Nuclear Fusion \textbf{2}(1-2), 23, (1962).

\bibitem{hazeltine2003plasma}
\textsc{R.~D. Hazeltine} and \textsc{J.~D. Meiss}.
\newblock \emph{Plasma confinement},  (Courier Corporation2003).

\bibitem{chew1956boltzmann}
\textsc{G.~Chew}, \textsc{M.~Goldberger} and \textsc{F.~Low}.
\newblock \emph{The Boltzmann equation and the one-fluid hydromagnetic
  equations in the absence of particle collisions}.
\newblock In \emph{Proceedings of the Royal Society of London A: Mathematical,
  Physical and Engineering Sciences}, volume 236, pp. 112--118,  (The Royal
  Society1956).

\bibitem{tronko2016second}
\textsc{N.~Tronko}, \textsc{A.~Bottino} and \textsc{E.~Sonnendruecker}.
\newblock \emph{Second order Gyrokinetic theory for Particle-In-Cell codes}.
\newblock arXiv preprint arXiv:1604.03538  (2016).

\bibitem{wimmel1970energy}
\textsc{H.~K. Wimmel}.
\newblock \emph{Energy-balance equation and enhanced collisional plasma
  diffusion}.
\newblock Nuclear Fusion \textbf{10}(2), 117, (1970).

\bibitem{zonca2006physics}
\textsc{F.~Zonca}, \textsc{S.~Briguglio}, \textsc{L.~Chen},
  \textsc{G.~Fogaccia}, \textsc{T.~Hahm}, \textsc{A.~Milovanov} and
  \textsc{G.~Vlad}.
\newblock \emph{Physics of burning plasmas in toroidal magnetic confinement
  devices}.
\newblock Plasma physics and controlled fusion \textbf{48}(12B), B15, (2006).

\bibitem{hasegawa1979nonlinear}
\textsc{A.~Hasegawa}, \textsc{C.~G. Maclennan} and \textsc{Y.~Kodama}.
\newblock \emph{Nonlinear behavior and turbulence spectra of drift waves and
  Rossby waves}.
\newblock Physics of Fluids (1958-1988) \textbf{22}(11), 2122--2129, (1979).

\bibitem{lin1998turbulent}
\textsc{Z.~Lin}, \textsc{T.~S. Hahm}, \textsc{W.~Lee}, \textsc{W.~M. Tang} and
  \textsc{R.~B. White}.
\newblock \emph{Turbulent transport reduction by zonal flows: Massively
  parallel simulations}.
\newblock Science \textbf{281}(5384), 1835--1837, (1998).

\bibitem{rosenbluth1998poloidal}
\textsc{M.~Rosenbluth} and \textsc{F.~Hinton}.
\newblock \emph{Poloidal flow driven by ion-temperature-gradient turbulence in
  tokamaks}.
\newblock Physical review letters \textbf{80}(4), 724, (1998).

\bibitem{hinton1999dynamics}
\textsc{F.~Hinton} and \textsc{M.~Rosenbluth}.
\newblock \emph{Dynamics of axisymmetric and poloidal flows in tokamaks}.
\newblock Plasma physics and controlled fusion \textbf{41}(3A), A653, (1999).

\bibitem{diamond2005zonal}
\textsc{P.~Diamond}, \textsc{S.~Itoh}, \textsc{K.~Itoh} and \textsc{T.~Hahm}.
\newblock \emph{Zonal flows in plasma—a review}.
\newblock Plasma Physics and Controlled Fusion \textbf{47}(5), R35, (2005).

\bibitem{itoh2006physics}
\textsc{K.~Itoh}, \textsc{S.-I. Itoh}, \textsc{P.~Diamond}, \textsc{T.~Hahm},
  \textsc{A.~Fujisawa}, \textsc{G.~Tynan}, \textsc{M.~Yagi} and
  \textsc{Y.~Nagashima}.
\newblock \emph{Physics of zonal flowsa)}.
\newblock Physics of Plasmas (1994-present) \textbf{13}(5), 055502, (2006).

\bibitem{antonsen1980kinetic}
\textsc{T.~M. Antonsen~Jr} and \textsc{B.~Lane}.
\newblock \emph{Kinetic equations for low frequency instabilities in
  inhomogeneous plasmas}.
\newblock Physics of Fluids (1958-1988) \textbf{23}(6), 1205--1214, (1980).

\bibitem{catto1981generalized}
\textsc{P.~Catto}, \textsc{W.~Tang} and \textsc{D.~Baldwin}.
\newblock \emph{Generalized gyrokinetics}.
\newblock Plasma Physics \textbf{23}(7), 639, (1981).

\bibitem{brizard1992nonlinear}
\textsc{A.~Brizard}.
\newblock \emph{Nonlinear gyrofluid description of turbulent magnetized
  plasmas}.
\newblock Physics of Fluids B: Plasma Physics (1989-1993) \textbf{4}(5),
  1213--1228, (1992).

\bibitem{chen2007nonlinear1}
\textsc{L.~Chen} and \textsc{F.~Zonca}.
\newblock \emph{Nonlinear equilibria, stability and generation of zonal
  structures in toroidal plasmas}.
\newblock Nuclear Fusion \textbf{47}(8), 886, (2007).

\bibitem{romanelli2013roadmap}
\textsc{F.~Romanelli}, \textsc{L.~H. Federici}, \textsc{R.~Neu},
  \textsc{D.~Stork} and \textsc{H.~Zohm}.
\newblock \emph{A roadmap to the realization of fusion energy}.
\newblock In \emph{Proc. IEEE 25th Symp. Fusion Eng}, pp. 1--4 (2013).

\bibitem{ihli2007recent}
\textsc{T.~Ihli}, \textsc{L.~Boccaccini}, \textsc{G.~Janeschitz},
  \textsc{C.~Koehly}, \textsc{D.~Maisonnier}, \textsc{D.~Nagy},
  \textsc{C.~Polixa}, \textsc{J.~Rey} and \textsc{P.~Sardain}.
\newblock \emph{Recent progress in DEMO fusion core engineering: improved
  segmentation, maintenance and blanket concepts}.
\newblock Fusion Engineering and Design \textbf{82}(15), 2705--2712, (2007).

\bibitem{garbet2010gyrokinetic}
\textsc{X.~Garbet}, \textsc{Y.~Idomura}, \textsc{L.~Villard} and
  \textsc{T.~Watanabe}.
\newblock \emph{Gyrokinetic simulations of turbulent transport}.
\newblock Nuclear Fusion \textbf{50}(4), 043002, (2010).

\bibitem{rutherford1970collisional}
\textsc{P.~H. Rutherford}.
\newblock \emph{Collisional diffusion in an axisymmetric torus}.
\newblock Physics of Fluids (1958-1988) \textbf{13}(2), 482--489, (1970).

\bibitem{spitzer1953transport}
\textsc{L.~Spitzer~Jr} and \textsc{R.~H{\"a}rm}.
\newblock \emph{Transport phenomena in a completely ionized gas}.
\newblock Physical Review \textbf{89}(5), 977, (1953).

\bibitem{angioni2000neoclassical}
\textsc{C.~Angioni} and \textsc{O.~Sauter}.
\newblock \emph{Neoclassical transport coefficients for general axisymmetric
  equilibria in the banana regime}.
\newblock Physics of Plasmas (1994-present) \textbf{7}(4), 1224--1234, (2000).

\bibitem{harvey1992cql3d}
\textsc{R.~Harvey} and \textsc{M.~McCoy}.
\newblock \emph{The cql3d fokker-planck code}.
\newblock Advances in Simulation and Modeling of Thermonuclear Plasmas p. 498
  (1992).

\bibitem{lin1995upper}
\textsc{Y.~Lin-Liu} and \textsc{R.~Miller}.
\newblock \emph{Upper and lower bounds of the effective trapped particle
  fraction in general tokamak equilibria}.
\newblock Physics of Plasmas (1994-present) \textbf{2}(5), 1666--1668, (1995).

\bibitem{krommes2002fundamental}
\textsc{J.~A. Krommes}.
\newblock \emph{Fundamental statistical descriptions of plasma turbulence in
  magnetic fields}.
\newblock Physics Reports \textbf{360}(1), 1--352, (2002).

\bibitem{kallenbach2013impurity}
\textsc{A.~Kallenbach}, \textsc{M.~Bernert}, \textsc{R.~Dux},
  \textsc{L.~Casali}, \textsc{T.~Eich}, \textsc{L.~Giannone},
  \textsc{A.~Herrmann}, \textsc{R.~McDermott}, \textsc{A.~Mlynek},
  \textsc{H.~M{\"u}ller} \emph{et~al.}
\newblock \emph{Impurity seeding for tokamak power exhaust: from present
  devices via ITER to DEMO}.
\newblock Plasma Physics and Controlled Fusion \textbf{55}(12), 124041, (2013).

\bibitem{janeschitz2001plasma}
\textsc{G.~Janeschitz}, \textsc{I.~JCT} \emph{et~al.}
\newblock \emph{Plasma--wall interaction issues in ITER}.
\newblock Journal of Nuclear Materials \textbf{290}, 1--11, (2001).

\bibitem{chang2004numerical}
\textsc{C.~Chang}, \textsc{S.~Ku} and \textsc{H.~Weitzner}.
\newblock \emph{Numerical study of neoclassical plasma pedestal in a tokamak
  geometry}.
\newblock Physics of Plasmas (1994-present) \textbf{11}(5), 2649--2667, (2004).

\bibitem{gravesinternal}
\textsc{J.~Graves}.
\newblock \emph{Kinetic stabilisation of the Internal Kink Mode for Fusion
  Plasmas}.
\newblock Ph.D. thesis, The University of Nottingham (1999).

\bibitem{graves2012control}
\textsc{J.~Graves}, \textsc{I.~Chapman}, \textsc{S.~Coda},
  \textsc{M.~Lennholm}, \textsc{M.~Albergante} and \textsc{M.~Jucker}.
\newblock \emph{Control of magnetohydrodynamic stability by phase space
  engineering of energetic ions in tokamak plasmas}.
\newblock Nature communications \textbf{3}, 624, (2012).

\bibitem{xu2007edge}
\textsc{X.~Xu}, \textsc{Z.~Xiong}, \textsc{M.~Dorr}, \textsc{J.~Hittinger},
  \textsc{K.~Bodi}, \textsc{J.~Candy}, \textsc{B.~Cohen}, \textsc{R.~Cohen},
  \textsc{P.~Colella}, \textsc{G.~Kerbel} \emph{et~al.}
\newblock \emph{Edge gyrokinetic theory and continuum simulations}.
\newblock Nuclear fusion \textbf{47}(8), 809, (2007).

\bibitem{parra2011phase}
\textsc{F.~I. Parra} and \textsc{I.~Calvo}.
\newblock \emph{Phase-space Lagrangian derivation of electrostatic gyrokinetics
  in general geometry}.
\newblock Plasma Physics and Controlled Fusion \textbf{53}(4), 045001, (2011).

\bibitem{cary1983noncanonical}
\textsc{J.~R. Cary} and \textsc{R.~G. Littlejohn}.
\newblock \emph{Noncanonical Hamiltonian mechanics and its application to
  magnetic field line flow}.
\newblock Annals of Physics \textbf{151}(1), 1--34, (1983).

\bibitem{onsager1931reciprocal}
\textsc{L.~Onsager}.
\newblock \emph{Reciprocal relations in irreversible processes. I.}
\newblock Physical review \textbf{37}(4), 405, (1931).

\bibitem{sonnino2007geometrical}
\textsc{G.~Sonnino} and \textsc{J.~Evslin}.
\newblock \emph{Geometrical thermodynamic field theory}.
\newblock International journal of quantum chemistry \textbf{107}(4), 968--987,
  (2007).

\bibitem{sonnino2007minimum}
\textsc{G.~Sonnino} and \textsc{J.~Evslin}.
\newblock \emph{The minimum rate of dissipation principle}.
\newblock Physics Letters A \textbf{365}(5), 364--369, (2007).

\bibitem{sonnino2014thermodynamic}
\textsc{G.~Sonnino} and \textsc{A.~Sonnino}.
\newblock \emph{The thermodynamic covariance principle}.
\newblock arXiv preprint arXiv:1403.0370  (2014).

\bibitem{PhysRevE.94.042103}
\textsc{G.~Sonnino}, \textsc{J.~Evslin}, \textsc{A.~Sonnino},
  \textsc{G.~Steinbrecher} and \textsc{E.~Tirapegui}.
\newblock \emph{Symmetry group and group representations associated with the
  thermodynamic covariance principle}.
\newblock Phys. Rev. E \textbf{94}, 042103, (2016).

\bibitem{chileproceeding}
\textsc{G.~Sonnino}, \textsc{M.~V. Falessi} and \textsc{F.~Zonca}.
\newblock \emph{Symmetry group and group representations associated with the
  thermodynamic covariance principle}.
\newblock Proceedings of 15th International Workshop on Instabilities and
  Non-Equilibrium Structures  (submitted).

\bibitem{glansdorff1973thermodynamic}
\textsc{P.~Glansdorff}, \textsc{I.~Prigogine} and \textsc{R.~N. Hill}.
\newblock \emph{Thermodynamic theory of structure, stability and fluctuations}.
\newblock American Journal of Physics \textbf{41}(1), 147--148, (1973).

\bibitem{catto2009extension}
\textsc{P.~J. Catto}.
\newblock \emph{Extension of gyrokinetics to transport time scales}.
\newblock Ph.D. thesis, Massachusetts Institute of Technology (2009).

\bibitem{krommes2009report}
\textsc{J.~Krommes} and \textsc{G.~Hammett}.
\newblock \emph{Report of the study group gk2 on momentum transport in
  gyrokinetics}.
\newblock PPPL Report PPPL-4945 (Princeton University, 2013)  (2009).

\bibitem{brizard2009orbit}
\textsc{A.~Brizard}, \textsc{J.~Decker}, \textsc{Y.~Peysson} and \textsc{F.-X.
  Duthoit}.
\newblock \emph{Orbit-averaged guiding-center Fokker--Planck operator}.
\newblock Physics of Plasmas (1994-present) \textbf{16}(10), 102304, (2009).

\bibitem{brizard2004guiding}
\textsc{A.~J. Brizard}.
\newblock \emph{A guiding-center Fokker--Planck collision operator for
  nonuniform magnetic fields}.
\newblock Physics of Plasmas (1994-present) \textbf{11}(9), 4429--4438, (2004).

\bibitem{burby2015energetically}
\textsc{J.~Burby}, \textsc{A.~Brizard} and \textsc{H.~Qin}.
\newblock \emph{Energetically consistent collisional gyrokinetics}.
\newblock Physics of Plasmas (1994-present) \textbf{22}(10), 100707, (2015).

\bibitem{idomura2003global}
\textsc{Y.~Idomura}, \textsc{S.~Tokuda} and \textsc{Y.~Kishimoto}.
\newblock \emph{Global gyrokinetic simulation of ion temperature gradient
  driven turbulence in plasmas using a canonical Maxwellian distribution}.
\newblock Nuclear Fusion \textbf{43}(4), 234, (2003).

\bibitem{grandgirard2007global}
\textsc{V.~Grandgirard}, \textsc{Y.~Sarazin}, \textsc{P.~Angelino},
  \textsc{A.~Bottino}, \textsc{N.~Crouseilles}, \textsc{G.~Darmet},
  \textsc{G.~Dif-Pradalier}, \textsc{X.~Garbet}, \textsc{P.~Ghendrih},
  \textsc{S.~Jolliet} \emph{et~al.}
\newblock \emph{Global full-f gyrokinetic simulations of plasma turbulence}.
\newblock Plasma Physics and Controlled Fusion \textbf{49}(12B), B173, (2007).

\bibitem{dimits1996scalings}
\textsc{A.~Dimits}, \textsc{T.~Williams}, \textsc{J.~Byers} and
  \textsc{B.~Cohen}.
\newblock \emph{Scalings of ion-temperature-gradient-driven anomalous transport
  in tokamaks}.
\newblock Physical review letters \textbf{77}(1), 71, (1996).

\bibitem{wang2007nonlocal}
\textsc{W.~Wang}, \textsc{T.~Hahm}, \textsc{W.~Lee}, \textsc{G.~Rewoldt},
  \textsc{J.~Manickam} and \textsc{W.~Tang}.
\newblock \emph{Nonlocal properties of gyrokinetic turbulence and the role of
  E$\times$ B flow shear}.
\newblock Physics of Plasmas (1994-present) \textbf{14}(7), 072306, (2007).

\bibitem{parker1993fully}
\textsc{S.~Parker} and \textsc{W.~Lee}.
\newblock \emph{A fully nonlinear characteristic method for gyrokinetic
  simulation}.
\newblock Physics of Fluids B: Plasma Physics (1989-1993) \textbf{5}(1),
  77--86, (1993).

\bibitem{sydora1996fluctuation}
\textsc{R.~Sydora}, \textsc{V.~Decyk} and \textsc{J.~Dawson}.
\newblock \emph{Fluctuation-induced heat transport results from a large global
  3D toroidal particle simulation model}.
\newblock Plasma Physics and Controlled Fusion \textbf{38}(12A), A281, (1996).

\bibitem{bottino2011global}
\textsc{A.~Bottino}, \textsc{T.~Vernay}, \textsc{B.~Scott},
  \textsc{S.~Brunner}, \textsc{R.~Hatzky}, \textsc{S.~Jolliet},
  \textsc{B.~McMillan}, \textsc{T.-M. Tran} and \textsc{L.~Villard}.
\newblock \emph{Global simulations of tokamak microturbulence: finite-$\beta$
  effects and collisions}.
\newblock Plasma Physics and Controlled Fusion \textbf{53}(12), 124027, (2011).

\bibitem{kotschenreuther1995comparison}
\textsc{M.~Kotschenreuther}, \textsc{G.~Rewoldt} and \textsc{W.~Tang}.
\newblock \emph{Comparison of initial value and eigenvalue codes for kinetic
  toroidal plasma instabilities}.
\newblock Computer Physics Communications \textbf{88}(2), 128--140, (1995).

\bibitem{jenko2000electron}
\textsc{F.~Jenko}, \textsc{W.~Dorland}, \textsc{M.~Kotschenreuther} and
  \textsc{B.~Rogers}.
\newblock \emph{Electron temperature gradient driven turbulence}.
\newblock Physics of Plasmas (1994-present) \textbf{7}(5), 1904--1910, (2000).

\bibitem{candy2003eulerian}
\textsc{J.~Candy} and \textsc{R.~Waltz}.
\newblock \emph{An eulerian gyrokinetic-maxwell solver}.
\newblock Journal of Computational Physics \textbf{186}(2), 545--581, (2003).

\bibitem{gorler2011global}
\textsc{T.~G{\"o}rler}, \textsc{X.~Lapillonne}, \textsc{S.~Brunner},
  \textsc{T.~Dannert}, \textsc{F.~Jenko}, \textsc{F.~Merz} and
  \textsc{D.~Told}.
\newblock \emph{The global version of the gyrokinetic turbulence code GENE}.
\newblock Journal of Computational Physics \textbf{230}(18), 7053--7071,
  (2011).

\bibitem{merz2008gyrokinetic}
\textsc{F.~Merz}.
\newblock \emph{Gyrokinetic simulation of multimode plasma turbulence}.
\newblock Ph.D. thesis, Universit{\"a}t M{\"u}nster (2008).

\bibitem{sugama2006collisionless}
\textsc{H.~Sugama} and \textsc{T.-H. Watanabe}.
\newblock \emph{Collisionless damping of zonal flows in helical systems}.
\newblock Physics of Plasmas (1994-present) \textbf{13}(1), 012501, (2006).

\bibitem{peeters2004effect}
\textsc{A.~Peeters} and \textsc{D.~Strintzi}.
\newblock \emph{The effect of a uniform radial electric field on the toroidal
  ion temperature gradient mode}.
\newblock Physics of Plasmas (1994-present) \textbf{11}(8), 3748--3751, (2004).

\bibitem{idomura2008conservative}
\textsc{Y.~Idomura}, \textsc{M.~Ida}, \textsc{T.~Kano}, \textsc{N.~Aiba} and
  \textsc{S.~Tokuda}.
\newblock \emph{Conservative global gyrokinetic toroidal full-f
  five-dimensional Vlasov simulation}.
\newblock Computer Physics Communications \textbf{179}(6), 391--403, (2008).

\bibitem{arnold2007mathematical}
\textsc{V.~I. Arnold}, \textsc{V.~V. Kozlov} and \textsc{A.~I. Neishtadt}.
\newblock \emph{Mathematical aspects of classical and celestial mechanics},
  volume~3,  (Springer Science \& Business Media2007).

\bibitem{littlejohn1983variational}
\textsc{R.~G. Littlejohn}.
\newblock \emph{Variational principles of guiding centre motion}.
\newblock J. Plasma Phys \textbf{29}(1), 111--125, (1983).

\bibitem{cary2009hamiltonian}
\textsc{J.~R. Cary} and \textsc{A.~J. Brizard}.
\newblock \emph{Hamiltonian theory of guiding-center motion}.
\newblock Reviews of modern physics \textbf{81}(2), 693, (2009).

\end{thebibliography}
